\documentclass[aps,prd,preprint,groupedaddress,nofootinbib]{revtex4}

\usepackage{amsmath}

\usepackage{hyperref}
\hypersetup{colorlinks=true,citecolor=blue}

\usepackage{slashed}

\usepackage{graphicx}

\usepackage[]{units}

\usepackage{bbold}

\usepackage{multirow}

\providecommand{\wz}{{\sc Whizard}}
\providecommand{\ii}{\text{i}}
\providecommand{\dd}{\text{d}}
\providecommand{\LL}{\mathcal{L}}
\providecommand{\LLt}[1]{\mathcal{L}_\text{#1}}
\providecommand{\OO}{\mathcal{O}}
\providecommand{\VO}{V_L^\text{off}}

\providecommand{\fierz}{\raisebox{-1ex}{\includegraphics[height=3ex]{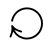}}}

\newcommand{\imgsize}{1.}

\begin{document}

\title{Anomalous top charged-current contact interactions in
single top production at the LHC}

\author{Fabian Bach}
\email[]{fabian.bach@desy.de}
\affiliation{DESY, Notkestra\ss e 85, 22607 Hamburg, Germany}

\author{Thorsten Ohl}
\email[]{ohl@physik.uni-wuerzburg.de}
\affiliation{Institut f\"ur Theoretische Physik und Astrophysik,
Universit\"at W\"urzburg, Hubland Nord, 97074 W\"urzburg, Germany}

\date{\today}

\begin{abstract}
In an effective theory approach, the full minimal set of leading
contributions to
anomalous charged-current top couplings comprises various new trilinear $tbW$
as well as quartic $tbf\! f^\prime$ interaction vertices,
some of which are related to one another by equations of motion.
While much effort in earlier work has gone into the extraction of the
trilinear couplings from single top measurements, we argue in this article
that these structures can be assessed independently by other observables,
while single top production forms a unique window to the four-fermion sector.
An effective theory approach is employed to infer and classify the minimal set
of such couplings from dimension~six operators
in the minimal flavor violation scheme.
In the phenomenological analysis,
we present a Monte~Carlo study
at detector level to quantify the expected performance
of the next LHC run
to bound as well as distinguish the various contact couplings.
Special attention is directed toward differential final
state distributions including detector effects as a means to
optimize the signal sensitivity
as well as the discriminative power with respect to the possible coupling
structures.
\end{abstract}

%\keywords{}

\maketitle

\section{Introduction}\label{intro}

It was with considerable public attention that the two LHC experiments
ATLAS~\cite{Aad2008,Aad2009} and CMS~\cite{Chatrchyan2008}
finally claimed their discovery of a Higgs-like resonance in summer
2012~\cite{Aad2012a,Chatrchyan2012a},
with subsequent investigations ever further backing up the picture
that this new particle might indeed be the standard model (SM) Higgs in its
simplest form~\cite{Chatrchyan2013,Aad2013,Aad2013a}.
However, this result, which at first glance appears as the
ultimate triumph of the theory, might turn out as a mixed blessing because
the more profound puzzle, namely the dynamical mechanism which triggers
electroweak symmetry breaking (EWSB)
and potentially stabilizes the Higgs mass beyond tree level, remains as
obscure as ever in the absence of any new physics (NP) beyond the SM (BSM).
From this point of view, even if no new degrees of freedom are in sight,
NP could manifest itself in slight deviations of SM predictions
in various parameters such as couplings or widths.
Revealing these effects requires two ingredients, namely the
identification and precise measurement of promising observables, but also
a solid theoretical description in order to parametrize their dependence
on NP scenarios in a consistent manner.

Apart from the Higgs particle itself the top sector
represents a natural arena to assess the puzzle of EWSB,
because of the large top mass of the order of the EWSB
scale itself, which has two important consequences:
First, the top has a robust $\OO(1)$ Yukawa coupling to the
symmetry breaking sector, making it a natural probe for its detailed layout.
Second, it is hard to access experimentally, so that its properties
have not been determined to very high precision until now by direct
measurements.  Even if top-Higgs couplings will remain poorly bounded
for quite a while yet, also its electroweak interactions such as
anomalous charged-current (CC) couplings $tbW$ and $tbf\!f^\prime$
may be driven by EWSB details.  These couplings
can already be constrained by direct measurements at the LHC via top decays
as well as single top production.  For example,
top pair production (forming the basis for decay measurements)
has been measured in various channels by the LHC multipurpose experiments
ATLAS~\cite{Aad:2010ey,ATLAS:2012aa,Aad:2012xh} and
CMS~\cite{Chatrchyan:2011ew,Chatrchyan:2011nb,Chatrchyan:2012vs}
with remarkable accuracy, while single top signals are
observed in the dominant $t$~channel production
$bq\to tq^\prime$~\cite{Chatrchyan:2011vp,Aad:2012ux,Chatrchyan1209,
Khachatryan2014}
and in associated $tW$ production
$bg\to tW$~\cite{Aad2012c,Chatrchyan2012,Chatrchyan:2014tua}.
With the spotlight now almost entirely directed at the LHC, it is
particularly remarkable that only recently
the Tevatron experiments were the first to each claim
evidence of an $s$~channel signal above the
$3\sigma$~level~\cite{Abazov2013,Aaltonen2014,CDF:2014uma}.

In view of the
lack of any BSM physics hinting
toward the larger theory, the standard procedure is to take the bottom-up
effective field theory (EFT) approach and to parametrize \emph{any} kind of
NP which might affect LHC observables in a model-independent way.
These effects are encoded in a systematic expansion of the effective
Lagrangian in terms of irrelevant operators of mass dimension~$d>4$ and
corresponding inverse powers of the heavy scale~$\Lambda$,
where NP effects should begin to dominate.
Based on the original classification of all effective
operators parametrizing the leading BSM terms of order $\Lambda^{-1}$ and
$\Lambda^{-2}$ in 1985~\cite{Buchmuller:1985jz},
considerable efforts have gone into the task of finding
an optimal---that is most general and yet minimal and consistent---operator
basis for an anomalous top sector within the effective theory
approach~\cite{Arzt:1994gp,Gounaris:1996vn,Gounaris:1996yp,
Brzezinski:1997av,Whisnant:1997qu,Yang:1997iv,Grzadkowski:2003tf,
AguilarSaavedra:2008zc,AguilarSaavedra:2009mx,AguilarSaavedra:2010zi,
Grzadkowski:2010es}.
At the heart of most of the arguments is the
theorem~\cite{Weinberg:1980wa,Gasser:1983yg,Georgi:1991ch,Rujula:1991se,
Arzt:1993gz}
that the equations of motion (EOM) may safely be applied
at a fixed order in~$\Lambda$, and thus be utilized to identify and eliminate
redundancies in the operator basis, with errors appearing only at higher
orders of~$\Lambda$.\footnote{
This procedure is systematically employed in~\cite{Grzadkowski:2010es}
to remove the remaining redundancies in the original
list~\cite{Buchmuller:1985jz} and present a conclusive operator list.}
However, as has been
pointed out e.g.~in~\cite{Grzadkowski:2003tf,AguilarSaavedra:2008zc},
the application of the EOM relates trilinear $tbW$ to four-fermion
$tbf\! f^\prime$ contact interactions,
which are often considered as independent from each other
in experimental analyses for the sake of simplicity.
Furthermore, EOM relations actually point toward interference terms
among trilinear and quartic couplings at the amplitude level, which are
neglected whenever just one of the interaction structures is considered.
The effects
on the total single top cross section results from adding
all interfering contact structures in a minimal way to the full set of
trilinear $tbW$~couplings have been presented
previously~\cite{Bach2012}.\footnote{
Note that this coupling basis was used recently for an analysis of the first
LHC runs at 7~and $\unit[8]{TeV}$~\cite{Fabbrichesi:2014wva}.}
On the other hand, with the $W$~helicity in the top decay there exist
LHC observables which are exclusively sensitive to the anomalous
helicity-changing trilinear $tbW$~couplings~\cite{Aad2012},
which can be exploited to make single top production a unique window
to measure anomalous four-fermion contact interactions including a top.
Therefore,
in this article we follow a complementary approach in the sense that our
attention is directed toward the full set of leading quartic $tbf\! f^\prime$
couplings
in the \emph{minimal flavor violation} (MFV)
scheme~\cite{Ali1999,Buras2001,DAmbrosio2002},
and including only the one trilinear coupling normalizing the
SM $tbW$ interaction which cannot be fixed by $W$~helicity fractions.

Single top production
might in fact present the only feasible window to assess such anomalous top
CC~contact couplings at the LHC.
Hence, the phenomenological part of this work
will be devoted to a Monte Carlo (MC) study of these final states,
including the modeling of the SM and BSM parts of these processes
at parton level with the leading order MC generator \wz~\cite{Kilian:2007gr}
as well as final state reconstruction at detector level,
using {\sc pythia~6}~\cite{Sjostrand2006} for showering and hadronization
and {\sc Delphes}~\cite{Ovyn2009,Favereau2013} for a fast detector simulation.
We also include leading order (LO) differential distributions of the
final state objects in single top events in order to  bound as well as
distinguish the different kinds of quartic contact interactions.
As a result, we argue that there are various sensitive distributions to
infer the chirality of the production vertex, while further ambiguities
could be resolved by a separate analysis of $s$ and $t$ channel bounds,
where special attention is paid to the crucial role of $b$~tagging
in order to discriminate the two final states at the detector level.

This article is organized as follows: in Sec.~\ref{theo}
we briefly review the effective theory approach and,
starting from the most general set of four-fermion operators as listed
in~\cite{Grzadkowski:2010es}, identify the dominant ones contributing
to anomalous top CC~contact interactions in the MFV scheme.
In Sec.~\ref{pheno} we discuss the LHC~phenomenology of single
top production as a window to assess these new coupling structures,
also including differential distributions of the final state objects at
detector level to discriminate various anomalous contact couplings.
A discussion and summary of the main statements and results
can be found in Sec.~\ref{sum}.

\section{Theoretical setup}\label{theo}

\subsection{Effective field theory}\label{eft}

The effective field theory paradigm~\cite{Weinberg:1980wa,Georgi:1991ch}
is to confront new physics going beyond a given well-tested theory
(e.~g.~the SM) completely unbiased with respect to the details of the
larger theory whose features become dominant at an energy scale~$\Lambda$
considerably above the scales accessible to current experiments.
In this setup, the new heavy degrees of freedom propagate only internally,
while external states are composed of the low energy particle spectrum,
so that the heavy propagators inside the full correlation functions
can be expanded as a power series in $1/\Lambda$.  The resulting pieces can
then be matched order by order in $\Lambda$ onto local irrelevant operators
parametrizing the NP effects at low energies.  Within the bottom-up approach,
one does not construct these operators from a specific UV~completion,
but rather considers \emph{all} operators compatible with the SM symmetries,
leading to an effective, nonrenormalizable Lagrangian
\begin{align}\label{L_eff}
 \LL_{\text{eff}} &= \LL_{\text{SM}}
  + \sum_{d>4,i} \frac{C_i^{(d)}}{\Lambda^{d-4}} O_i^{(d)} + \text{H.c.}\;,
\end{align}
where naively the Wilson coefficients~$C_i$ are expected to be of $\OO(1)$.
However,
it turns out that the operator set straightforwardly constructed out of the SM
fields contains redundancies, which can be removed by application of the
classical EOM~\cite{Buchmuller:1985jz,Grzadkowski:2010es}.
Obviously, the choice of an operator basis is not
unique, and EOM relations further
complicate the choice of a minimal basis.

Apart from this formal procedure supported by the equivalence
theorem~\cite{Georgi:1991ch,Arzt:1993gz},
there are some more heuristic arguments to introduce further hierarchies
among the $C_i$, which may have to be considered once a phenomenological
study is afflicted by an unmanageable number of free parameters.
For instance, a popular ordering principle is
MFV~\cite{Ali1999,Buras2001,DAmbrosio2002},
imposing the flavor structure of the SM on any NP operator containing
fermion fields.  This is phenomenologically well motivated by
flavor observables confirming the CKM structure to a very high precision,
and thus driving many nonminimal flavor changing NP effects to
$\Lambda\sim\OO(10\text{--}\unit[100]{TeV})$ and above.
Technically, one postulates a global flavor symmetry
(concentrating on the quarks here)
\begin{align}\label{GF}
 G_F &\sim \mathbf{SU}(3)_{q_L} \otimes \mathbf{SU}(3)_{u_R}
  \otimes \mathbf{SU}(3)_{d_R}
\end{align}
under which the SM gauge representations of quark fields, namely the
left-handed doublet $q_L$ as well as the right-handed singlets $u_R$
and $d_R$ are separately charged,
and which is
assumed to be
broken \emph{only} by the SM Yukawa matrices $Y_{u,d}$
even in the NP contributions.
This is achieved by promoting the $Y_i$ to spurion fields
with the usual Yukawa couplings as vacuum expectation values,
where the spurion
representation under~\eqref{GF} can be read off from the SM Yukawa terms.
NP operators are then made invariant by inserting the minimal
number of $Y_i$ required by the specific fermion content.
After rotating to the mass basis, all flavor indices are contracted with
products of the CKM matrix~$V$ and diagonal mass
matrices~$M_{u,d}$.  This way, charged currents are still governed by~$V$,
and flavor changing neutral currents (FCNC)
explicitly remain suppressed by the GIM mechanism~\cite{Glashow1970}.

\subsection{Operator basis}\label{op_basis}

Homing in now on anomalous top contact couplings,
the leading contributing operators
are of mass dimension $d=6$, for which an exhaustive and minimal list was
presented in~\cite{Grzadkowski:2010es}.
Concerning the appropriate basis of four-fermion operators for our analysis,
we will hence refer to this list as well as to the one given
in~\cite{AguilarSaavedra:2010zi} concentrating on four-fermion operators.
Both operator bases are completely equivalent and minimal in the sense that
they exhaust all possibilities to combine Fierz reorderings and completeness
relations of the $\mathbf{SU}(3)_C$ and $\mathbf{SU}(2)_L$ gauge group
generators to eliminate redundant operators, however without assuming
further structure such as MFV.
Rather than repeating here both versions of the complete list of 11
$B$-conserving
operators potentially relevant for single top $\Delta T=1$ transitions
(neglecting all quark-lepton operators as they are not important for
LHC~production and heavily suppressed in the
decay~\cite{AguilarSaavedra:2010zi}),
we simply give here the classification principle in terms of SM quantum
numbers, so that looking them up
in~\cite{AguilarSaavedra:2010zi,Grzadkowski:2010es} is straightforward:
the four fermion fields can be arranged into two bilinears of definite
Lorentz and gauge quantum numbers contracted with each other,
where hypercharge singlets come as Lorentz vectors of definite chirality
$(\overline LL)(\overline LL)$, $(\overline RR)(\overline RR)$ and
$(\overline LL)(\overline RR)$ while hypercharged bilinears are arranged
into products of chirality flipping scalars $(\overline LR)(\overline RL)$
and $(\overline LR)(\overline LR)$, cf.~Table~\ref{mfv_coeffs}.
The two differences between~\cite{AguilarSaavedra:2010zi}
and~\cite{Grzadkowski:2010es} are as follows:
\begin{enumerate}

\item The parametrization of the additional gauge group structure,
namely which of the two terms on the right-hand side of the $\mathbf{SU}(N)$
completeness relation
\begin{align}\label{completeness}
 \frac{1}{N} \delta_{12}\delta_{34} &= \delta_{14}\delta_{32}
  -2\left(T^a\right)_{12}\left(T^a\right)_{34}
\end{align}
is dropped in favor of the other two.
The generators $T^a$ are summed over in the adjoint
representation and the subscript numbers label the fundamental group indices
of the four fermion field slots within any operator.

\item The choice of dropping either
$(\overline LL)(\overline RR)$~\cite{AguilarSaavedra:2010zi}
or $(\overline LR)(\overline RL)$~\cite{Grzadkowski:2010es},
as both versions result from each other via Fierz rearrangements.
\end{enumerate}
At this point, an important remark must be made with respect to the
present analysis, which uses the MFV~structure to impose a hierarchy
on the operator basis and then
studies kinematical distributions of single top production
final states:
Fierz reordering generally does not commute with the application of the MFV
scheme.
MFV is sensitive to the chiral representation of the fermions in each
of the two bilinears forming any four-fermion operator, while Fierz
reorderings obviously interchange them.
For instance, the vector currents of mixed chirality
$(\overline LL)(\overline RR)$
are transformed into a superposition of scalars and pseudoscalars
$(\overline LR)(\overline RL)$, which receive MFV~weights
different from
the vectors.
Similarly, the $(\overline RR)(\overline RR)$
operators $O_{ud^{(\prime)}}$ in~\cite{AguilarSaavedra:2010zi} (purely NC)
could be Fierz'ed into CC versions, which
again receive different MFV~prefactors.
Of course, it is by construction of the MFV scheme that
FCNC transitions are much more suppressed than the CC ones,
so rather than
simply picking just one list out of~\cite{AguilarSaavedra:2010zi}
or~\cite{Grzadkowski:2010es}, we classify \emph{all} possibilities to obtain
NC or CC $\Delta T=1$ flavor transitions in the mass eigenstates out of the
underlying operators at hand,
which are formulated in terms of the weak eigenstates.
In this way, the Fierz ambiguity, which is a mere consequence
of the ignorance about the explicit realization of the heavy underlying physics
rather than a fundamental ordering principle, does not affect the actual
classification in terms of MFV coefficients. All operators for which a
potentially less suppressed CC version exists are thus being accounted for.

\begin{table}%[H] add [H] placement to break table across pages
\begin{tabular}{|rl|c|cl|cl|}
 \hline
 \multicolumn{3}{|c|}{four-fermion operators} & \multicolumn{4}{|c|}{MFV factors} \\
 \multicolumn{2}{|c|}{chirality} & spin & \multicolumn{2}{|c|}{CC} & \multicolumn{2}{|c|}{NC} \\
 \hline
  $\left( \overline{L}L \right)\left( \overline{L}L \right)$ & \fierz
   & vector & 1                  &           & $y_b^2 V_{3i}$         & $\sim 10^{-6}$  \\
  $\left( \overline{L}R \right)\left( \overline{L}R \right)$ & \fierz
   & scalar & $y_s$ or $y_b y_c$ & $\sim 10^{-3}$ & $y_{c,s} y_b^2 V_{3i}$ & $\lesssim 10^{-8}$ \\
  $\left( \overline{L}R \right)\left( \overline{R}L \right)$ &
  \multirow{2}{*}{$\;$\raisebox{-1.3ex}{\includegraphics[height=4ex]{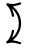}}}
   & scalar & $y_s$ or $y_b y_c$ & $\sim 10^{-3}$ & $y_{c,s} y_b^2 V_{3i}$ & $\lesssim 10^{-8}$ \\
  $\left( \overline{L}L \right)\left( \overline{R}R \right)$ &
   & vector & ---                &        & $y_b^2 V_{3i}$         & $\sim 10^{-6}$  \\
% \hline
  $\left( \overline{R}R \right)\left( \overline{R}R \right)$ & \fierz
   & vector & $y_b y_c y_s$      & $\sim 10^{-6}$ & $y_b^2 V_{3i}$         & $\sim 10^{-6}$  \\
 \hline
\end{tabular}
\caption{MFV suppression of
$\Delta T=1$ processes
from four-fermion operators with spinors in the various SM representations
$q_L$ (``$L$'') and $u_R$, respecitvely, $d_R$ (``$R$'').
The $y_i$ are Yukawa couplings, and $V_{3i}$ denotes off-diagonal CKM matrix
elements with $i\neq 3$ (setting $y_t\sim V_{tb}\sim 1$).
Fierz relations among operators are indicated by the arrows.
\label{mfv_coeffs}}
\end{table}

The result is displayed in Table~\ref{mfv_coeffs},
with operators sorted in decreasing order of relevance from top to bottom rows,
and Fierz relations indicated by the arrows.
Starting with the CC~column, beneath the $\OO(1)$
left-handed vector transition in the top row which has the
same quantum numbers as the SM~$W$ exchange (a heavy Fermi interaction),
one finds the next-to-leading scalar CC~transitions
which are suppressed by a single light Yukawa
coupling, $y_s$ or $y_b y_c$, each of $\OO(10^{-3})$.
The right-handed vector CC~transition (bottom row) collects at least
two light Yukawa insertions~$y_b y_c y_s\sim\OO(10^{-6})$.
Moving to the NC~column, the general expectation is confirmed
that the GIM mechanism remains active as any $\Delta T=1$ transition
gets suppressed at least by a factor
$(VM_d^2V^\dagger)_{3i}\sim y_b^2V_{3i}\sim\OO(10^{-6})$,
which is driven by the down-type mass splitting $\Delta y_d\sim y_b$.
Note that the right-handed vector CC~transition is numerically of the same
MFV~order as the leading NC~transitions.
Using this MFV hierarchy, one must choose
how far to go down in the MFV~order for the analysis,
where the leading order is made up of merely one NP~parameter,
namely the real $(\overline LL)(\overline LL)$ operator coefficient.
While focusing on this left-handed vector transition might be
natural as a first step in an actual experimental analysis,
we include here also the subleading scalars, because
it broadens the scope of NP~effects in the observables
at the cost of only two more real NP~parameters (without CP~violation),
as will be shown below in Sec.~\ref{cc_contact}.
Besides, as already addressed
in~\cite{Cao:2007ea,Bach2012} and also discussed in more detail
in Sec.~\ref{pheno}, the left-handed vector has a
unique interference pattern with the SM piece,
so including the scalar couplings
adds the leading noninterfering directions to the NP~parameter space.

To summarize, in our basis we will consider the dominant vector operator
$(\overline LL)(\overline LL)$ along with the scalar
operators $(\overline LR)(\overline RL)$ and $(\overline LR)(\overline LR)$.
The corresponding operators read
\begin{subequations}\label{op_4f}
\begin{align}
 (\overline LL)(\overline LL):&&
 O_{qq}^{(ijkl)} &= \left( \bar q_{Li}\gamma^\mu\tau^I q_{LJ} \right)
  \left( \bar q_{Lk}\gamma_\mu\tau^I q_{Ll} \right)\;, \label{Oqq} \\
 (\overline LR)(\overline RL):&&
 O_{qu}^{(ijkl)(\prime)} &= \left( \bar q_{Li} u_{Rj} \right)
  \left( \bar u_{Rk} q_{Ll} \right)\;, \label{Oqu} \\
 (\overline LR)(\overline RL):&&
 O_{qd}^{(ijkl)(\prime)} &= \left( \bar q_{Li} d_{Rj} \right)
  \left( \bar d_{Rk} q_{Ll} \right)\;, \label{Oqd} \\
 (\overline LR)(\overline LR):&&
 O_{quqd}^{(ijkl)(\prime)} &= \left( \bar q_{Li}^p u_{Rj} \right)\epsilon_{pr}
  \left( \bar q_{Lk}^r d_{Rl} \right)\;, \label{Oquqd}
\end{align}
\end{subequations}
where $\tau^I$ are the three $\mathbf{SU}(2)_L$ generators, $ijkl$ are
flavor labels in the mass basis, and $\epsilon_{pr}$ is the antisymmetric
tensor with fundamental $\mathbf{SU}(2)_L$ indices.
The primed versions of the scalar operators~\eqref{Oqu}--\eqref{Oquqd}
look exactly the same, but with additional $\mathbf{SU}(3)_C$
generators~$\lambda^a$ inserted into each bilinear,
or equivalently twisted color flows among the
fermions, according to Eq.~\eqref{completeness}.
As will be further clarified below in Sec.~\ref{pheno}, the following study
builds on the fact that the trilinear $tbW$ couplings $V_R$ and $g_{L,R}$
can be fixed independently of the single top production.
Conversely, the normalization of the SM vertex $V_L$ remains unbounded,
and is therefore included along with the contact couplings here.
The respective Hermitian $d=6$ operator generating this term
reads~\cite{AguilarSaavedra:2008zc,AguilarSaavedra:2009mx}
\begin{align}\label{Opq3}
 O_{\phi q}^{(3,ij)} &=
  \big( \phi^\dagger \ii\overleftrightarrow{D}_{\!\mu}^I \phi \big)
  \big( \bar{q}_{Li} \gamma^\mu \tau^I q_{Lj} \big)\,,
\end{align}
where $\overleftrightarrow{D}^I_{\!\mu}\equiv\tau^I D_\mu-\overleftarrow{D}_{\!\mu}\tau^I$
with the second derivative acting on the left.\footnote{
The operator basis~\eqref{Oqq} and~\eqref{Opq3}
generating the vector couplings is not unique because there is an EOM relation
connecting them to a third operator,
called $O_{qW}^{ij}$ in~\cite{AguilarSaavedra:2008zc},
but  we choose to eliminate this one,
as advocated in~\cite{AguilarSaavedra:2008zc}.}

In the next section, we will analyze the operator basis given by
Eqs.~\eqref{op_4f} and~\eqref{Opq3} to find the
corresponding interaction terms in the Lagrangian.

\subsection{Charged-current contact interactions}\label{cc_contact}

With the list of $d=6$ operators in Eq.~\eqref{op_4f}, the
interaction part of the effective Lagrangian is found by extracting the
$\Delta T=1$ four-fermion transitions,
\begin{subequations}\label{L_cc}
\begin{align}
 \Delta\LL_{\text{CC},4f} =& \frac{1}{\Lambda^2} \Big[
    V^{4f}_{L}  \big( \bar b \gamma_\mu P_L t \big)
             \big( \bar u_k \gamma^\mu P_L d_k \big)
   +\text{h.c.} \label{L_V} \\
  &\quad+\sum_{A,B=L,R}\left( S_{AB} \big( \bar b P_A t \big)
         \big( \bar u_k P_B d_k \big)
     +S_{AB}^\prime \big( \bar b \lambda^a P_A t \big)
         \big( \bar u_k \lambda^a P_B d_k \big) \right)
     +\text{h.c.} \Big] \label{L_S}
\end{align}
\end{subequations}
summing over light flavors~$k$ for simplicity\footnote{
Of course, there might be an additional hierarchy between the first and
second generation, according to the MFV classification, which then interplays
with respective proton pdf suppressions depending on the production channel,
but resolving this would
require sensitivity to the light jet flavor in single top $t$~channel
production, which is clearly not feasible.}
and keeping possible MFV factors implicit
in the scalar couplings~$S_{AB}^{(\prime)}$.

This leaves us with one vector and eight scalar couplings in total where,
as has been discussed e.~g.~in~\cite{Cao:2007ea,Bach2012},
$V_L^{4f}$ stands out as the only structure interfering with the SM piece of
the amplitude, which was precisely the reason to include it also in the
parameter space of Ref.~\cite{Bach2012}. In addition,
it turns out that there is no sizable interference direction among any of
the remaining contact couplings, because generally
the vectors (including the SM piece) do not interfere with any scalars, and
likewise the scalar-scalar interferences either vanish due to different
color structures, or are negligible compared to
the squared part because one
always collects at least two additional mass insertions and hence at least a
suppression $y_b\sim\OO(10^{-2})$, or much more from the light
lines.
Furthermore, with the lack of any remaining interference term, the squared
$2\to 2$ production matrix elements $\propto S_{AB}^2$, respectively,
$\propto S^{\prime 2}_{AB}$ are identical up to a global color factor.
Analytically, averaging over initial and summing over final state colors,
this factor amounts to
\begin{align}\label{cc_scalar_color}
 A_C = \left(\frac{1}{3}\right)^2\text{tr}[\lambda^a \lambda^b]\,
  \text{tr}[\lambda^a \lambda^b]
 &= \frac{1}{9}\cdot\frac{1}{4}\cdot \text{tr}\,\mathbb{1}_{adj}
  = \frac{2}{9}\;.
\end{align}
It follows that the scalar directions $S_{AB}$ and $S^\prime_{AB}$ are
kinematically degenerate because of identical differential matrix elements
(up to negligible effects from the decay vertex),
so in our approach based
on binned distributions, one may just as well drop the primes and
work in the singletlike normalization for the scalar couplings from now on.
Respective bounds can be interchanged to the octet normalization simply by
multiplying a factor of~$\sqrt{A_C}$, cf.~Eq.~\eqref{cc_scalar_color}.
Moreover, with absent scalar interferences one also loses any sensitivity to
the helicity in the light fermion line, which makes $S_{LL}$
and $S_{LR}$, respectively, $L\leftrightarrow R$ combination degenerate as well
with respect to final state distributions.

This eventually leads us to the parametrization of single top
charged-current contact interactions at the Lagrangian level,
in its final form for this analysis:
\begin{subequations}\label{L_pheno}
\begin{align}
 \Delta\LLt{CC} =
  & \left( V_L + \frac{q^2-m_W^2}{m_W^2}\VO \right)
    \left(\bar b \gamma^\mu P_L t \right)\,W_\mu^-
   +\text{h.c.} \label{L_vector} \\
  &+\frac{1}{\Lambda^2} \Big[
    S_L \big( \bar b P_L t \big) \big( \bar u_k \Gamma d_k \big)
   +S_R \big( \bar b P_R t \big) \big( \bar u_k \Gamma^\prime d_k \big)
   +\text{H.c.} \Big] \label{L_scalar}
\end{align}
\end{subequations}
with $V_L\simeq1$ and all other couplings vanishing in the SM,
where $\Gamma^{(\prime)}$ is any normalized superposition of
$\mathbb{1}$ and $\gamma_5$,
controlled by the relative admixture of the various operators.
We explicitly keep the $\VO$~normalization of~\cite{Bach2012} here
for the vector coupling
in order to facilitate the comparison
with the results thereof, and highlighting again the fact that this particular
contact interaction is related to a trilinear $tbW$ coupling via
the EOM.
The respective coupling relation with Eq.~\eqref{L_V} reads
\begin{align}
 \VO &= \frac{\upsilon^2}{2\Lambda^2} V_L^{4f}\;.
\end{align}
The scalar couplings $S_{L,R}$ in Eq.~\eqref{L_scalar} are
normalized by $\Lambda=\unit[1]{TeV}$, so any respective numerical values
quoted later can be understood as being multiplied
by~$(\Lambda/\text{TeV})^2$.
Finally, the mapping of the couplings in Eq.~\eqref{L_pheno} onto
operator coefficients of Eq.~\eqref{op_4f} is
\begin{subequations}\label{op_cpl}
\begin{align}
 \delta V_L &= C_{\phi q}^{(3,33)}\frac{\upsilon^2}{\Lambda^2}\;,\label{VL} \\
 \VO &= 2 C_{qq}^{(33kk)} \frac{\upsilon^2}{\Lambda^2}\;, \label{VO} \\
 S_L &= \alpha\,C_{quqd}^{(33kk)*}
  + \beta\,C_{qd}^{(33kk)*} \;, \label{SL} \\
 S_R &= \alpha\,C_{quqd}^{(kk33)} + \beta\,C_{qu}^{(33kk)} \;, \label{SR}
\end{align}
\end{subequations}
where all  vector couplings in~\eqref{VL} and \eqref{VO} are
real valued because the generating operators are hermitian.
The coefficients $\alpha$ and $\beta$ in \eqref{SL} and \eqref{SR} are
arbitrary mixing factors, resolved only by
measuring the light fermion helicity
(the corresponding octet versions with primed coefficients are also possible
for the scalars).

\section{LHC phenomenology}\label{pheno}

As already mentioned in Sec.~\ref{op_basis}, top charged-current interactions
with $\Delta T=1$ are generally parametrized within the EFT approach by
a set of four trilinear $tbW$ couplings $V_{L,R}$ and $g_{L,R}$ plus
$tbf\!f^\prime$ contact interactions, namely $\VO$ and $S_{L,R}$
in our basis, cf.~\eqref{L_pheno}.
At the LHC, there are two main classes of observables which are directly
sensitive to such electroweak BSM
contributions, i.e.~single top production and top decay properties.
Since the contact interactions are heavily suppressed in the
latter~\cite{AguilarSaavedra:2010zi}, we will argue now that the $W$~helicity
fractions of the top decay provide a clean handle on the trilinear
subset $V_R$, $g_L$ and $g_R$---indeed, respective limits have already
been published~\cite{Aad2012}.
On the other hand, single top cross sections do receive sizable contributions
from both trilinear and contact interactions stemming from the
production insertion~\cite{Cao:2007ea,AguilarSaavedra:2008gt,Bach2012}.
So rather than considering single top cross sections as additional
observables for the trilinear couplings, one could also employ top decay
results to \emph{fix} those beforehand\footnote{
Nonzero values would add to the reference offset of the single top cross
sections and also change the spin analyzers discussed below,
but in the absence of any experimental hint toward NP in
these couplings, we just use the SM values $V_R=g_{L,R}=0$ as
a~reference point.}
and then use the single top channels to
cleanly constrain the four-fermion couplings, as parametrized in
Eq.~\eqref{L_pheno}:  this is the idea of the study
to be presented in this section.

\subsection{Technical setup}\label{pheno_contact_setup}

\begin{figure*}
 \includegraphics[trim = 99mm 234mm 92mm 41mm, clip, scale=1]{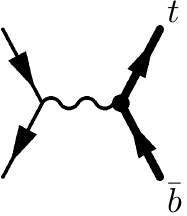}
 \hspace{2.cm}
 \includegraphics[trim = 64mm 234mm 56mm 41mm, clip, scale=1]{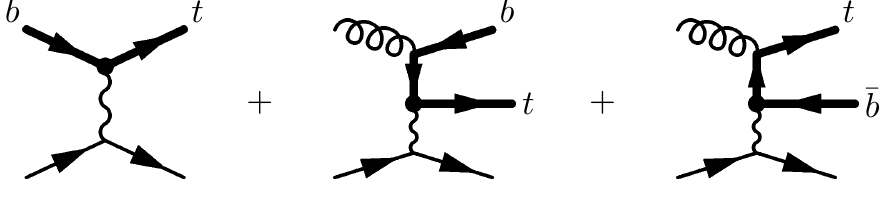}
 \vspace{1.cm} \\
 \includegraphics[trim = 99mm 234mm 92mm 41mm, clip, scale=1]{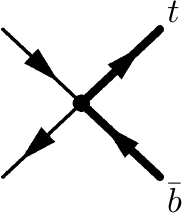}
 \hspace{2.cm}
 \includegraphics[trim = 64mm 234mm 56mm 41mm, clip, scale=1]{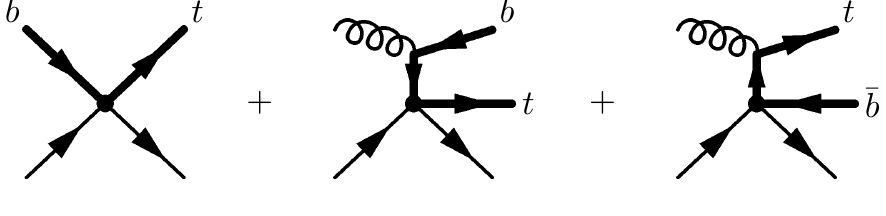}
 \caption{
Tree level diagrams contributing to on-shell single top production in the
$s$~channel (left column) and in the $t$~channel (right column):
SM pieces (top row) and anomalous $tbf\!f^\prime$~contact insertions
(bottom row).
\label{diags}}
\end{figure*}

In order to assess the contact interactions, we make use of the kinematics
which can be expected to be quite different from the SM because of the
absence of the $W$~propagator. Therefore, rather than considering just the
total cross sections as observables,
we perform a binned likelihood analysis
over various distributions which are either sensitive to the general admixture
of contact interactions or disambiguate the various anomalous directions
contained in Eq.~\eqref{L_pheno}.
As mentioned before,
and in sharp contrast to the situation with trilinear
couplings~\cite{Bach2012},
the effect on the top decay insertion is in this case negligible already at
the amplitude level.
This simplifies the parametric dependence on $\vec g=\{V_L,\VO,S_L,S_R\}$,
because a simple quadratic form in the
single
top production insertions (cf.~Fig.~\ref{diags}) now manifestly
holds even with full matrix elements including top production and decay
as well as acceptance cuts on the final states.
The parton level differential cross section in each production
channel---namely the $s$ or $t$~channel to be sensitive on $tbf\!f^\prime$
insertions---is then given as a function
of $\vec g$ as
\begin{align}\label{ds_g}
 \dd\sigma(\vec g,\Phi) &= \sigma_\text{SM}
  \sum_{i,j} \dd\kappa_{ij}(\Phi)\,g_i\,g_j
\end{align}
with LO differential cross section templates
$\dd\kappa_{ij}(\Phi)$ as a function of the phase space point~$\Phi$,
which are normalized to the total SM cross section~$\sigma_\text{SM}$.
Neglecting the tiny $\vec g$-constant irreducible backgrounds,
Eq.~\eqref{ds_g} contains just five independent directions,
i.e.~four squared ones $\dd\kappa_{ij}=\delta_{ij} \dd\hat\kappa_i$
plus the $V_L$-$\VO$~interference (cf.~Sec.~\ref{cc_contact}).

The analysis strategy is to produce parton level event samples
corresponding to all five parameter directions, identify kinematic
distributions which can discriminate either the contact couplings from
the SM part or several different contact couplings from each other, and
perform a binned likelihood test on the $\dd\kappa_{ij}(\Phi)$ templates
to derive bounds on~$\vec g$.
This is done as a first step at parton level in Sec.~\ref{pheno_p}
to discuss the observables and
compare the sensitivities of the different production channels.
In Sec.~\ref{pheno_d},
we will then include detector effects accounting also for the
relative admixture of the partonic processes in the final state selection.
The aim is to describe differential cross sections at the detector level as
\begin{align}\label{ds_det}
 \dd\sigma_{\text{det},i}(\vec g,\Phi) &=
  \sum_j \epsilon_{ij}(\Phi)\,\dd\sigma_j(\vec g,\Phi)\;,\qquad i,j=s,t,
\end{align}
where $\epsilon_{ij}(\Phi)$ is the detector efficiency matrix to be determined,
and $\dd\sigma_i(\vec g,\Phi)$ is given by Eq.~\eqref{ds_g} in each
partonic channel.
To that end, binned detector responses are inferred for $s$~and
$t$~channel selection along each (one-dimensional) kinematic direction
in~$\Phi$ considered
in the likelihood function with $\vec g$ set to the SM.
Each bin entering the analysis can then be mapped onto the detector level by
multiplying the number of parton level entries with the respective
mean efficiency matrix~$\epsilon_{ij}$ inferred for this bin
covering a region $\left[\Phi,\Phi+\Delta\Phi\right]$.
Note that this entails
leaving the spectator jet (a~$b$~quark in the $s$~channel, respectively,
a light quark
in the $t$~channel) untagged at parton level in order to remain inclusive
for the imperfect $b$-tagging performance of the detector
(the point will be taken up in more detail in Sec.~\ref{pheno_d}).

At parton level, the production channels are specified by their respective
final states, namely (assuming leptonic top decay) $\ell\nu bb$ in the
$s$~channel and $\ell\nu bj(b)$ in the $t$~channel, with $\ell=e,\mu$
and a light quark jet~$j$. As indicated by the optional second $b$ jet in the
$t$~channel, respective samples are obtained by matching the LO $2\to 2$
process to the $2\to 3$ process with a gluon
splitting~\cite{Sullivan:2004ie,Boos:2006af}.
Technically, this is done in this case
by subtracting the LO contribution to the $g\to bb$ splitting from the
$b$~pdf to be convoluted with the $2\to 2$ matrix
element~\cite{AguilarSaavedra:2008gt},
a procedure automatically provided by \wz~\cite{Kilian:2007gr}
once it is linked against {\sc Hoppet}~\cite{Salam2009}.
In order to produce samples which sufficiently populate the relevant
signal regions in phase space, we apply basic acceptance cuts already on
the hard partonic matrix elements.  As any channel specific selection
must go into $\epsilon$ later on, cf.~\cite{Bach2012},
these cuts should be completely inclusive
with respect to the production channels:
\begin{subequations}\label{cuts_p}
\begin{align}
   p_T\left(\ell,\nu\right) > \unit[25]{GeV}\quad
  & \text{and}\quad \left| \eta\left(\ell\right) \right|<3\,,\label{phi_l} \\
   p_T\left(j,b\right) > \unit[30]{GeV}\quad
  & \text{and}\quad \left| \eta\left(j,b\right) \right|<3\,,\label{phi_j} \\
   \unit[150]{GeV} < m_{b\ell\nu} < \unit[225]{GeV}\quad
  & \text{and}\quad \sqrt s > \unit[400]{GeV}\,, \label{phi_m}
\end{align}
\end{subequations}
where
Eq.~\eqref{phi_j} is required for only one of the two $b$s in the $2\to 3$
process to be inclusive.
The observable $\sqrt s$ here is the invariant mass of the reconstructed top
and the hardest
spectator jet in the event ($b$~tagged or not).
This together with the rather tight hadronic $|\eta|$ cut in~\eqref{phi_j}
accounts for
the altered kinematics of the contact diagrams,
which is expected to be shifted toward the high energy tails and hence also
more central compared to the SM especially
in the $t$~channel because of the absent propagator (there is not necessarily
a distinct ``forward tagging jet'' any more in the NP event topology).
These two cuts in particular increase the sensitivity to the NP~contributions
with respect to the SM single top contribution $\sim\delta V_L$,
but in a more involved study addressing the
reducible backgrounds in more detail, it must be checked to what extent
the forward tagging can be relaxed without spoiling the
$t$~channel signal altogether.

With the stated acceptance cuts,
\wz\ is employed to produce LHC samples at
$\unit[14]{TeV}$
(using CTEQ6L1 proton pdfs~\cite{Pumplin:2002vw})
with $10^6$~events per
production channel and charge state for each NP direction,
respecitvely, $2\times 10^6$~events per channel,
lepton flavor ($\ell=e,\mu$) and charge
for the SM part
(statistics is increased for the SM samples because
they form the basis for the detector response discussed later on).
The squared NP~directions are produced by setting only one of the couplings
in~$\vec g$ to unit at the top production vertex
while keeping the dominant SM contribution to the decay.
The one remaining interference direction is obtained from a fifth sample
where both $V_L$ and $\VO$ are set to unit.
These samples then form the templates for normalized differential
matrix element shapes~$\dd\kappa_{ij}(\Phi)$,
while the integrated cross sections of the SM samples ($\sim V_L^2$) also
convey the normalizations $\sigma_\text{SM}$ for Eq.~\eqref{ds_g} at LO
including partonic acceptance efficiencies~\eqref{cuts_p}.
They get multiplied by channel-specific $K$~factors to
account for higher order QCD corrections, where we employ NNLL~numbers
from~\cite{Kidonakis:2010tc,Kidonakis:2011wy,Kidonakis:2010ux}.
Particularly, NLO
results~\cite{Sullivan:2004ie,Boos:2006af,Falgari:2011qa,Campbell:2012uf}
indicate that single top differential distributions in the $s$ and
$t$~channels are only marginally distorted by QCD effects,
at the~$\OO(\unit[1]{\%})$ level,
and can thus be readily accounted for by overall $K$~factors.
Conversely, sizable distortions of event shapes, as produced by the
contact interactions, do present a robust sign of NP, and suffer from fewer
theoretical uncertainties than total cross sections.

For the binned likelihood test, we introduce a $\chi^2$~function
\begin{align}\label{chi2_cc_contact}
 \chi^2(\vec g) &= \sum_i \left(
 \frac{w^\text{exp}_i -w^\text{th}_i(\vec g)}{\delta_i} \right)^2
\end{align}
with the sum running over all bins of all \emph{normalized} histograms
included in the likelihood test.\footnote{
Obviously, total cross sections must still be included to constrain the
SM normalization $\sim V_L$, which does not distort the shapes;
respective sensitivities are carried over
from~\cite{AguilarSaavedra:2008gt,Bach2012}.}
The combined statistical and systematic uncertainty~$\delta_i$ of each
bin weight is
\begin{align}\label{bin_error}
 \delta_i &= \sqrt{ \frac{1}{N}\left( w_i +w_i^2 \right)
  +\delta_\text{sys}^2\,w_i^2 } \;,
\end{align}
with the total number of events in a given final state
$N=\int\! L\cdot\sigma_\text{tot}$ normalized to
$\int\! L=\unit[100]{fb^{-1}}$, and a tentative
systematic error assumed to be $\delta_\text{sys}=\unit[3]{\%}$
(the number is varied in the Appendix to estimate the impact on the bounds
obtained later).
The higher order dependence of the statistical error on~$w_i$ in
Eq.~\eqref{bin_error} simply comes
from the fact that we are considering normalized bin weights rather than
absolute entries in order to reduce the systematic uncertainty.
The sample sizes stated above are
chosen such that the MC~statistics is at least 10~times the
expected~$N_\text{exp}$,
while the binning is adjusted such that each bin contains
at least $\OO(10)$~events at the SM point so that the Poisson
statistics is well approximated by the Gaussian assumed in the
$\chi^2$~function.\footnote{
This criterion is not always strictly fulfilled, particularly in the small
$s$~channel, but it was checked that the results obtained are stable against
rebinnings and also against replacement of Eq.~\eqref{chi2_cc_contact} by the
full Poisson statistics.}
The $1\sigma$~limits on the couplings are then inferred by scanning the
parameter space~$\vec g$ and accepting all points within the respective
$\Delta\chi^2$ interval, assuming an experimental confirmation of the SM
with a typical minimum at $\chi^2_\text{min}=1$.

\subsection{Partonic level}\label{pheno_p}

At parton level, it is at first interesting to examine the sensitivities of
each production channel to the various couplings.
To that end, we examine several kinematic distributions of the final state
objects. Specifically, in both channels
one-dimensional distributions
of the following nine observables
are considered:
\begin{equation}
\label{dist_obs}
 \sqrt s,\; \slashed E_T,\; p_T(b),\; |\eta(b)|,\; p_T(\ell),\;
 |\eta(\ell)|,\; p_T(j),\; |\eta(j)|,\; \cos\theta_\ell
\end{equation}
with $\sqrt s\equiv m_{tj}$ as defined in Eq.~\eqref{cuts_p}.
Note that the label $b$ only refers to the $b$~jet reconstructing the top,
while $j$ denotes the hardest spectator jet in the event, irrespective
of $b$ flavor, in order to stay inclusive with respect to channel impurity
at detector level; cf.~details in Sec.~\ref{pheno_d}.
The top spin analyzer angle $\theta_\ell$ is the angle between the
charged lepton and the hardest spectator jet in the top rest
frame~\cite{Jezabek1994,AguilarSaavedra:2006fy,Aguilar-Saavedra2010}.
The cross section follows the distribution
\begin{align}
 \frac{1}{\sigma} \frac{\dd\sigma}{\dd\cos\theta_X} &=
  \frac{1}{2} \left( 1 +\rho_z \alpha_X \cos\theta_X \right)
\end{align}
with top spin analyzers $\alpha_X(V_R,g_L,g_R)$, $X=b,\ell,\nu$,
and the top polarization $\rho_z$ along an arbitrary axis~$z$.
Within the SM, $\alpha_\ell=1(0.998)$ at LO~\cite{Jezabek1994a}
(NLO~\cite{Czarnecki1995,Brandenburg2002,Bernreuther2004})
in~$\alpha_s$ ($X=b,\nu$ being less sensitive, and harder
to handle experimentally), and the
polarization is $\rho\gtrsim 0.9$ if the spectator jet direction
is employed as reference axis~$z$ in the top
frame~\cite{Mahlon2000}.
This is due to the left-handed production vertex~$\sim V_L$, so
once $\alpha_\ell$ is fixed by $W$~helicities, $\cos\theta_\ell$
is cleanly sensitive to $\rho_z$ and hence the top production
polarization.

As shown in Fig.~\ref{cc_contact_tb_part} for the $s$~channel,
respecitvely, \ref{cc_contact_tj_part} for the $t$~channel,
there are many distributions [most prominently $\sqrt s$ and $p_T(j)$,
but also $|\eta(j)|$] that
are very sensitive to anomalous contact contributions in general but mostly
blind to their relative admixture.  The former is no surprise given the very
different energy scaling of the contact terms with respect to the SM piece.
However, the spin analyzer $\cos\theta_\ell$ but also $p_T(\ell)$ indeed
turn out to be discriminative observables in both channels for an anomalous
right-handed top production mode, as parametrized here by~$S_R$.
On the other hand, it should be much harder
to tell the anomalous left-handed couplings $\VO$ and $S_L$ apart
from normalized matrix element shapes only,
with the most promising observables being the pseudorapidity distributions
$|\eta(b)|$ and $|\eta(j)|$ in both channels.
As a side remark, note that the $|\eta(j)|$ histogram in
Fig.~\ref{cc_contact_tj_part} (bottom left)
justifies the relaxed forward tagging of the
spectator jet in the $t$~channel introduced before,
as restricting to the SM-like forward phase space $|\eta|\gtrsim 2.5$
would indeed kill most of the signal one is looking for.

\renewcommand{\imgsize}{0.4}
\begin{figure*}
% \vspace{-2mm}
 \includegraphics[scale=\imgsize]{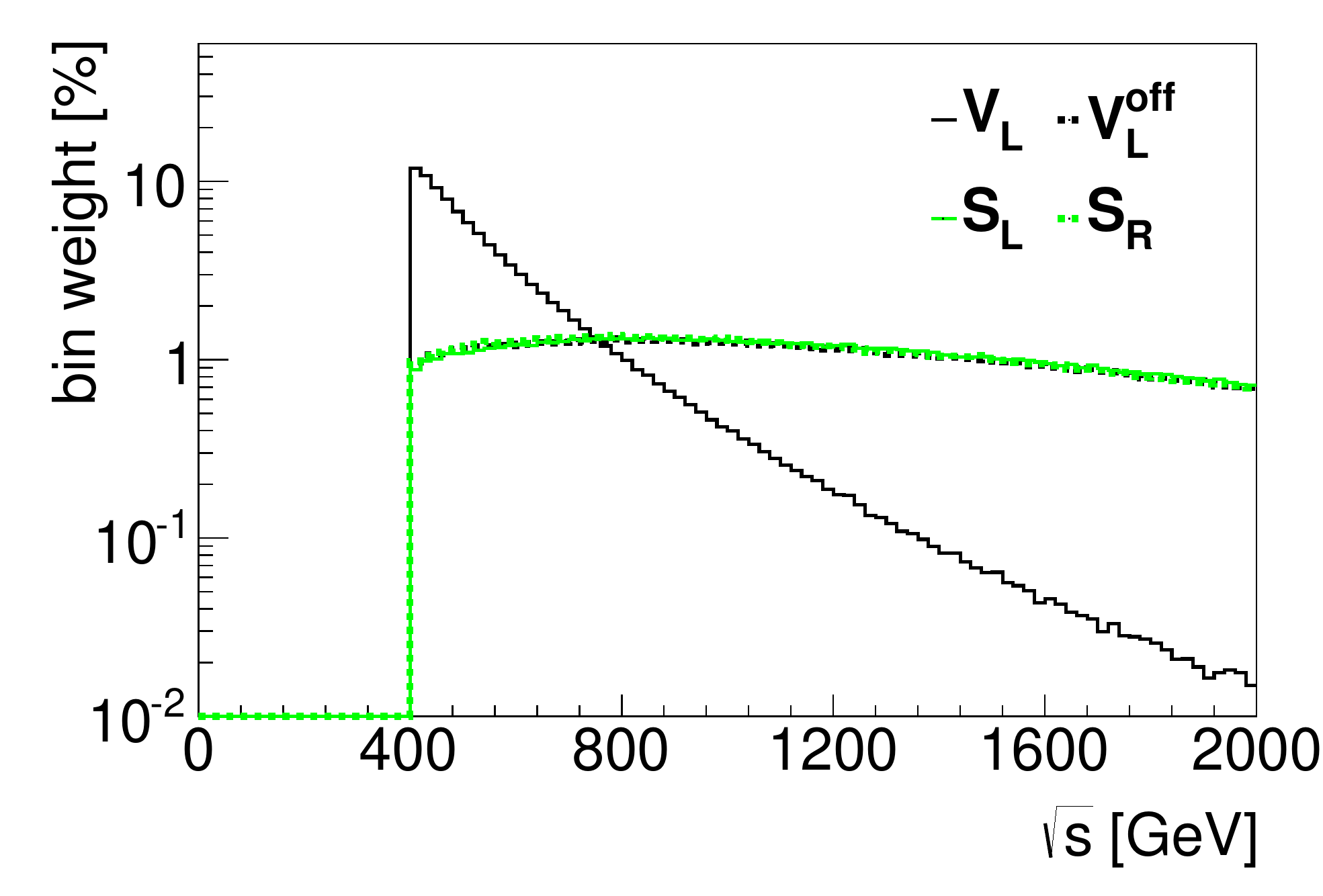}
 \hfill{}
% \vspace{0.5cm}
 \includegraphics[scale=\imgsize]{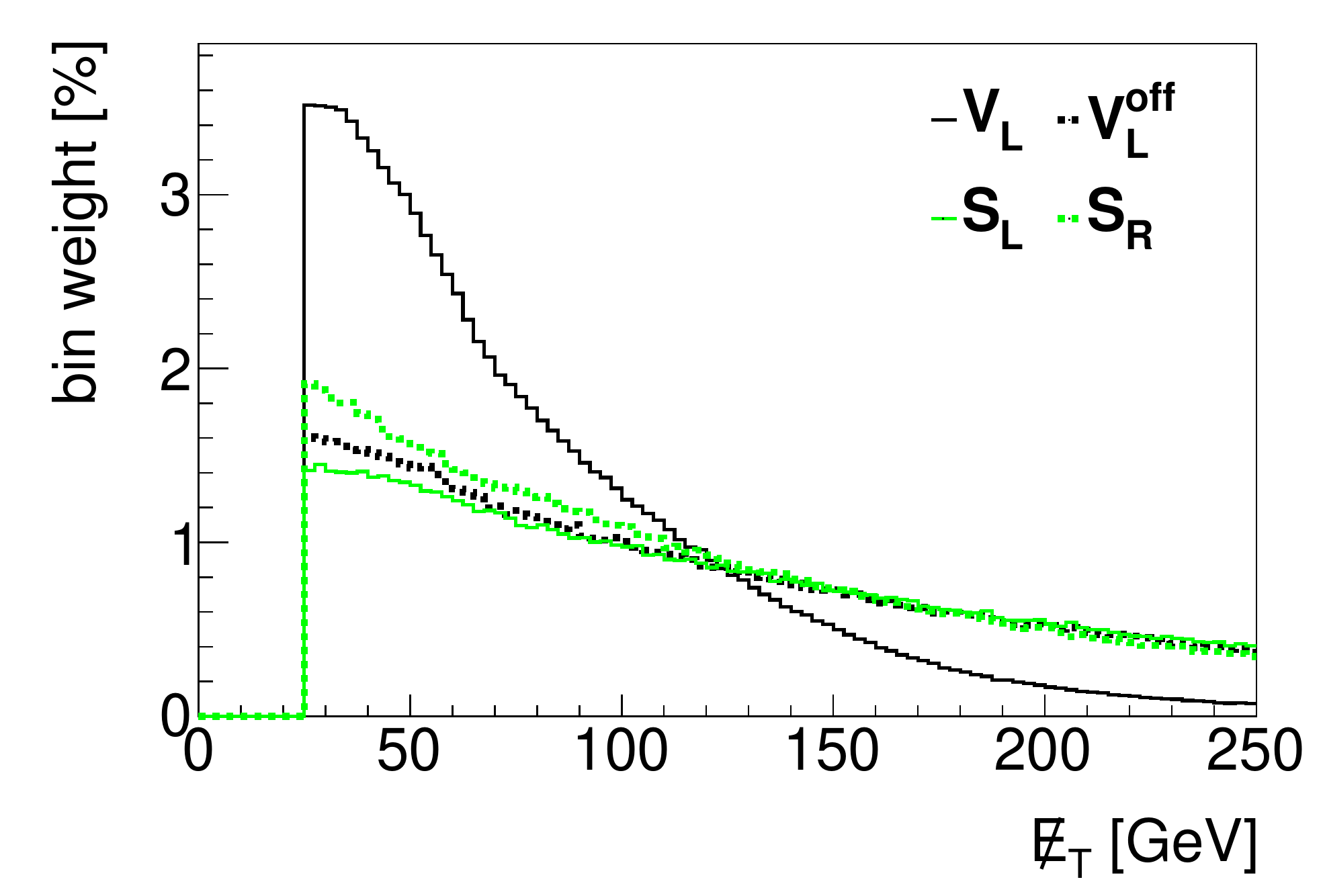}
 \includegraphics[scale=\imgsize]{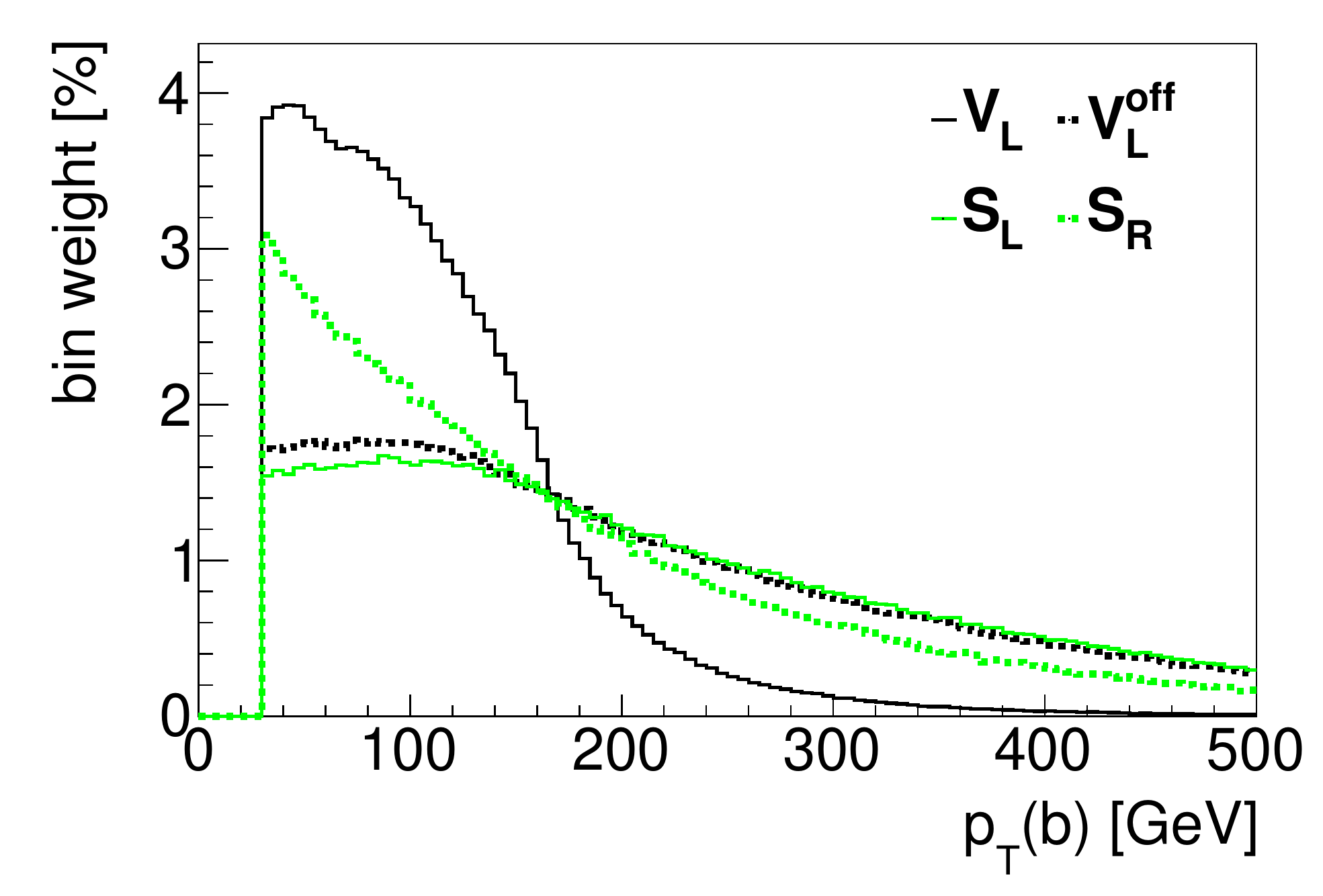}
 \hfill{}
% \vspace{0.5cm}
 \includegraphics[scale=\imgsize]{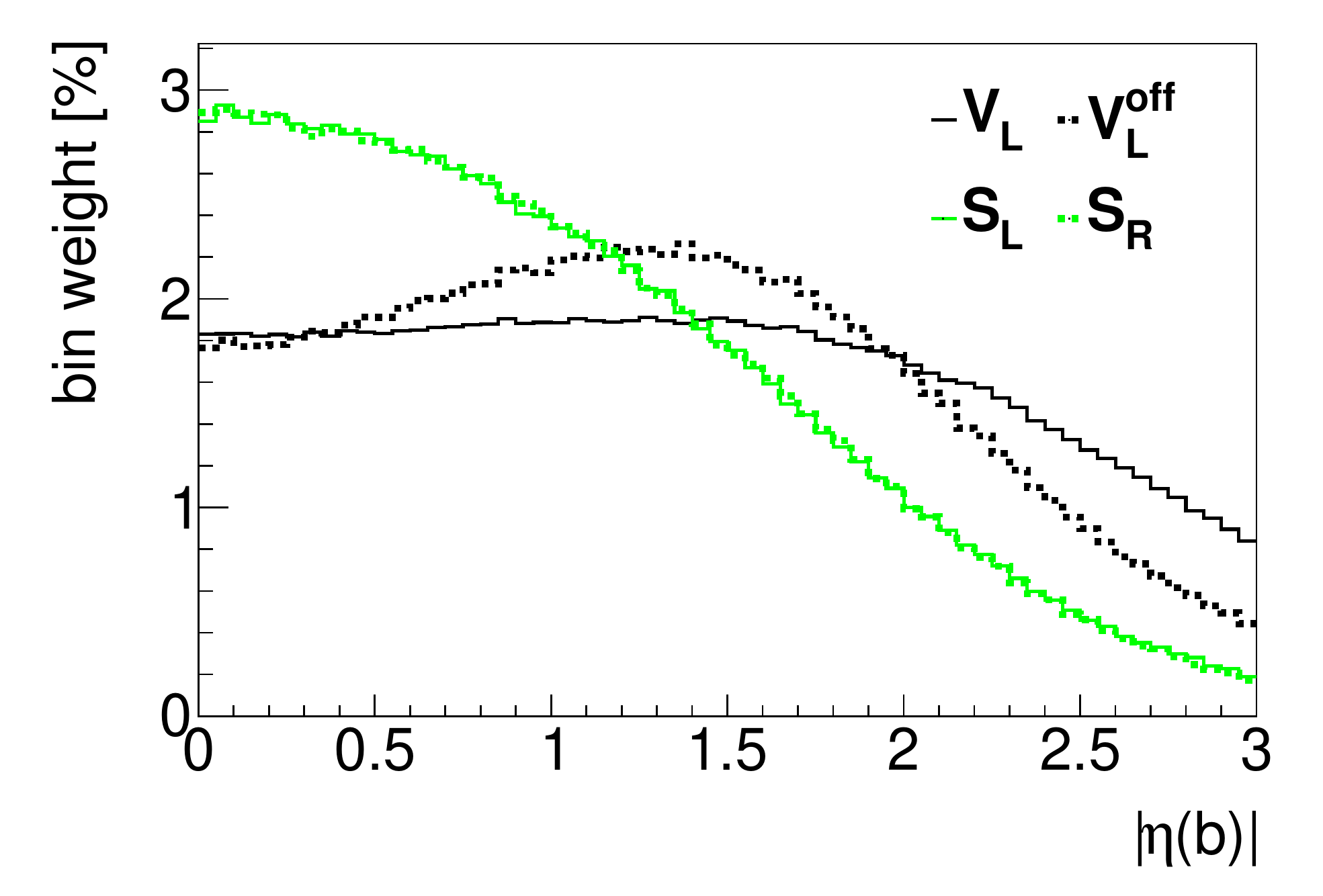}
 \includegraphics[scale=\imgsize]{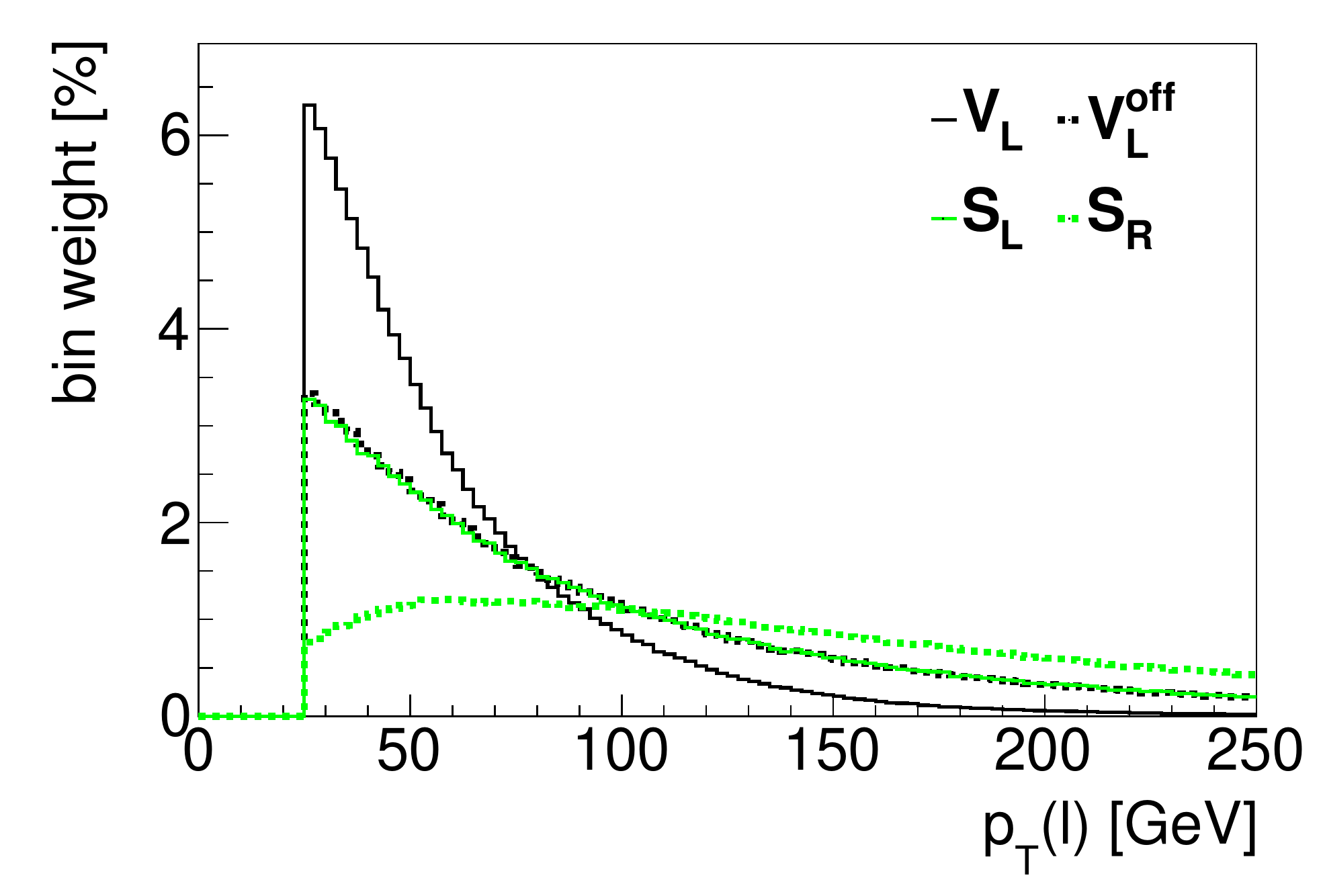}
 \hfill{}
% \vspace{0.5cm}
 \includegraphics[scale=\imgsize]{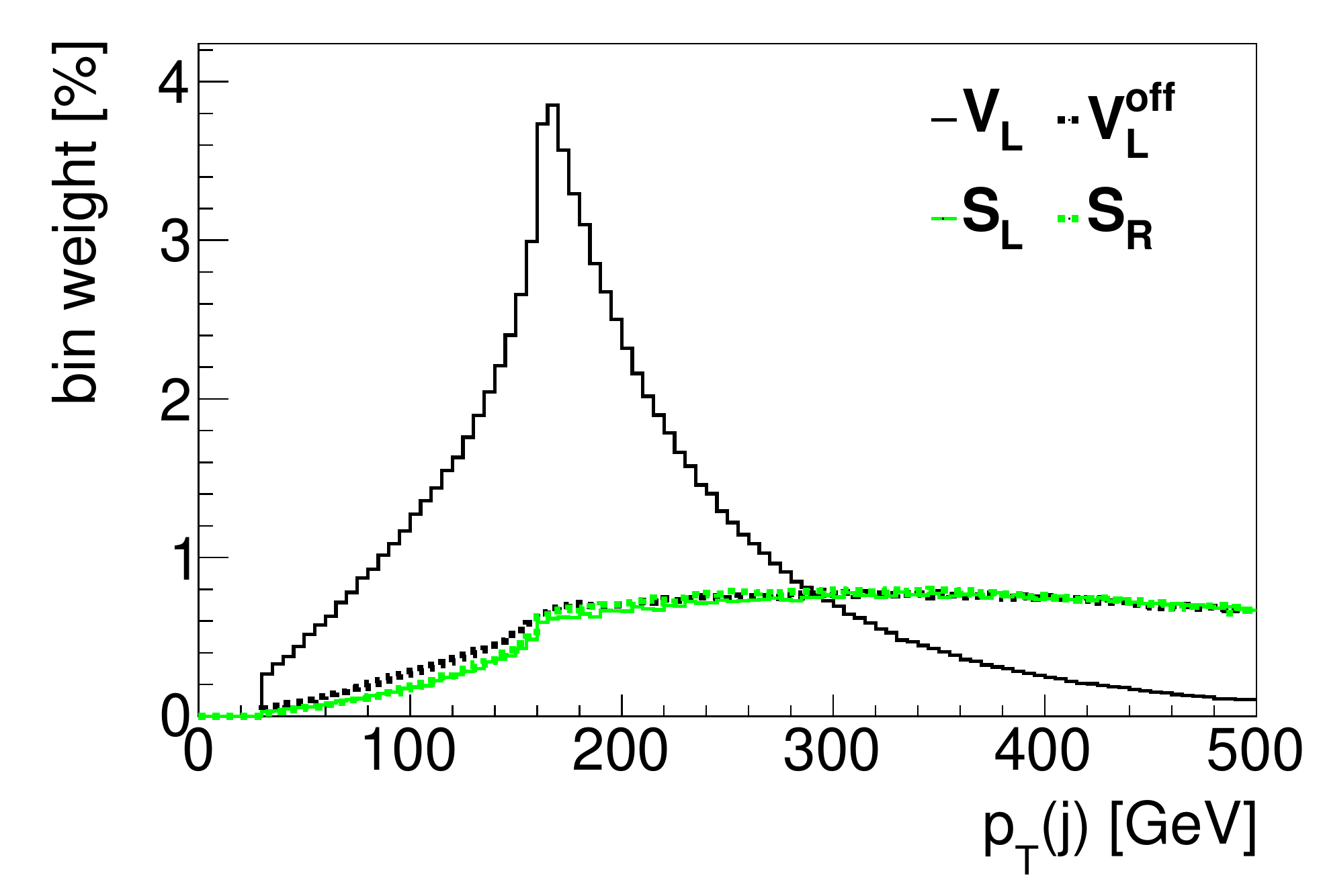}
 \includegraphics[scale=\imgsize]{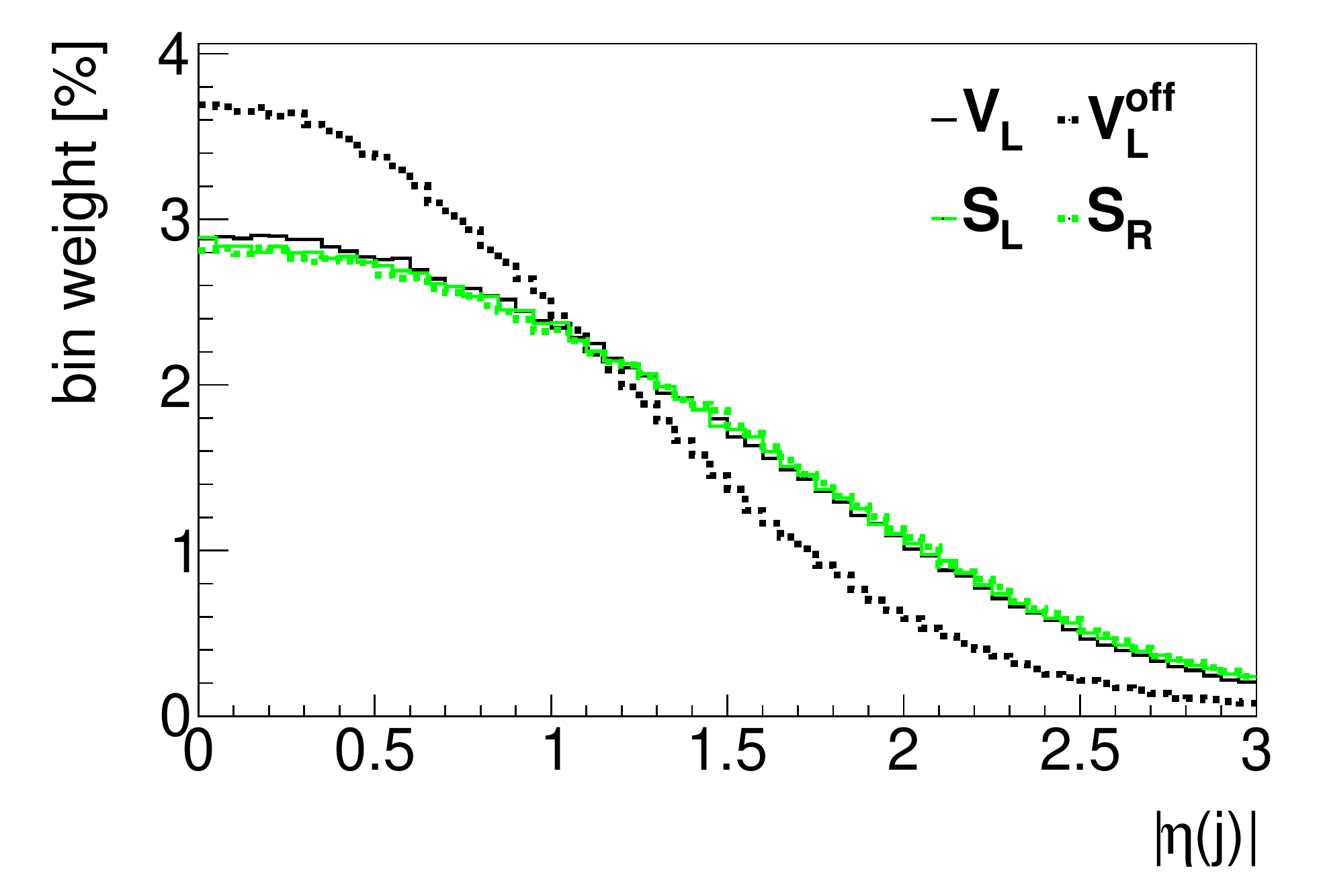}
% \vspace{-5mm}
 \hfill{}
% \vspace{0.5cm}
 \includegraphics[scale=\imgsize]{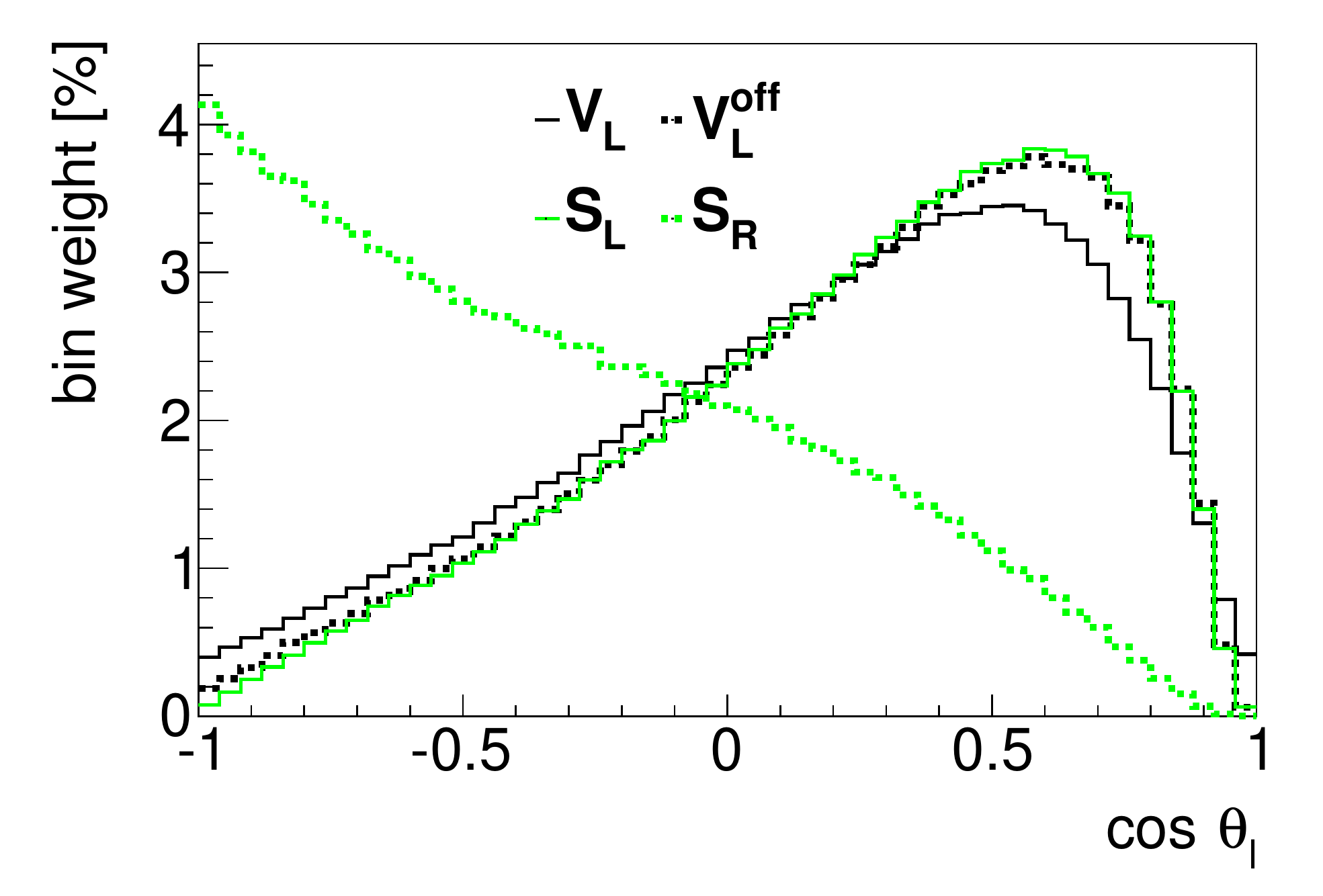}
 \caption{Normalized $s$~channel distributions at parton level.
 \label{cc_contact_tb_part}}
\end{figure*}

\renewcommand{\imgsize}{0.4}
\begin{figure*}
% \vspace{-2mm}
 \includegraphics[scale=\imgsize]{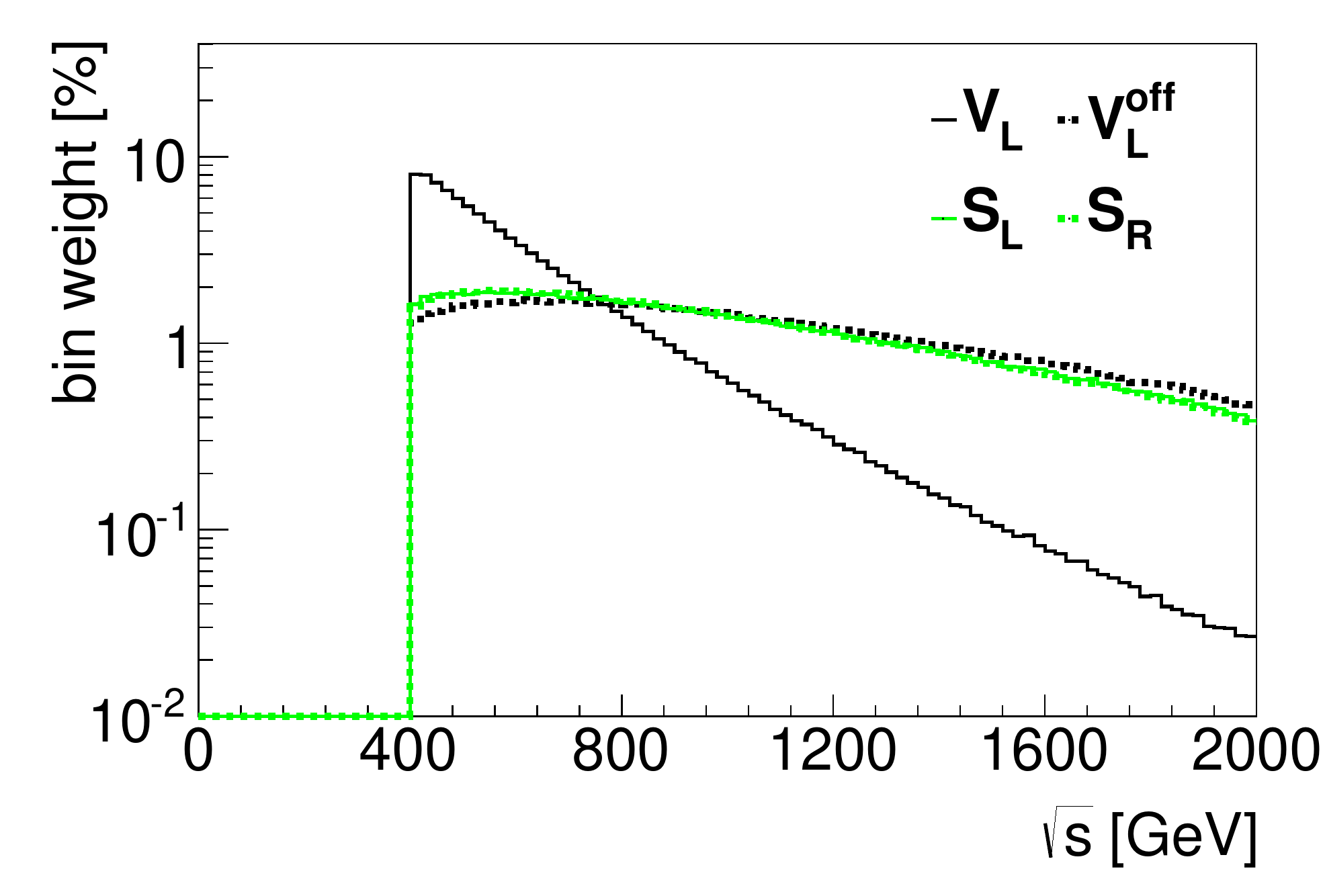}
 \hfill{}
% \vspace{0.5cm}
 \includegraphics[scale=\imgsize]{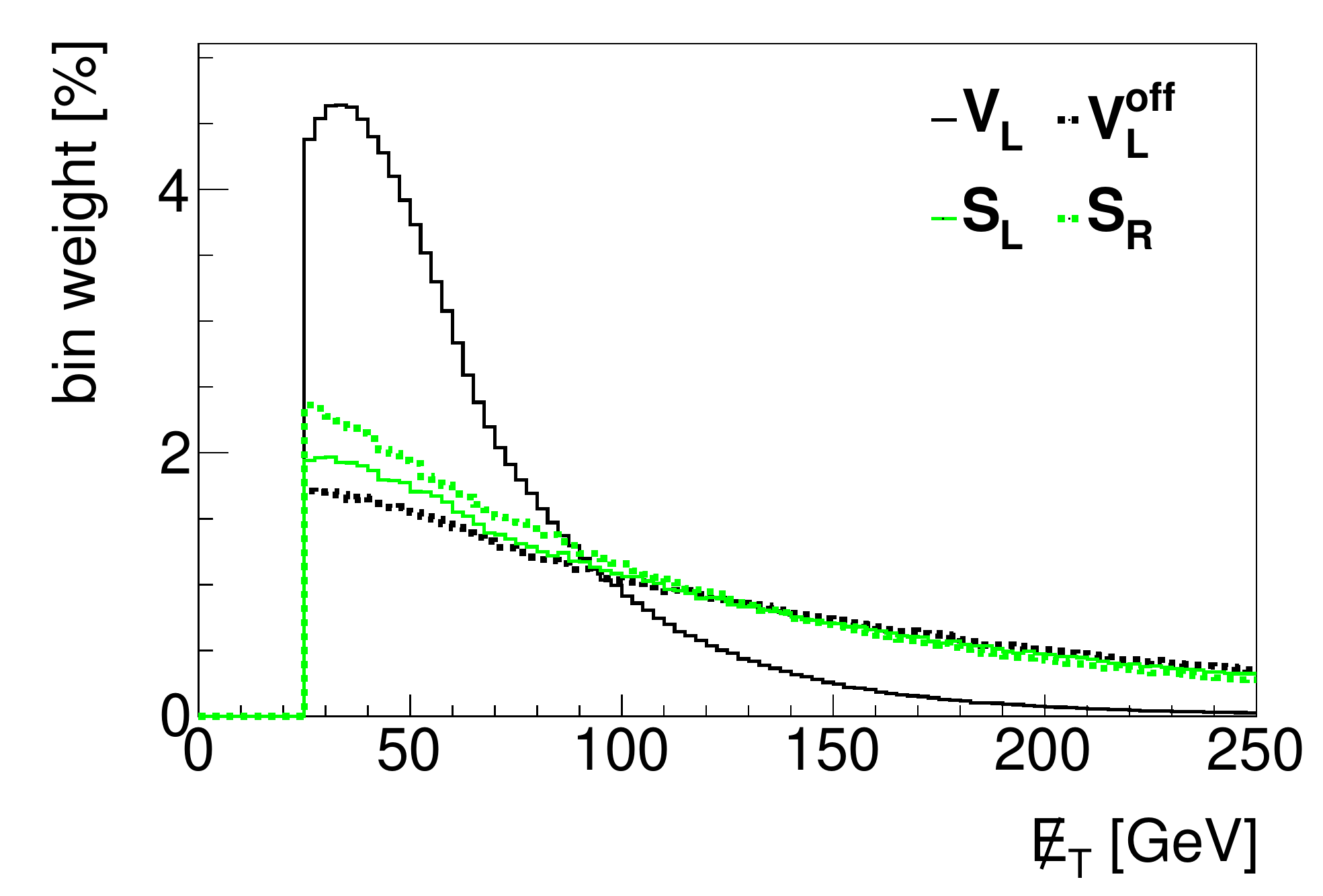}
 \includegraphics[scale=\imgsize]{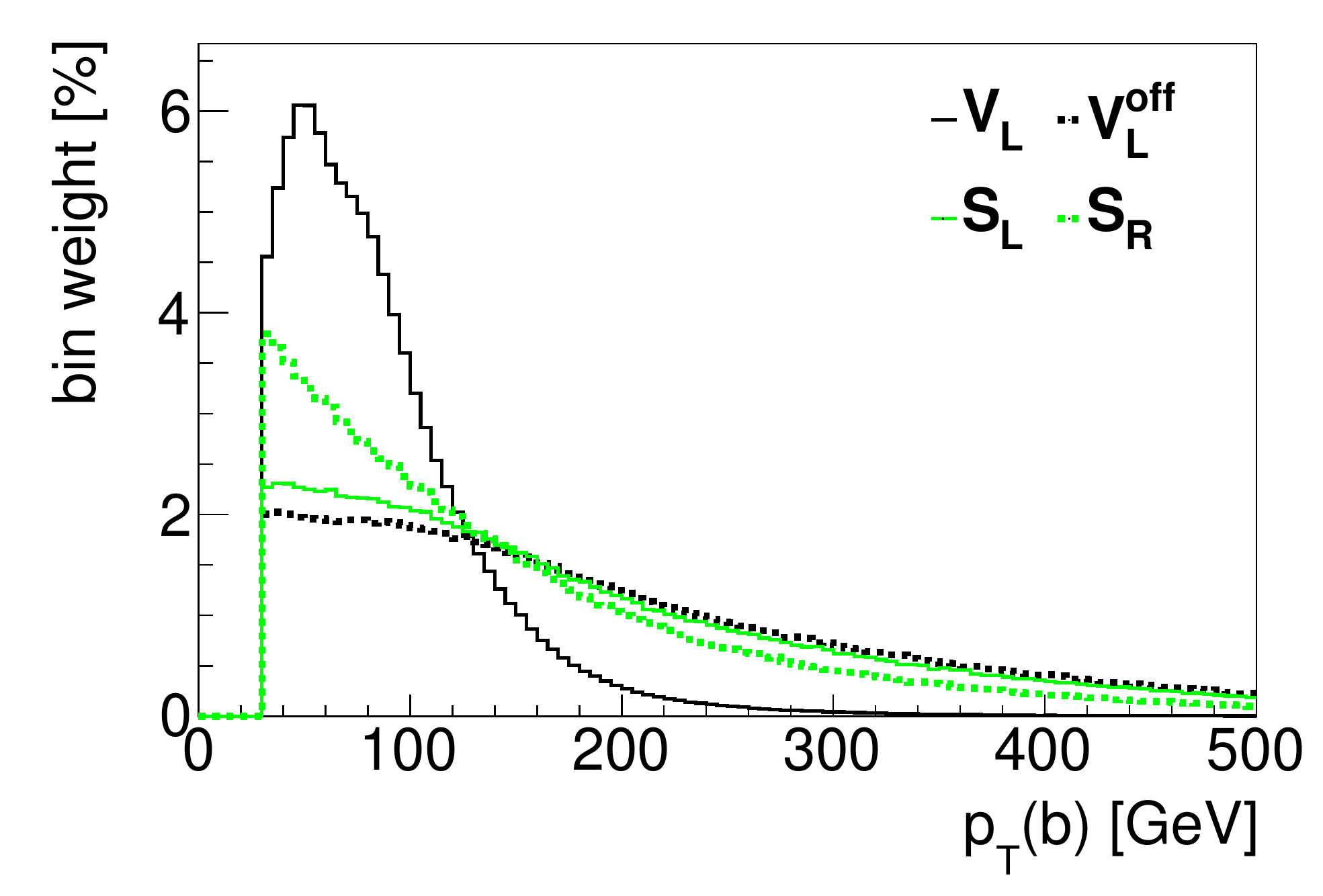}
 \hfill{}
% \vspace{0.5cm}
 \includegraphics[scale=\imgsize]{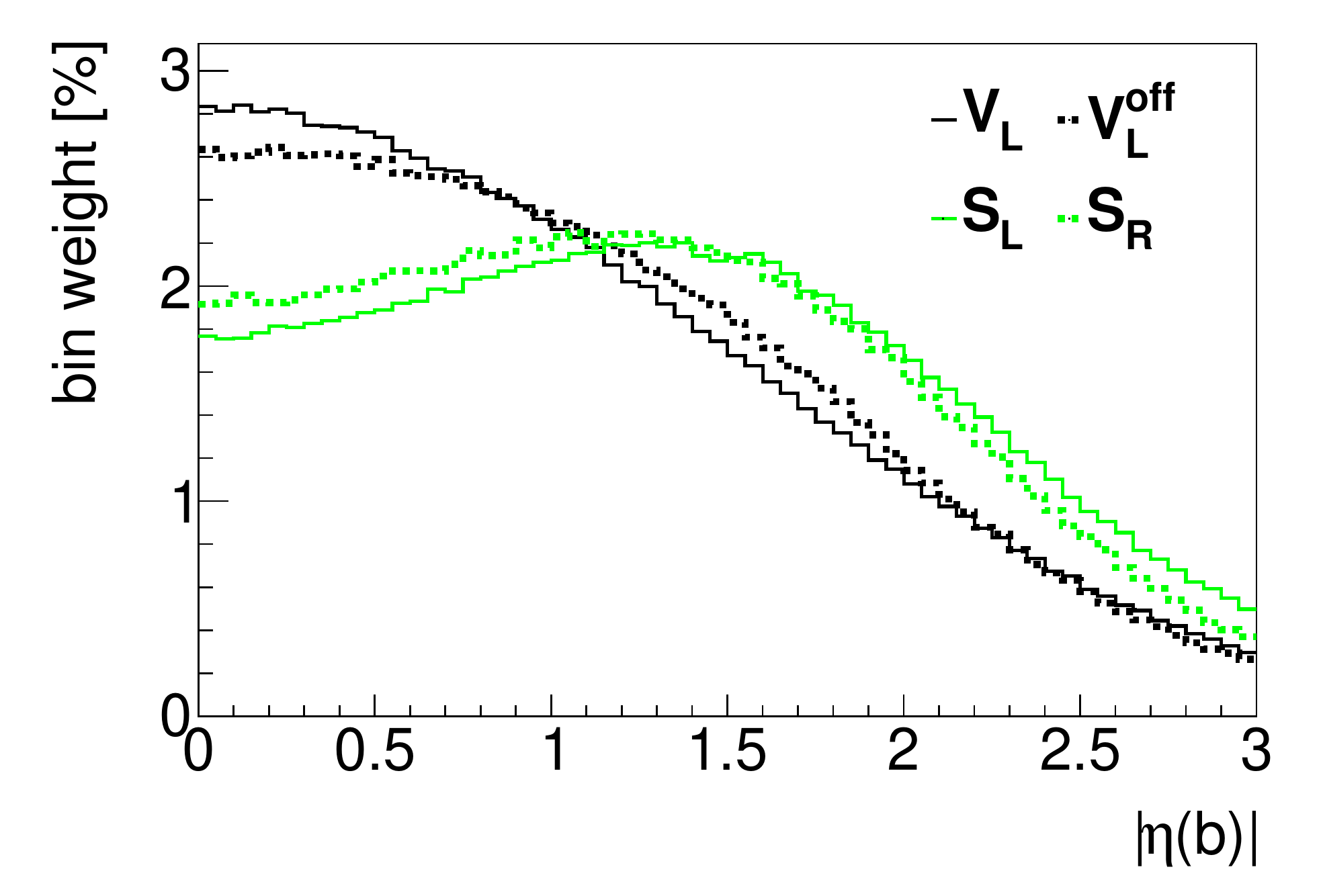}
 \includegraphics[scale=\imgsize]{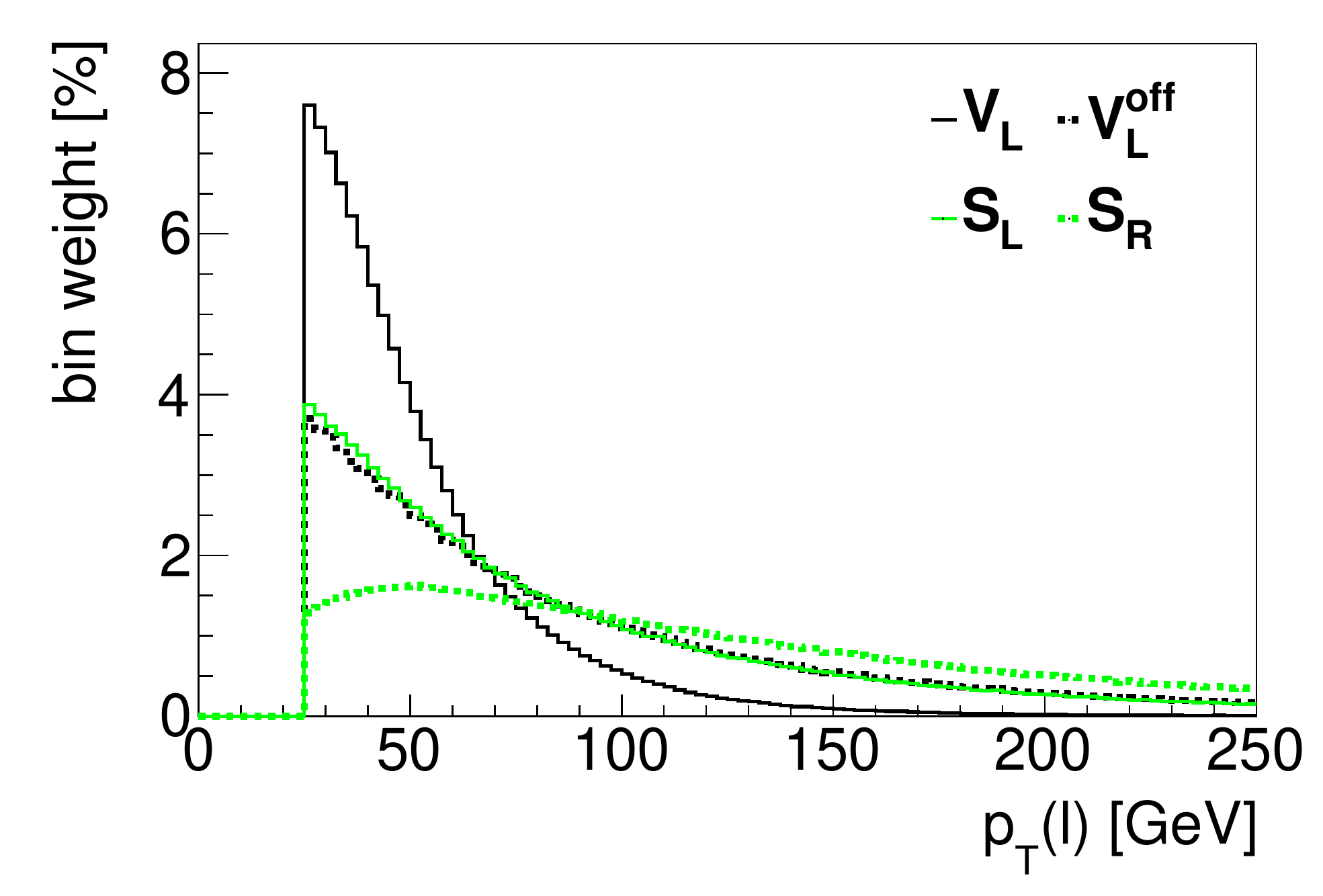}
 \hfill{}
% \vspace{0.5cm}
 \includegraphics[scale=\imgsize]{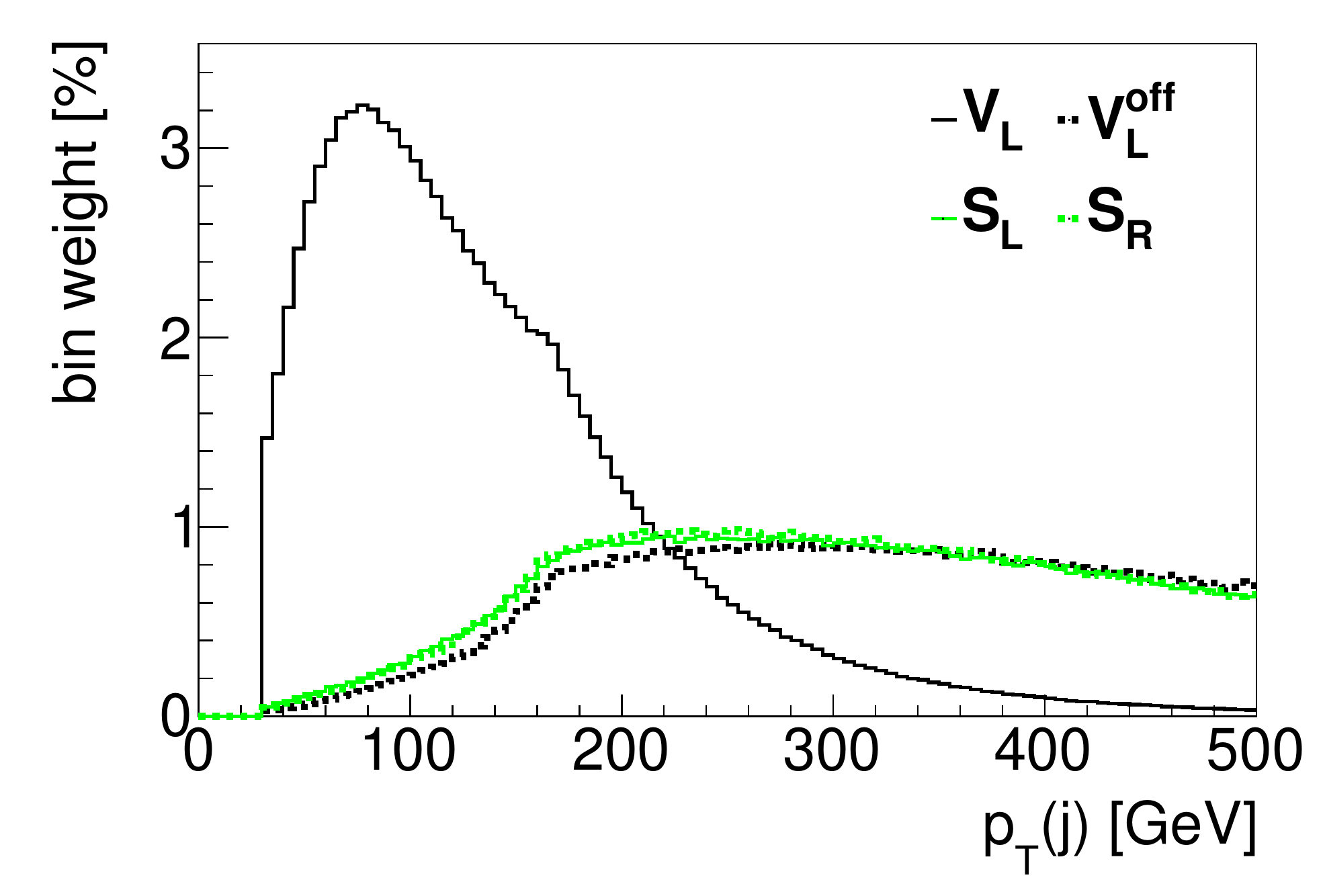}
 \includegraphics[scale=\imgsize]{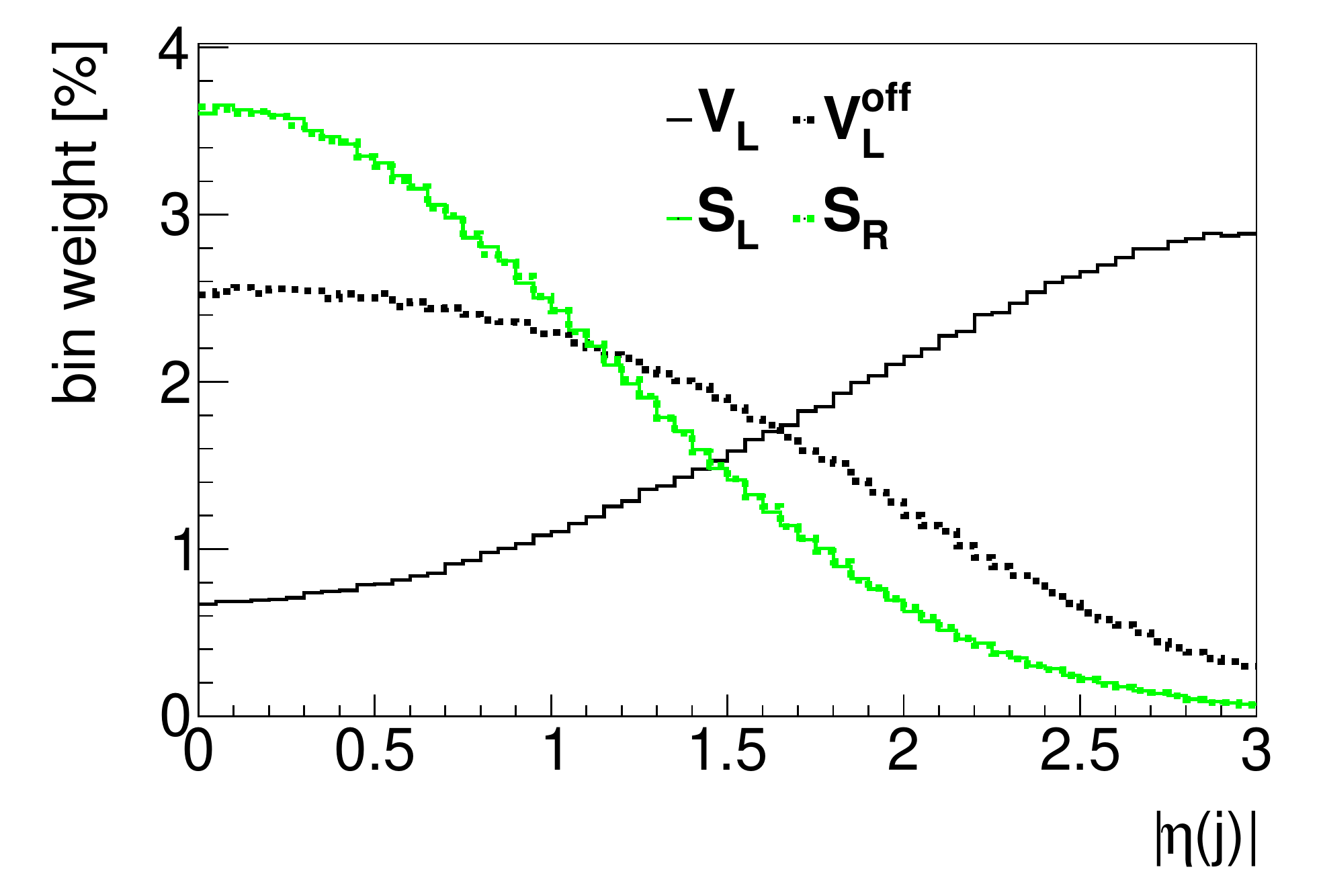}
% \vspace{-5mm}
 \hfill{}
% \vspace{0.5cm}
 \includegraphics[scale=\imgsize]{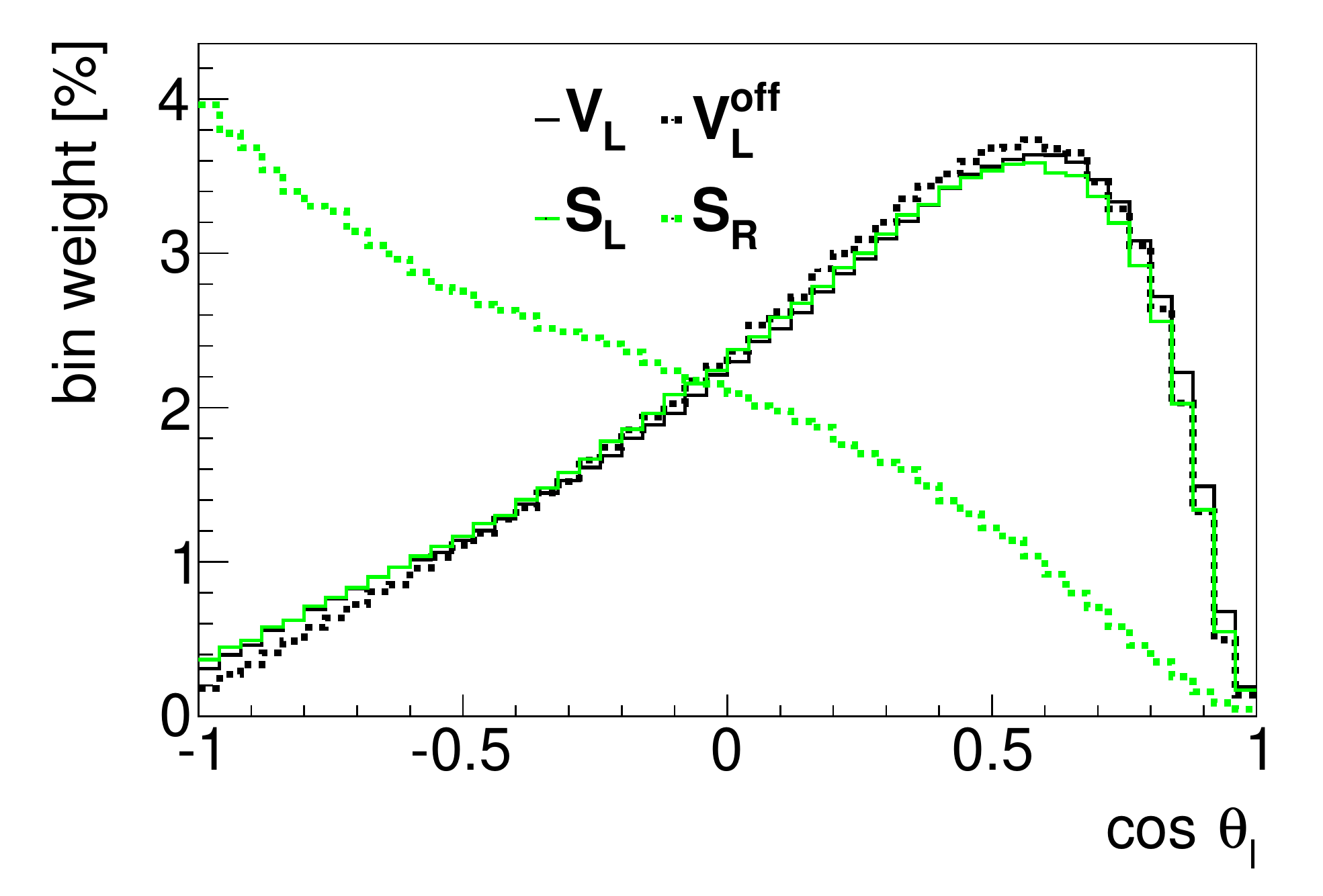}
 \caption{Normalized $t$~channel distributions at parton level.
 \label{cc_contact_tj_part}}
\end{figure*}

\renewcommand{\imgsize}{0.62}
\begin{figure*}[t]
 \includegraphics[scale=0.63]{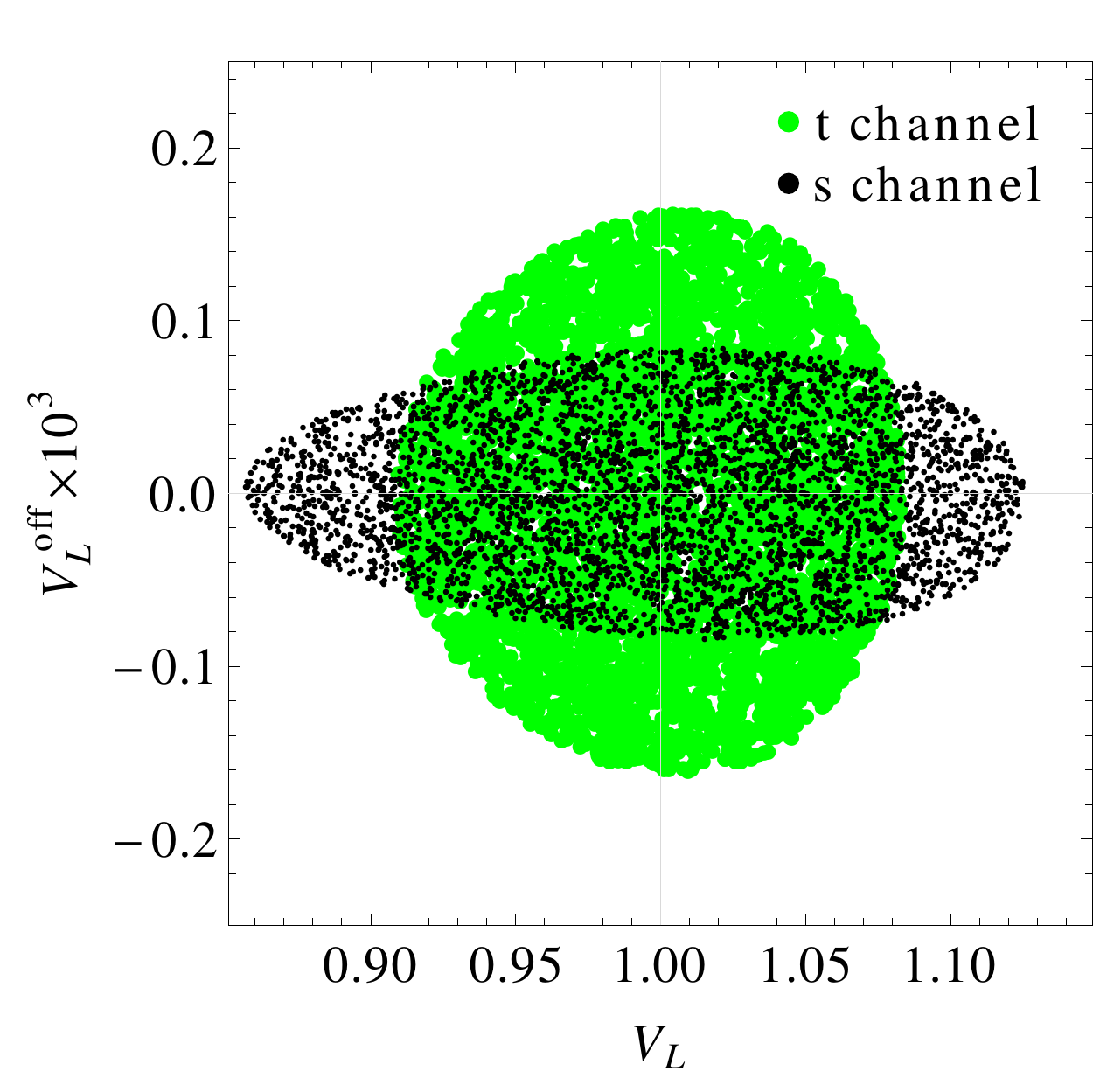}
% \hspace{0.3cm}
 %\vspace{0.5cm}
 \hfill{}
 \includegraphics[scale=\imgsize]{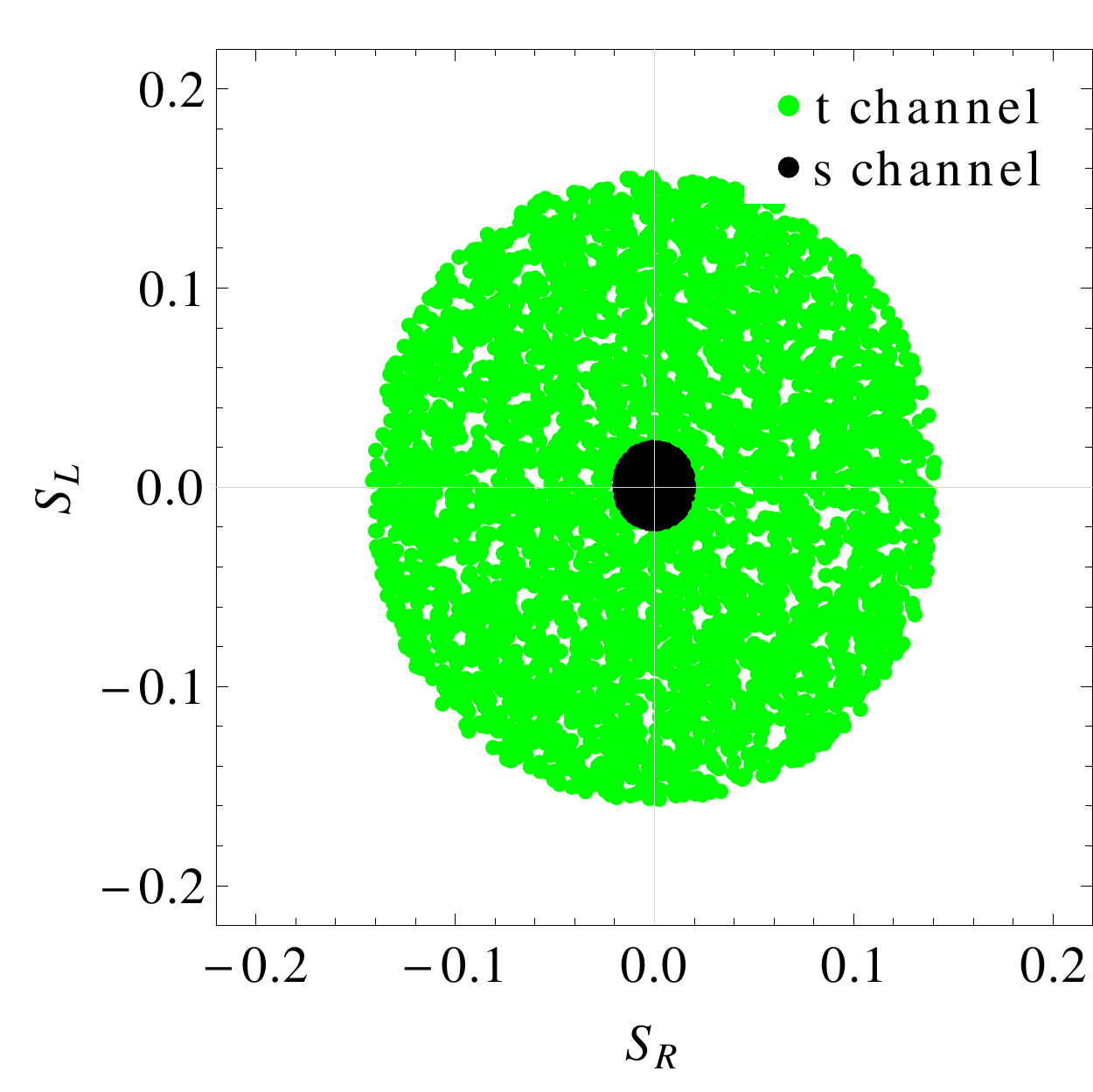}
 %\hspace{0.5cm}
 \includegraphics[scale=\imgsize]{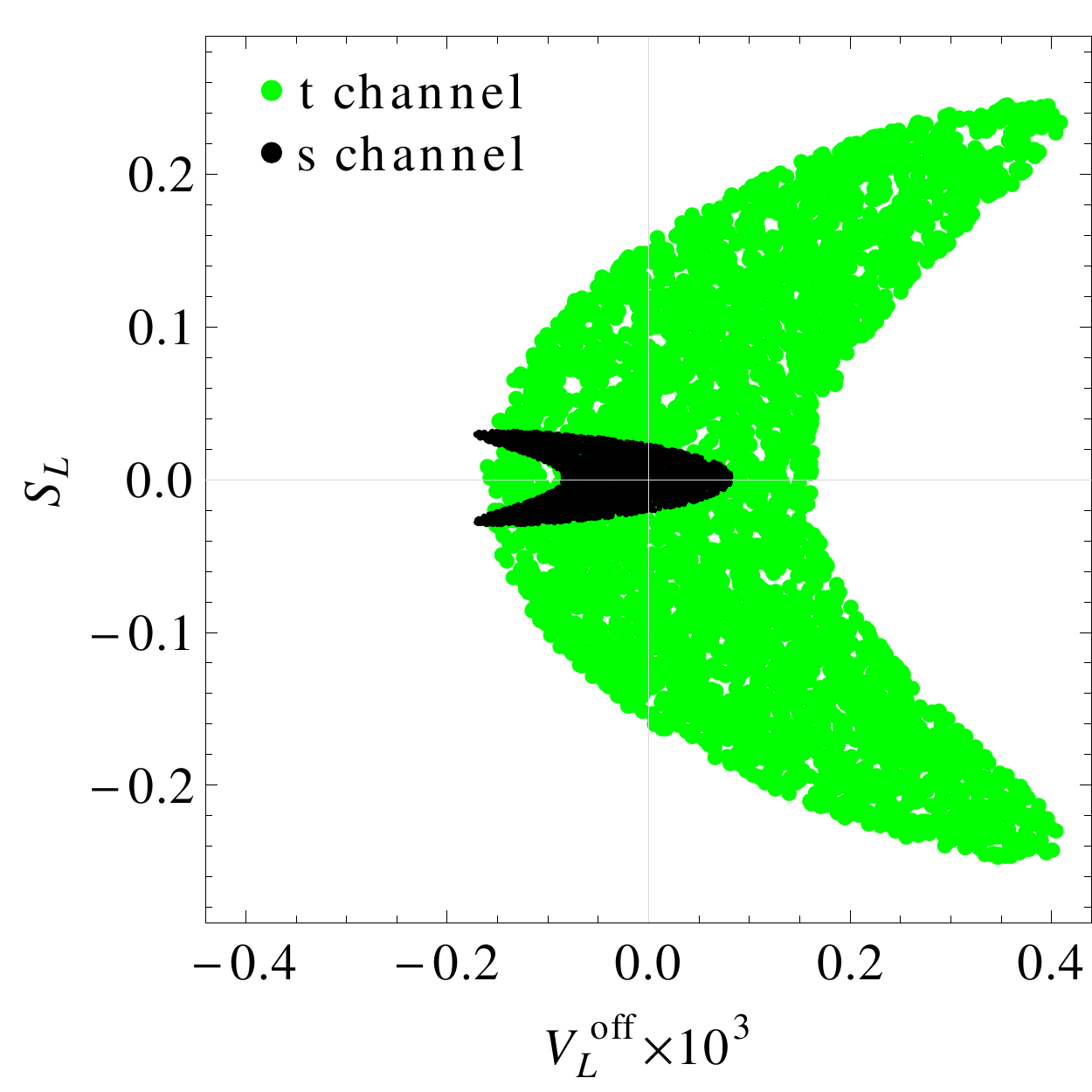}
% \hspace{0.3cm}
 \vspace{-5mm}
 \hfill{}
 \includegraphics[scale=\imgsize]{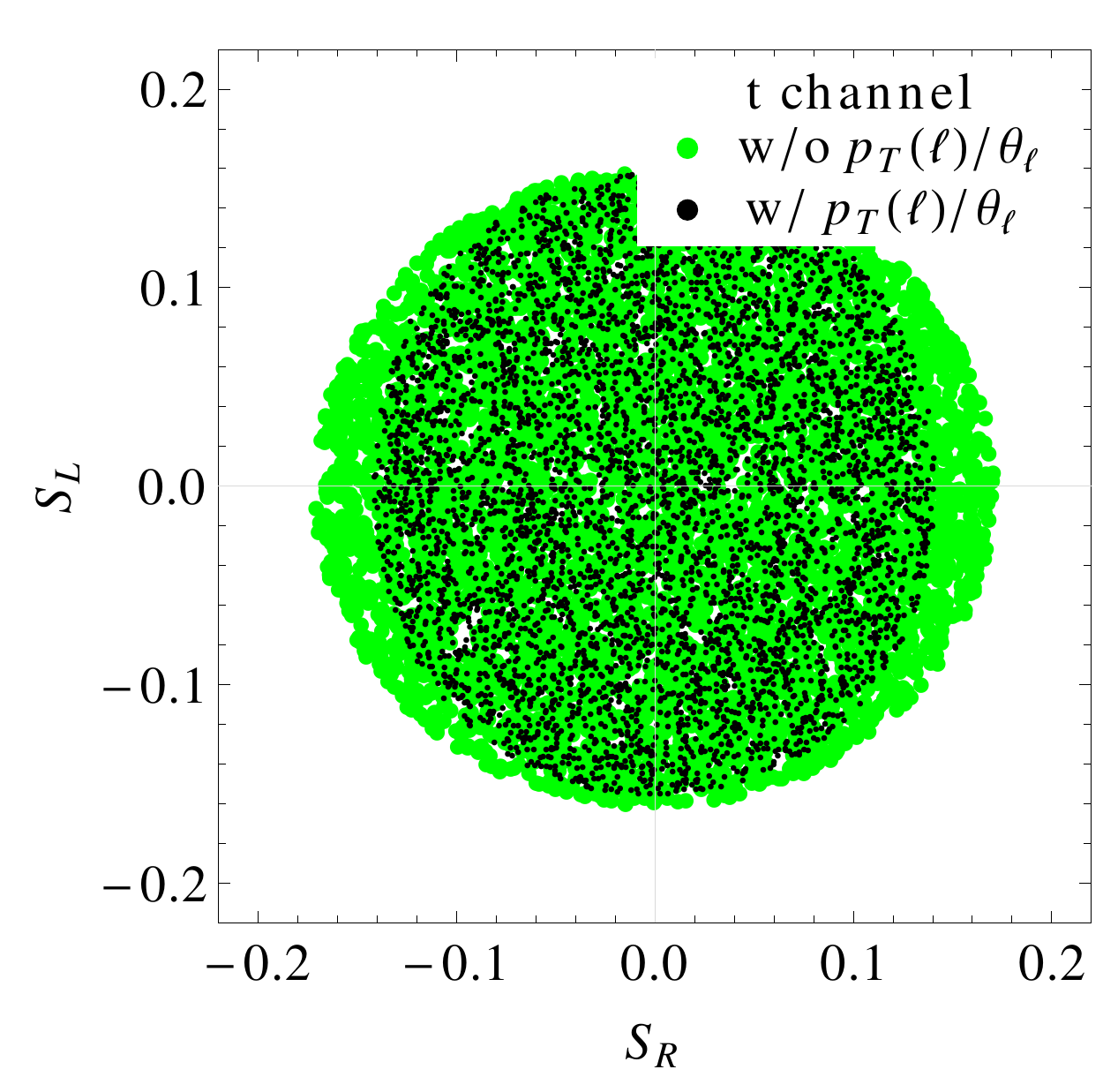}
% \vspace{-5mm}
 \caption{Channel specific sensitivities in various coupling planes
 (top and bottom left) and effect
 of the $S_R$ sensitive observables $p_T(\ell)$ and $\cos\theta_\ell$
 in the $t$~channel (bottom right) at parton level.
 \label{cc_contact_scatter_part}}
\end{figure*}
Finally, in Fig.~\ref{cc_contact_scatter_part} we compare the
$1\sigma$~contours in the two different production channels at parton level,
finding that the $s$~channel generally is much more sensitive to the contact
couplings than the $t$~channel.
This could be expected from the fact that the $s$~channel is maximally
sensitive to the high energy tails of the contact interactions, which is
particularly evident in the $S_L$-$S_R$ plane.  The fact that the $s$~channel
sensitivity to the scalars supersedes the $t$~channel one by roughly an order
of magnitude is not only due to more sensitive shapes, but also due to the
much larger overall sizes of the matrix elements,
a feature which is not
displayed in the normalized shapes of Fig.~\ref{cc_contact_scatter_part}.
Note that either the different relative sensitivities in the two channels when
comparing the scalar plane to the vector plane
or the sign change in the $V_L$-$\VO$ interference
(cf.~Fig.~\ref{cc_contact_scatter_part} bottom left)
can in principle be used
to distinguish $\VO$ from the scalar couplings by a separate analysis of
each individual channel.
However, the detector level analysis below will address the question whether
the production channels can be separated at all with a sufficient purity.
Moreover, if one compares the sensitivity to $V_L$ with the one obtained
in~\cite{Bach2012}, there is obviously no significant gain from the
binned analysis:
in fact, this is no surprise, since $V_L$ only sets
the overall normalization of the SM shapes but does not distort them,
while on the other hand, the sensitivity to $\VO$ increases by more than
2~orders of magnitude.
Finally, as also illustrated in Fig.~\ref{cc_contact_scatter_part},
the observables $p_T(\ell)$ and $\cos\theta_\ell$ are indeed specifically
sensitive to $S_R$, potentially amounting to a $\sim\unit[20]{\%}$ gain on the
$S_R$ limit with respect to the $S_L$ one, and could hence be employed to
distinguish any anomalous right-handed part in single top production.
To conclude this paragraph, one may state that the LHC sensitivity reach to the
four-fermion interactions vitally depends on the capability to cleanly separate
the $s$~channel signal at the detector level, a task that shall be addressed
now in more detail.

\subsection{Detector level}\label{pheno_d}

Detector effects are being accounted for by processing the SM~samples with
{\sc pythia~6}~\cite{Sjostrand2006} and
{\sc Delphes}~\cite{Ovyn2009,Favereau2013}
to obtain detector level samples,
where in the latter we assume by default a global $b$-tagging efficiency
of~$\unit[60]{\%}$ with corresponding impurity from charm and light
flavors taken from~\cite{Aad2009}.
On these samples the following final state selections are applied:
in addition to an isolated lepton
with $p_T>\unit[25]{GeV}$ and missing transverse energy
$\slashed E_T>\unit[25]{GeV}$,
the selection criteria for the two final state signatures are, respectively,
\begin{enumerate}
 \item $s$~channel or ``$tb$'' selection:
  exactly two $b$~tagged jets with $p_T>\unit[30]{GeV}$,
  and neither central nor forward light jets with $p_T>\unit[15]{GeV}$.
  Furthermore, the top momentum
  is reconstructed from one of the $b$~jets together with the charged lepton
  and~$\slashed E_T$ (identified with the neutrino~$p_T$), by applying
  the on-shell constraint $\left( p_\ell+p_\nu \right)^2=m_W^2$ and picking
  the smaller of the two solutions for the longitudinal component of $p_\nu$.
  The resulting top mass is required to be within
  150 and $\unit[225]{GeV}$.
 \item $t$~channel or ``$tj$'' selection:
  one or more $b$~jets with $p_T>\unit[30]{GeV}$
  (one of them reconstructing the top together with the leptons as before),
  one light forward jet with $p_T>\unit[50]{GeV}$ and
  $0<\left|\eta\right|<3$ and at most one additional light central jet,
  which may have $p_T<\unit[30]{GeV}$ only.
\end{enumerate}
Finally, the overall invariant mass of the event given by the reconstructed
top and the hardest spectator jet must be $m_{tj}>\unit[400]{GeV}$.
Note that the universal partonic acceptance cuts applied on the
matrix elements, Eq.~\eqref{cuts_p},
have been deliberately designed such that they contain the
entire phase space regions of the final state selections stated here for
\emph{any} of the two channels.

\renewcommand{\imgsize}{0.38}
\begin{figure*}
% \vspace{-2mm}
 \includegraphics[scale=\imgsize]{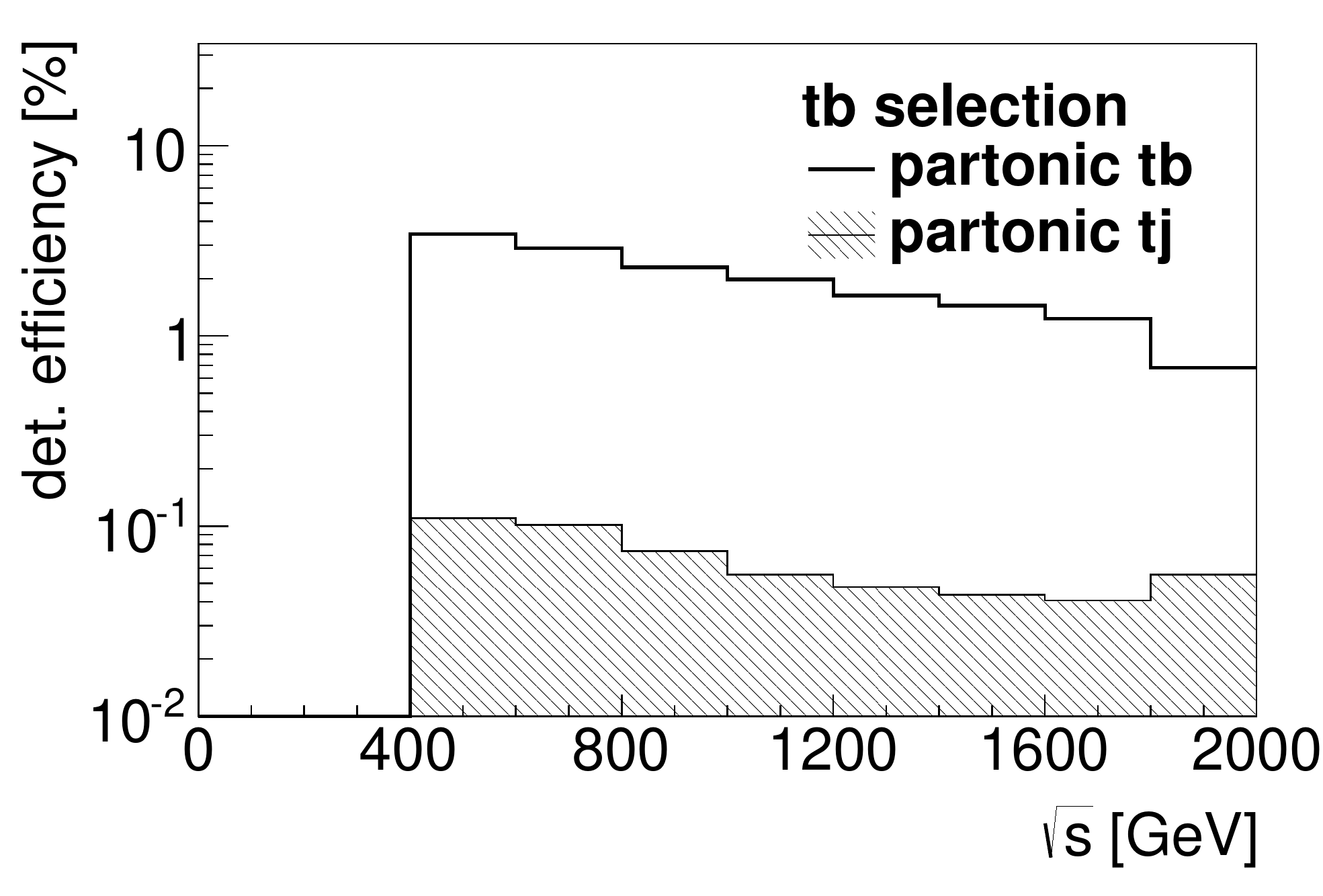}
 \hfill{}
% \vspace{0.5cm}
 \includegraphics[scale=\imgsize]{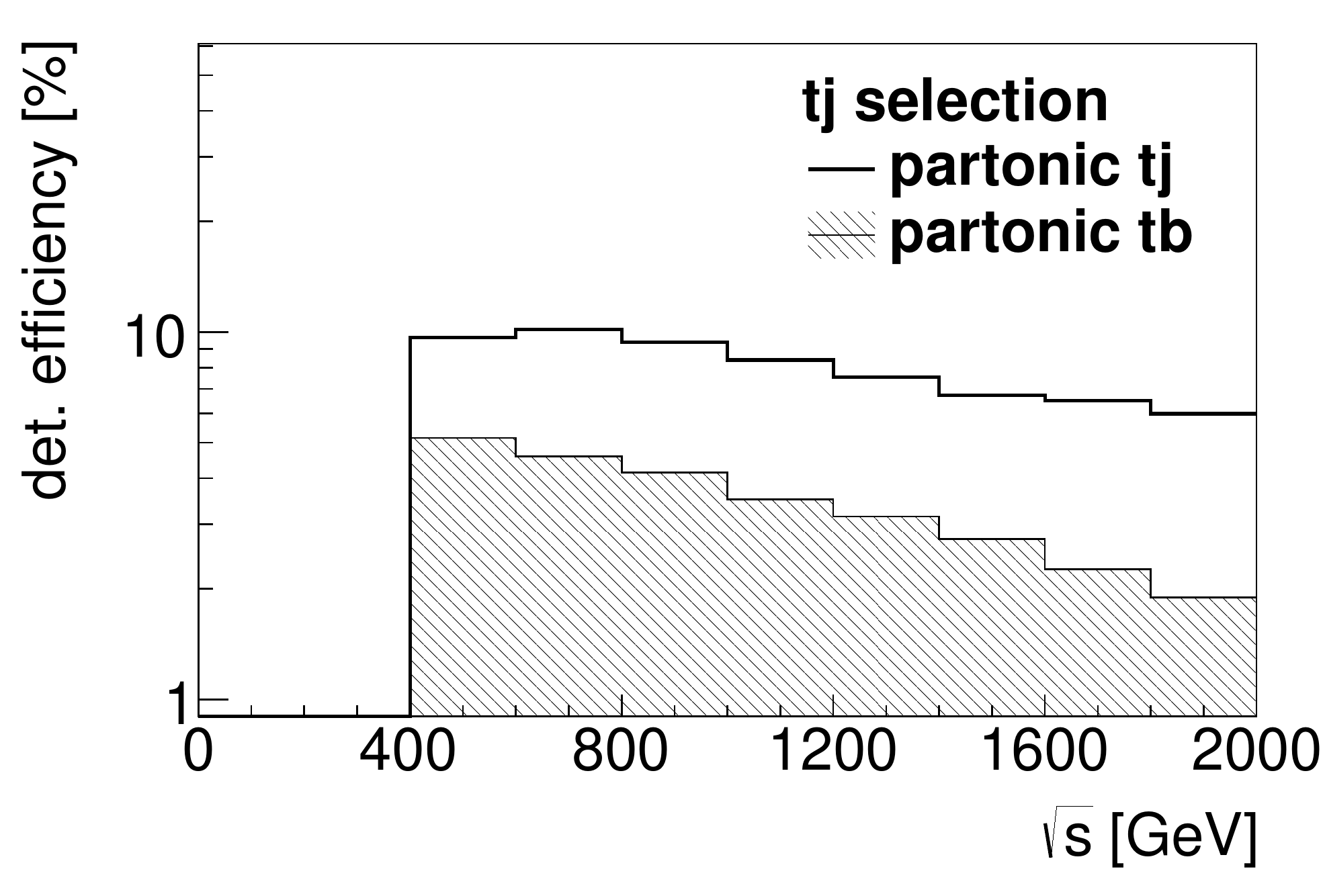}
 \includegraphics[scale=\imgsize]{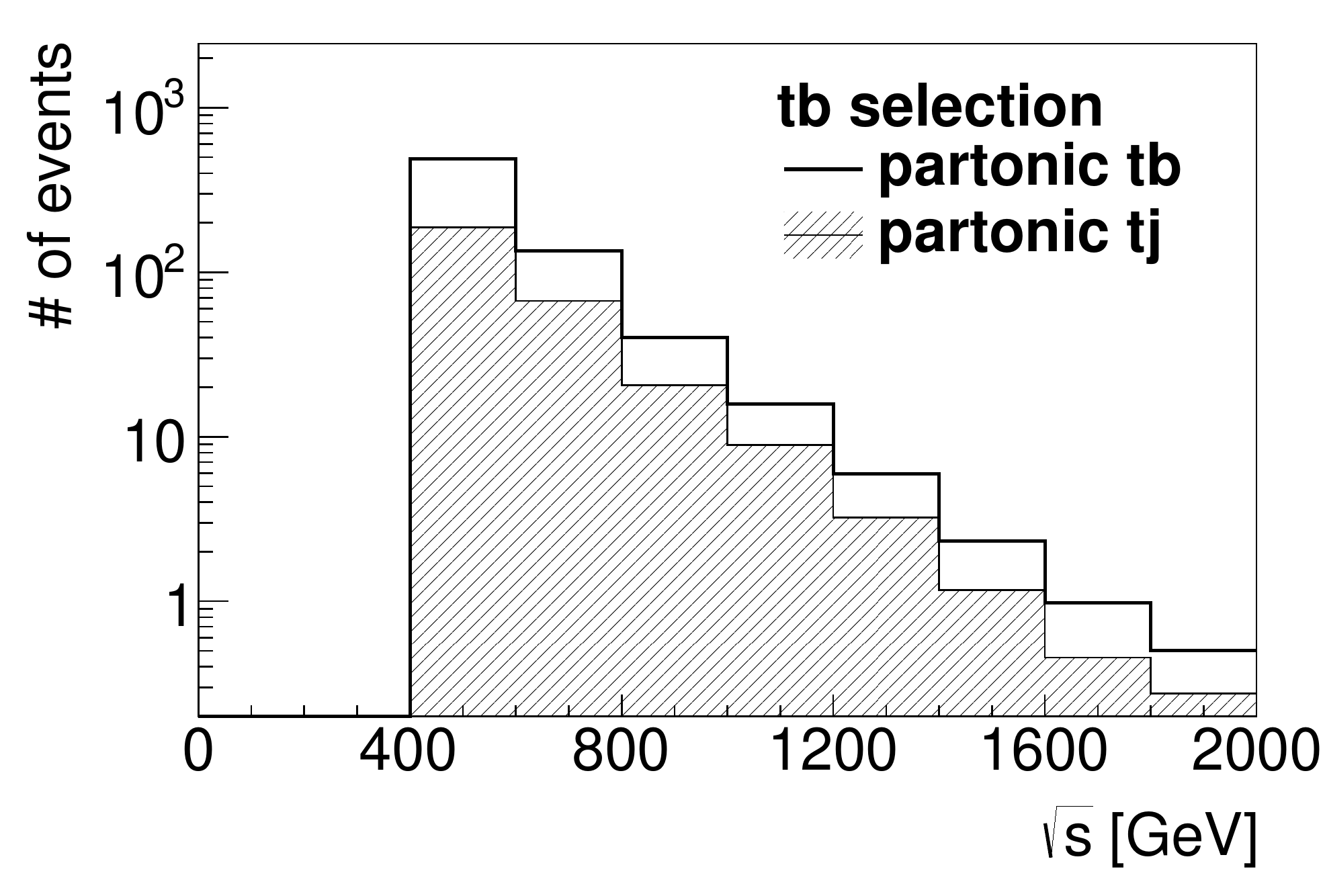}
 \hfill{}
% \vspace{0.5cm}
 \includegraphics[scale=\imgsize]{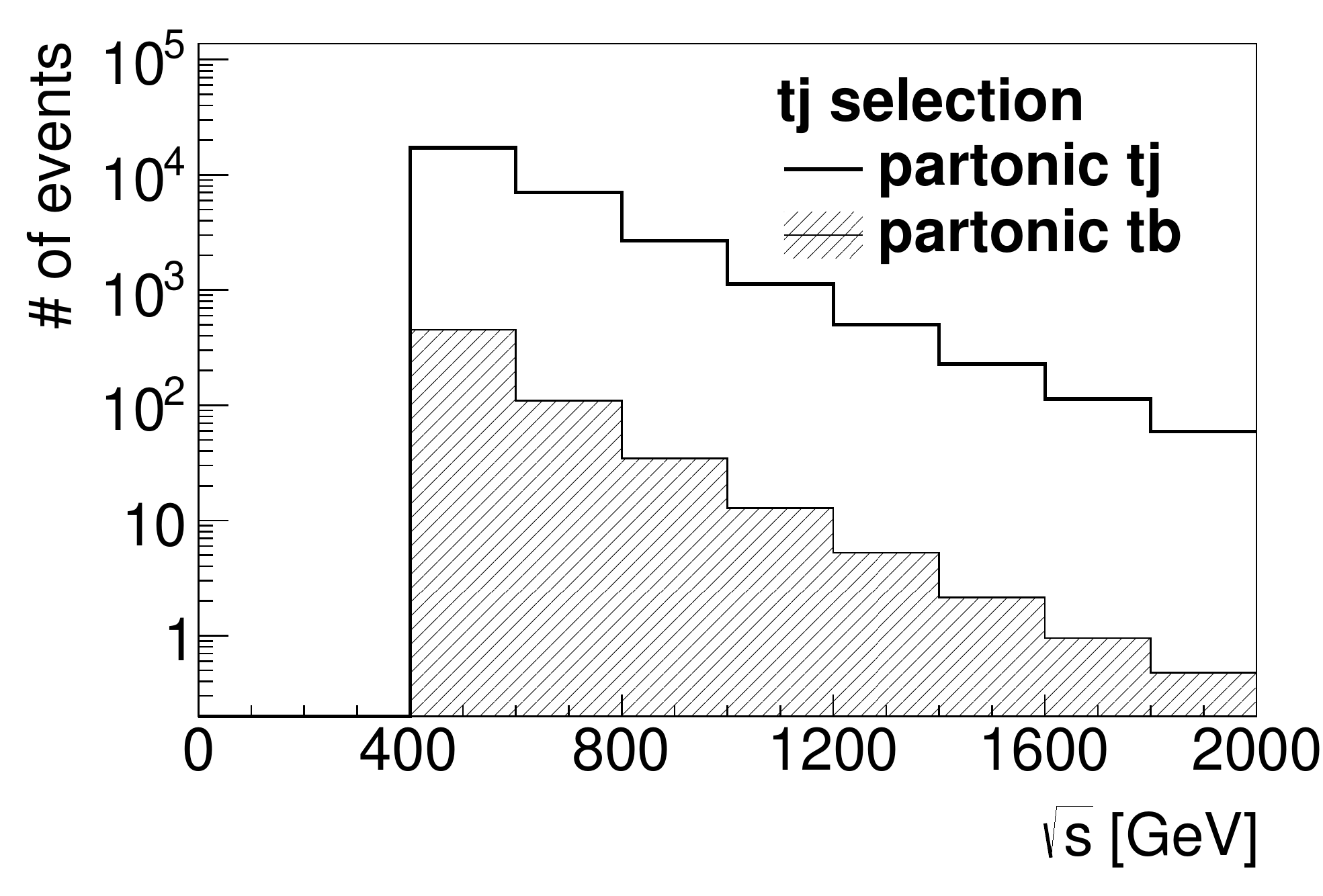}
 \includegraphics[scale=\imgsize]{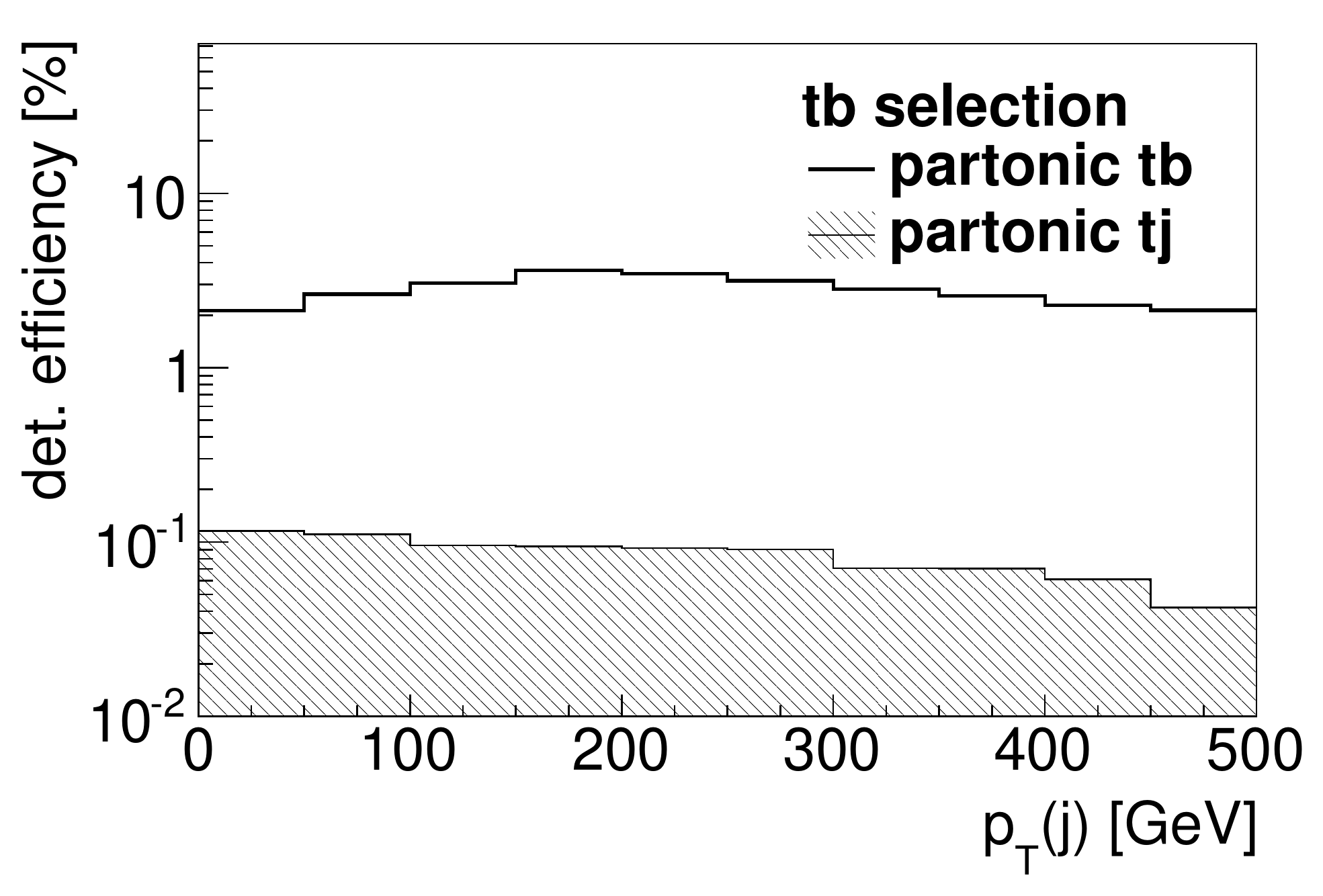}
 \hfill{}
% \vspace{0.5cm}
 \includegraphics[scale=\imgsize]{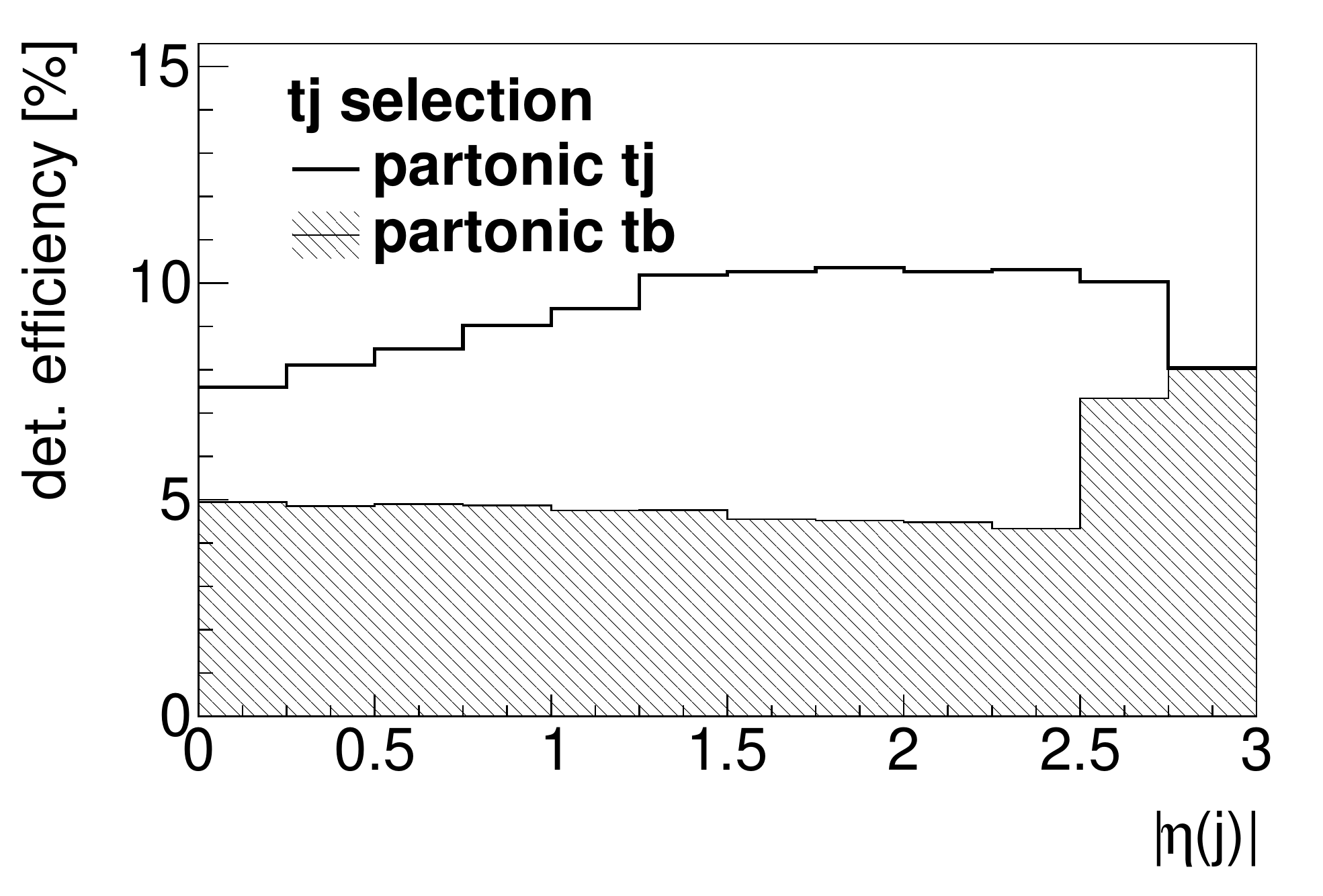}
 \includegraphics[scale=\imgsize]{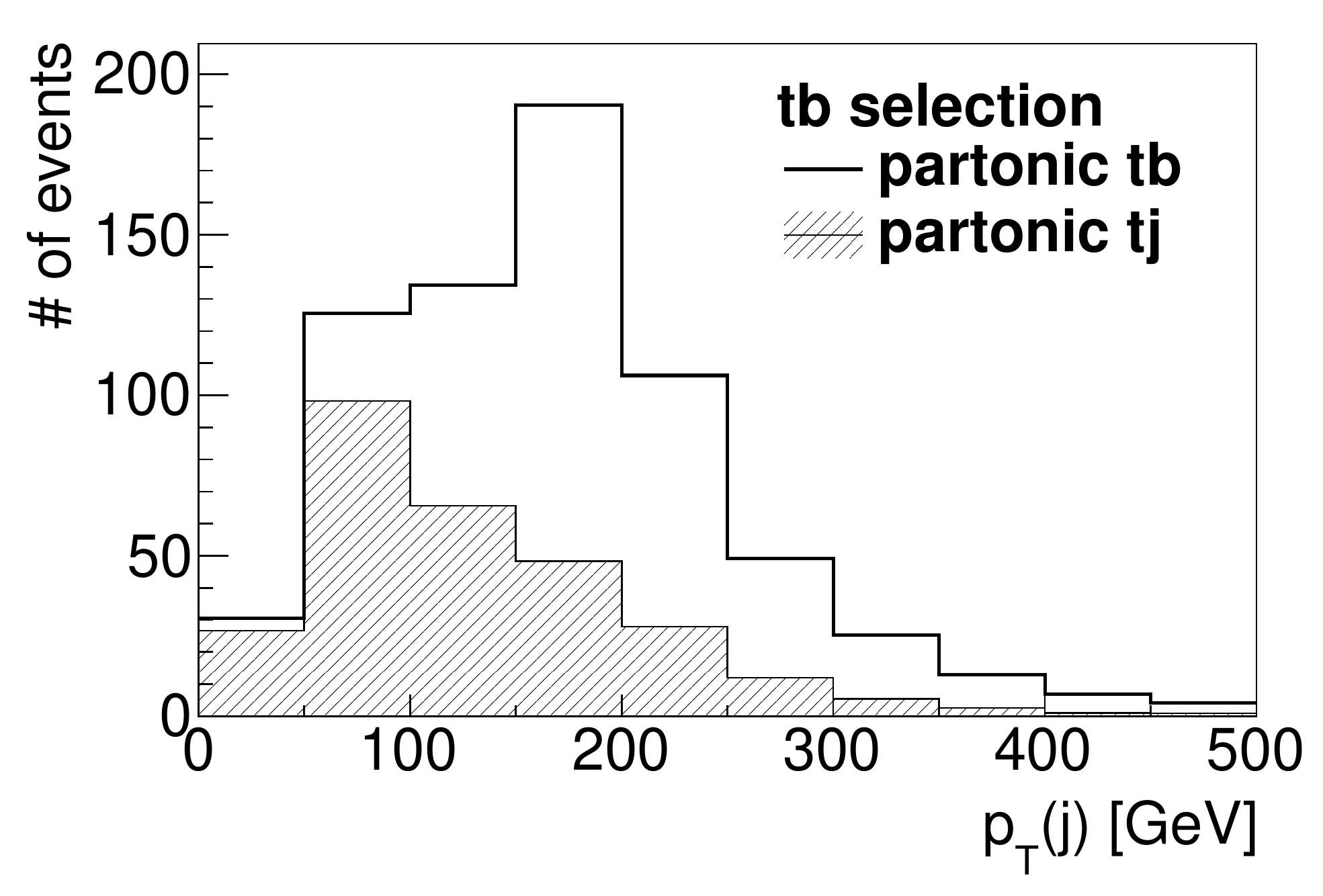}
 \vspace{-5mm}
 \hfill{}
% \vspace{0.5cm}
 \includegraphics[scale=\imgsize]{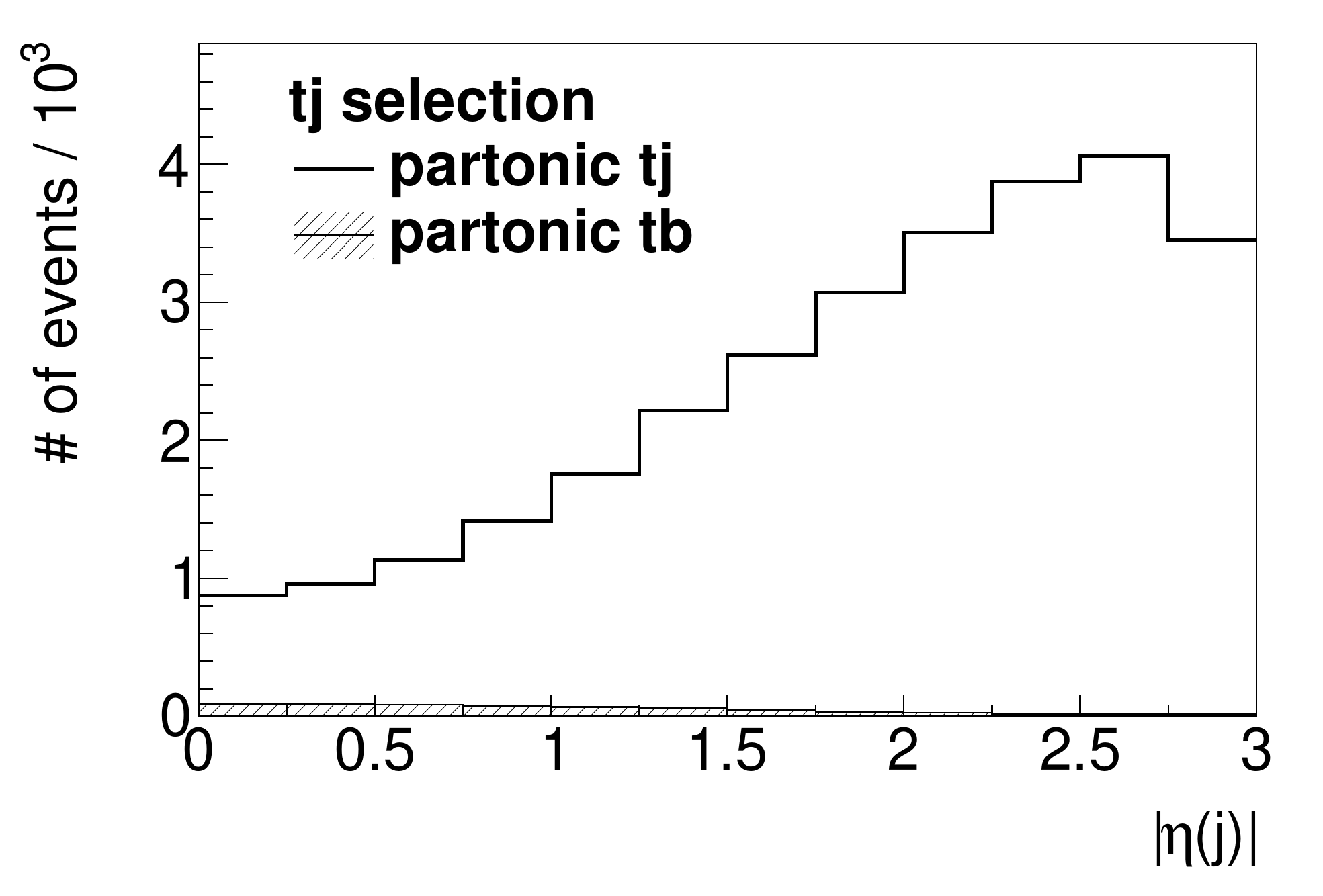}
 \caption{The detector efficiency matrix entries $\epsilon_{ij}$
 ($tb$ selection on each partonic input process left, $tj$ selection right)
 and resulting binned event numbers
 corresponding to $\int\! L=\unit[100]{fb^{-1}}$,
 for various kinematic distributions.
 \label{cc_contact_sm_d}}
\end{figure*}
Within each of the two selections, we fill the histograms of the
kinematic observables introduced at parton level in Eq.~\eqref{dist_obs}.
All histograms are then rebinned appropriately to match parton and detector
levels, so that the efficiency matrix~$\epsilon_{ij}$
is given by the ratio of detector level events in selection~$i$ over
the partonic ones in production channel~$j$,
for each bin in the analysis.
However, there remains a subtlety in the approach:  $b$~tagging should be part
of the detector response, so in a strict sense one cannot use
flavor information at parton level.  On the other hand, the spectator
kinematics (a $b$~jet in the $s$~channel, respecitvely, a light jet in the
$t$~channel)
is a vital property of the final states, so it should be accounted
for in the analysis, including off-diagonal elements of the detector response.
This is resolved by leaving the spectator jet untagged in both channels
\emph{after} the final state selection (i.e.~demanding only one or
exactly two $b$~jets in the $t$, respecitvely, $s$~channel). The spectator, globally
denoted ``$j$'', is then
simply identified in both selections as the hardest jet remaining in the event
after the $b$~tagged one reconstructing the top momentum has been removed.

The result of the procedure is displayed in
Fig.~\ref{cc_contact_sm_d} for some characteristic observables in each
final state, showing the binned entries of the detector efficiency matrix
$\epsilon_{ij}$ in percent as well as
resulting absolute event counts at detector level, normalized to the reference
luminosity $\int\! L=\unit[100]{fb^{-1}}$.  Figure~\ref{cc_contact_sm_d}
illustrates how the $s$~channel selection efficiently suppresses the
$t$~channel sample, whereas the $t$~channel selection is less restrictive
against an  $s$~channel admixture.  However, this high $s$~channel purity is
mandatory because the total cross sections deviate by more than an order of
magnitude in the two production channels, so that the $s$~channel still
suffers from sizable $t$~channel pollution, as reflected in the total event
numbers resolved by the partonic sample input.  On the other hand, for the same
reason the $t$~channel
selection is indeed very clean, despite the fact that the efficiencies of
the partonic input samples only deviate roughly by a factor of two.
In the last bin of the $|\eta(j)|$ spectator histogram, the $tj$
selection efficiencies even become identical on both inputs, which is actually
a consistency requirement:  since $b$~tagging does not work any more above
$|\eta|\gtrsim 2.5$, the hard $b$ in the $s$~channel mimics the hard light
jet in the $t$~channel and hence passes the respective selection criteria
with practically the same efficiency as an original partonic $tj$ event.
\renewcommand{\imgsize}{0.4}
\begin{figure*}
 \includegraphics[scale=\imgsize]{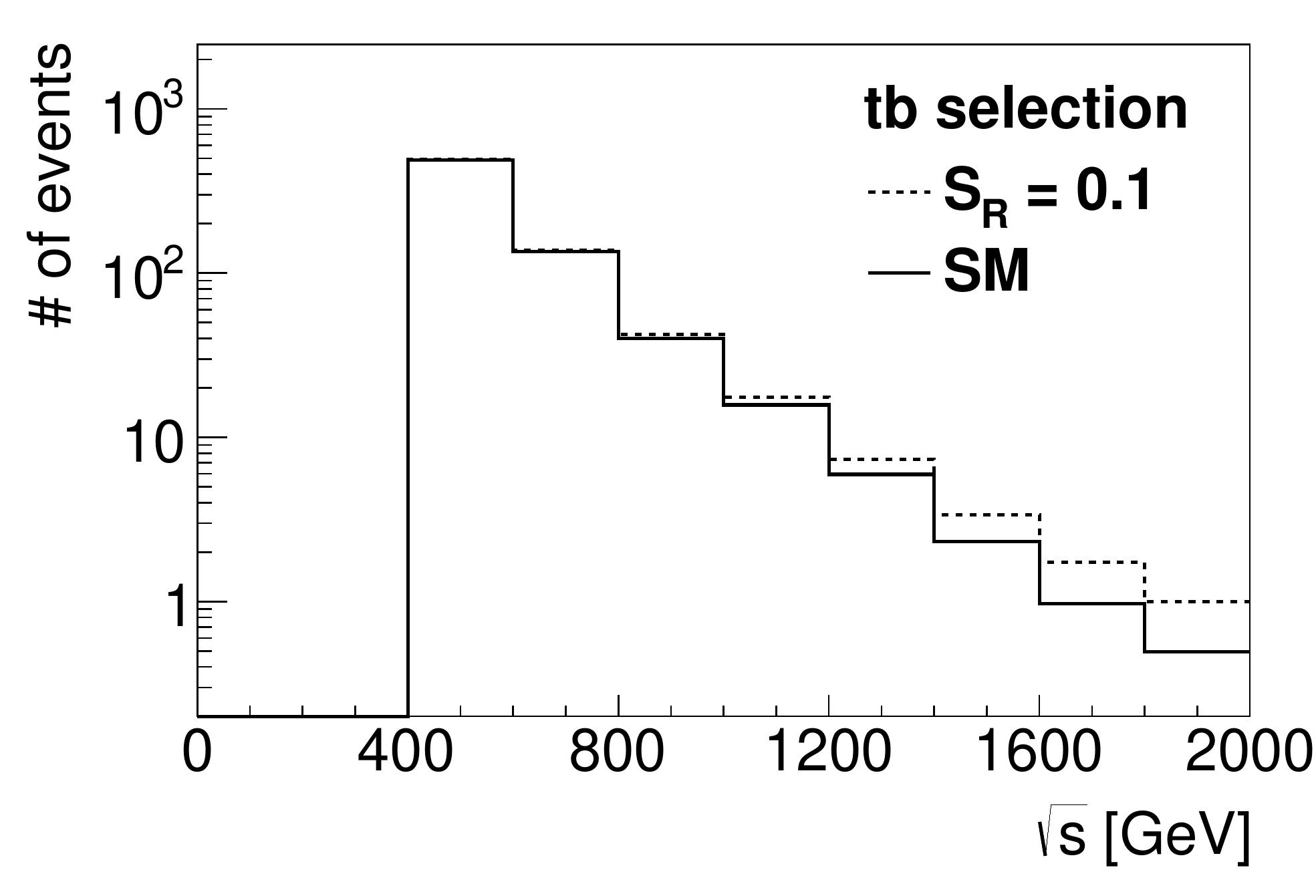}
% \vspace{-5mm}
 \hfill{}
% \vspace{0.5cm}
 \includegraphics[scale=\imgsize]{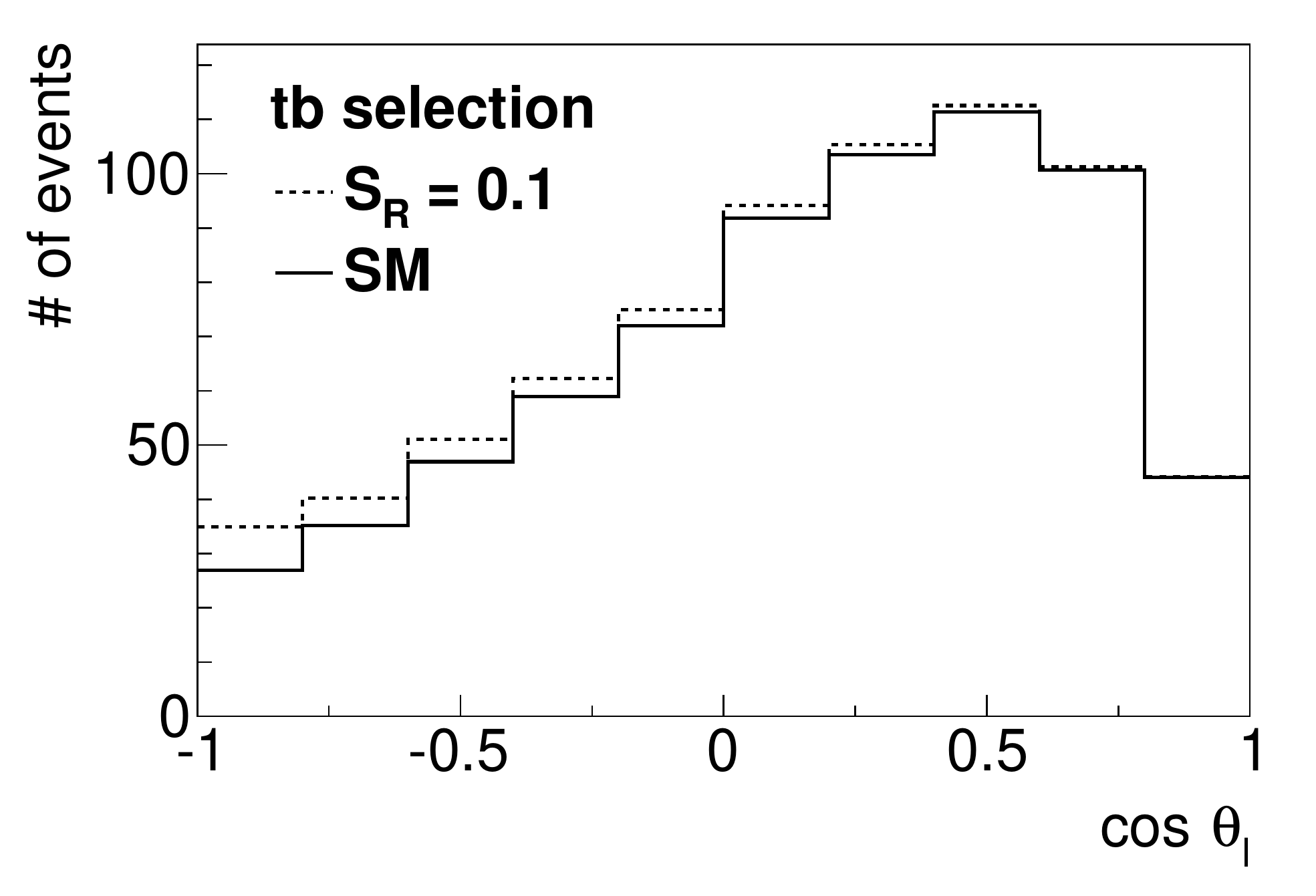}
 \caption{Detector level histograms of $\sqrt s$ (left) and
 $\cos\theta_\ell$ (right) after $tb$~selection at the reference point
 $S_R=0.1$ (event numbers normalized to $\int\!L=\unit[100]{fb^{-1}}$).
 \label{cc_contact_rp_d}}
\end{figure*}

\renewcommand{\imgsize}{0.56}
\begin{figure*}
 \vspace{-3mm}
 \includegraphics[scale=\imgsize]{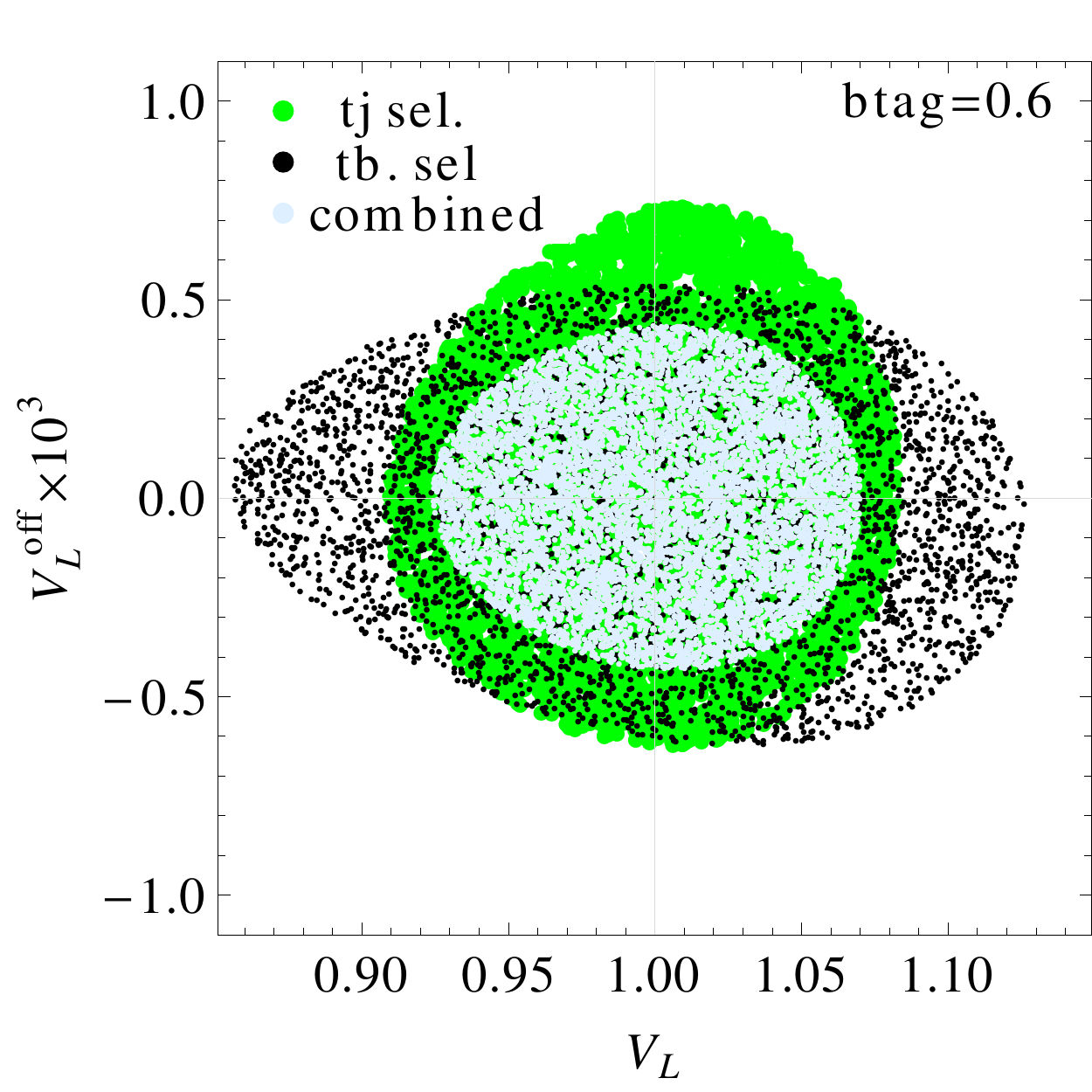}
% \hspace{0.3cm}
 \vspace{-1.3mm}
 \hfill{}
 \includegraphics[scale=\imgsize]{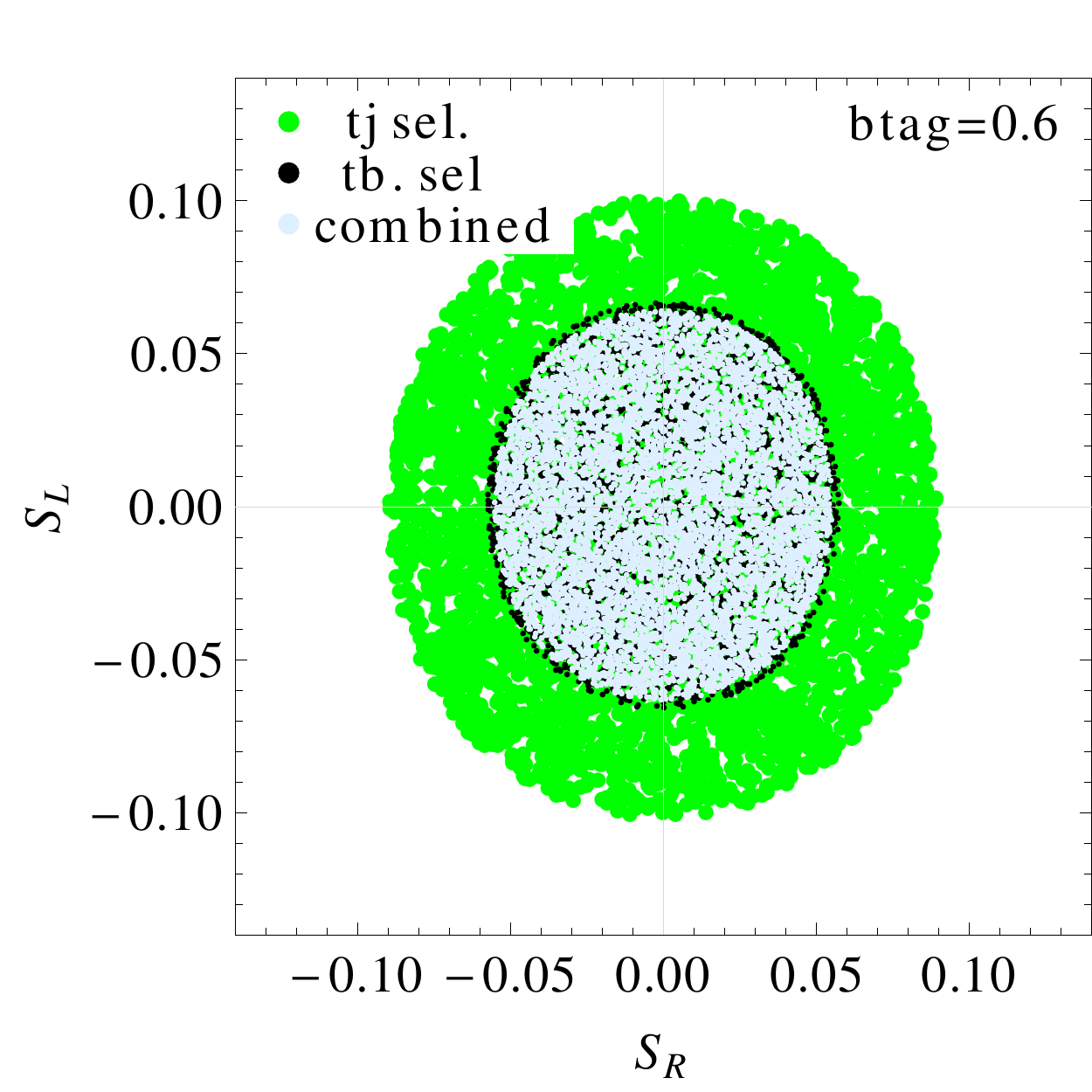}
% \vspace{-8mm}
 \includegraphics[scale=\imgsize]{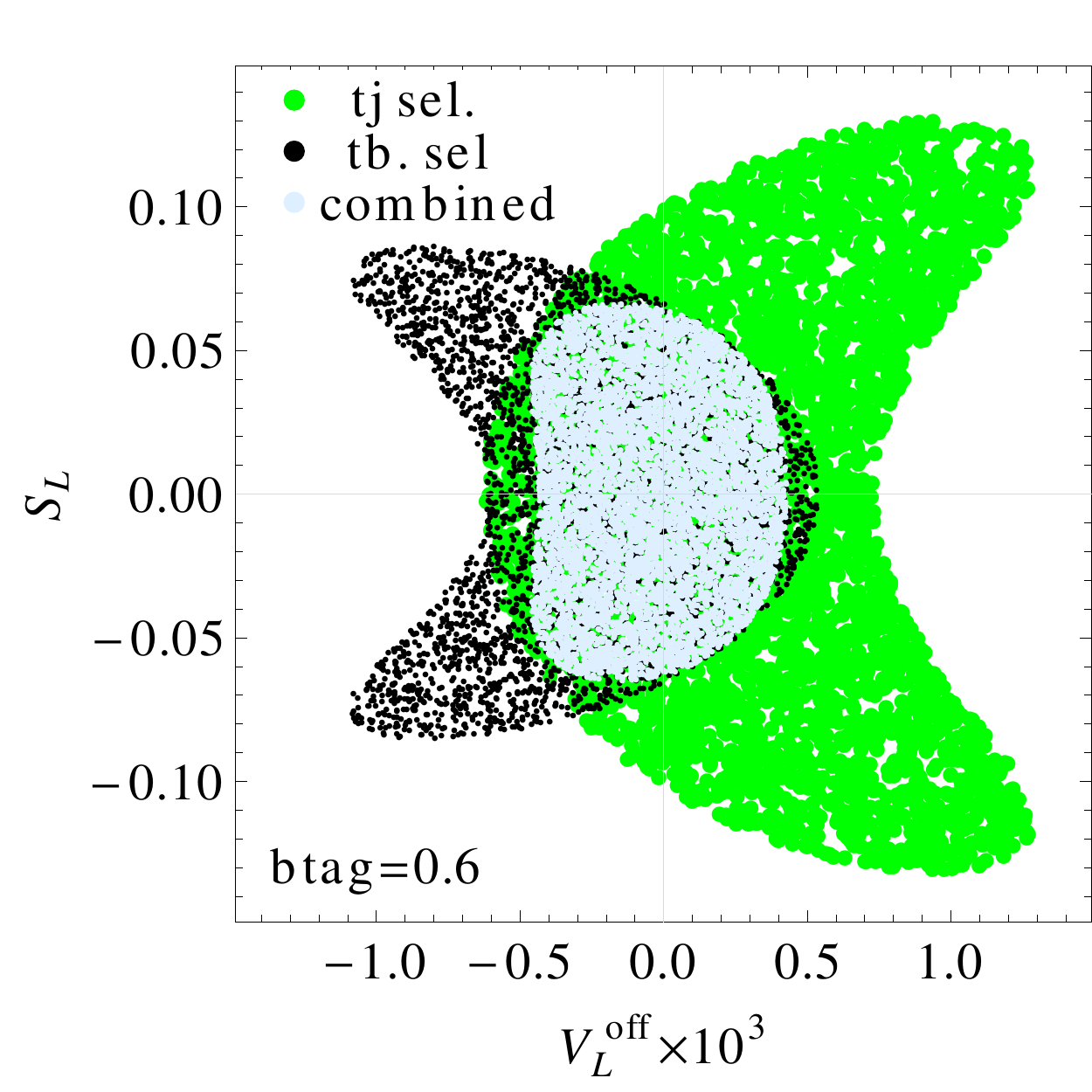}
% \hspace{0.3cm}
 \vspace{-1.5mm}
 \hfill{}
 \includegraphics[scale=\imgsize]{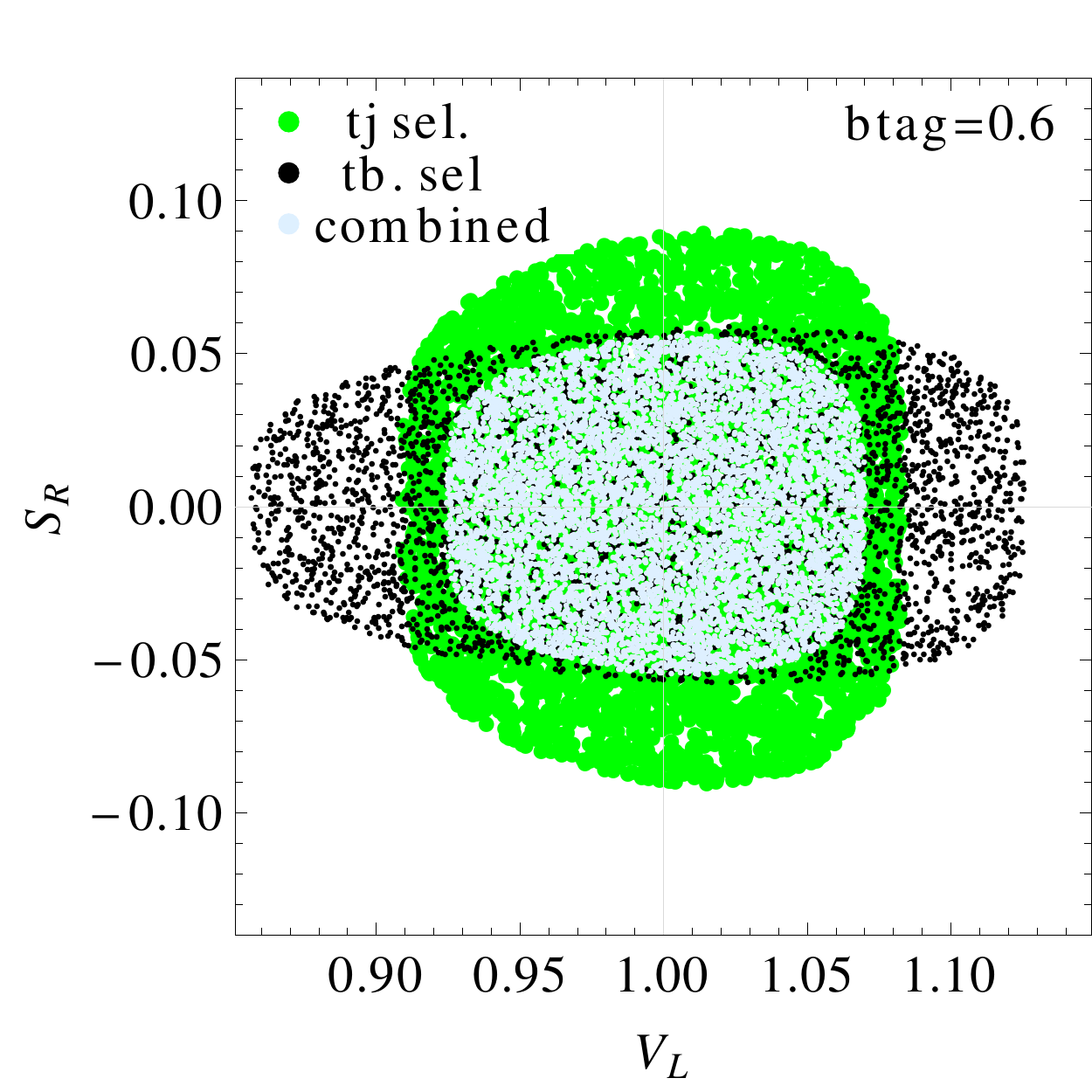}
 %\hspace{0.5cm}
 \includegraphics[scale=\imgsize]{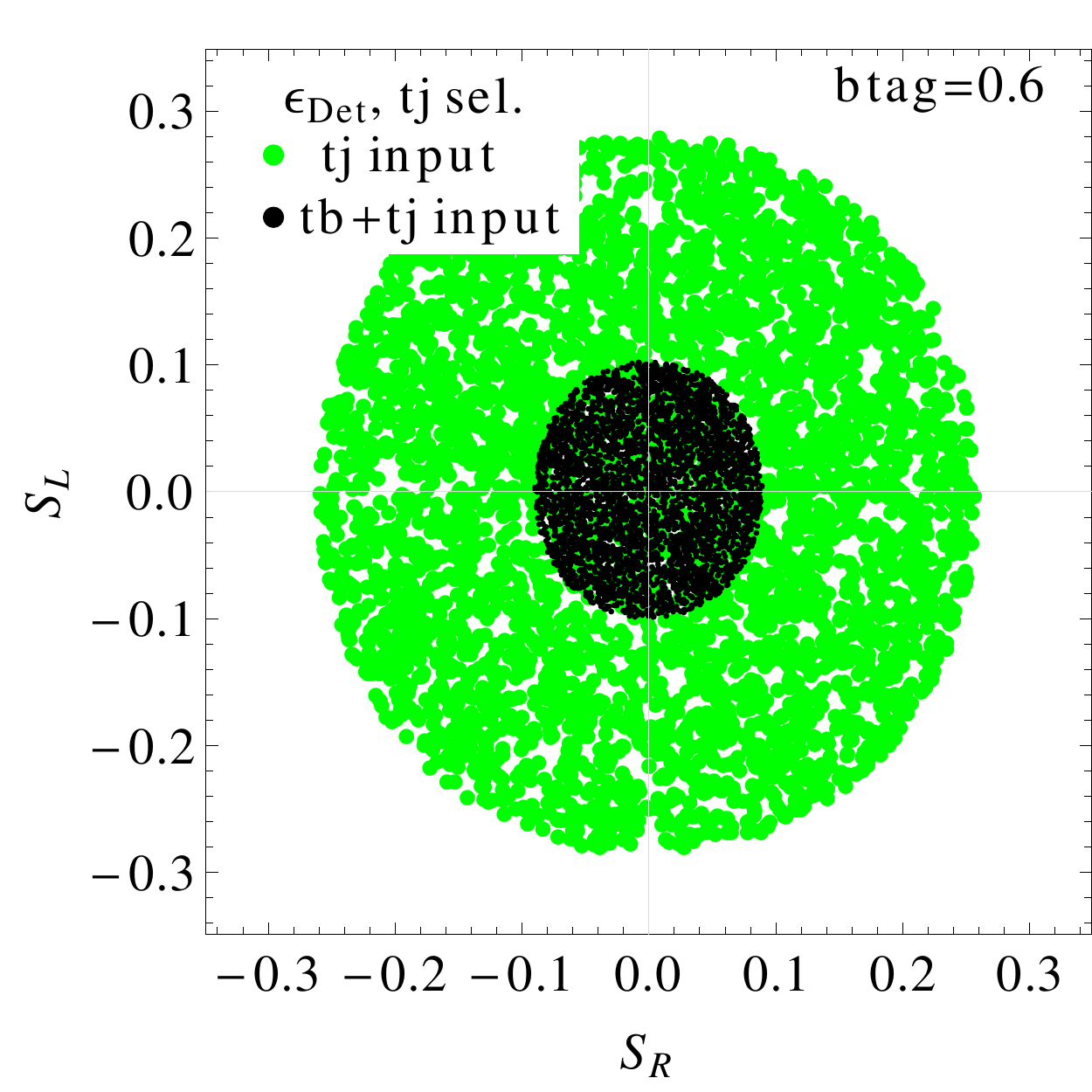}
% \hspace{0.3cm}
 \vspace{-5mm}
 \hfill{}
 \includegraphics[scale=\imgsize]{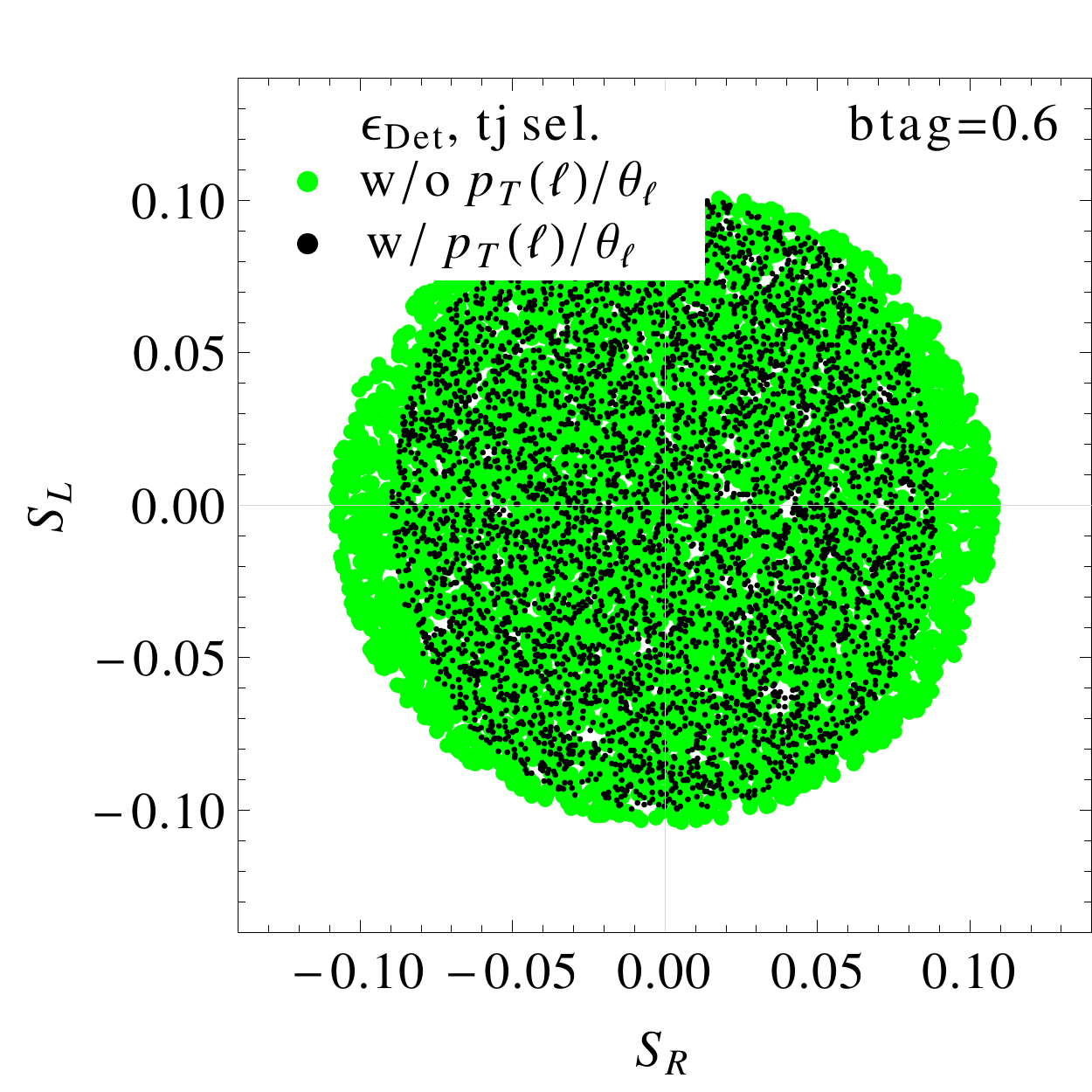}
% \vspace{-5mm}
 \caption{
 The $1\sigma$ limits resolved by final state selections at detector level
 in various coupling planes, always setting the other
 couplings to their SM values (top and center rows).
 The bottom row shows additional features in the $S_L$-$S_R$ plane (see text).
 \label{cc_contact_scatter_w}}
\end{figure*}
Going on now to the results based on the binned likelihood test which are
summarized in Fig.~\ref{cc_contact_scatter_w}
(cf.~the Appendix for systematic effects from variations of various analysis
parameters),
let us first compare the
$1\sigma$ bounds resolved by final state selections displayed in the top row
with the ones on the partonic level (top row in
Fig.~\ref{cc_contact_scatter_part}).  It is no surprise that generally the
limits tend to get worse at the detector level, especially in the $s$~channel
which gets considerably diluted by $t$~channel events.
However, in the
scalar plane (top right in Fig.~\ref{cc_contact_scatter_w})
one observes the counterintuitive result that the
$t$~channel sensitivity appears to \emph{increase} at the detector level,
but in truth
this is also explained by the selection impurity, however tiny in the
$t$~channel at the SM~point.  Once the scalar NP couplings are turned on,
the $s$~channel events also significantly populate the $t$~channel bins at the
detector level, due to the fact that the integrated NP cross sections differ
by orders of magnitude, as already pointed out in the parton level discussion
above.  A simple cross-check is given by switching off the $s$~channel
partonic input to the $t$~channel selection, as was done in the bottom left
plot in Fig.~\ref{cc_contact_scatter_w}, highlighting that the
$t$~channel sensitivity to the scalar couplings is indeed governed by the
$s$~channel admixture.  It follows that the best limits on the scalar
couplings are entirely driven by the $s$~channel sensitivity, numerically
amounting to $|S_L|\lesssim 0.075$, respecitvely, $|S_R|\lesssim 0.065$
in this study (varying all four couplings independently),
which translates into a sensitivity on the underlying Wilson coefficients
already of $\OO(0.1\text{--}1)$
when the naive NP scale estimate $\Lambda\sim\unit[3]{TeV}$ is employed.
The enhanced sensitivity to the right-handed part is again due to the
discriminative observables $p_T(\ell)$ and $\cos\theta_\ell$ that were
already discussed at the parton level.
To further quantify this observation, in Fig.~\ref{scalar_cont} we show the
relative $\chi^2$~value
$\chi^2|_{p_T(\ell),\cos\theta_\ell}/\chi^2_\text{tot}$ collected in these
observables inside the $1\sigma$ region obtained from the $tj$~selection.
One finds that along the $S_R$ direction the $\chi^2$ is indeed
dominated ($\gtrsim\unit[50]{\%}$) by $p_T(\ell)$ and $\cos\theta_\ell$,
while along $S_L$ the contribution is moderate ($\lesssim\unit[10]{\%}$).
For illustration, Fig.~\ref{cc_contact_rp_d} shows the effect of an NP
contribution corresponding to $S_R=0.1$, which is in tension with the SM at
the $\sim 3\sigma$~level in our analysis (varying all couplings), to the
$tb$~channel selection.
\renewcommand{\imgsize}{0.63}
\begin{figure*}
% \vspace{-2mm}
% \centerline{
 \includegraphics[scale=\imgsize]{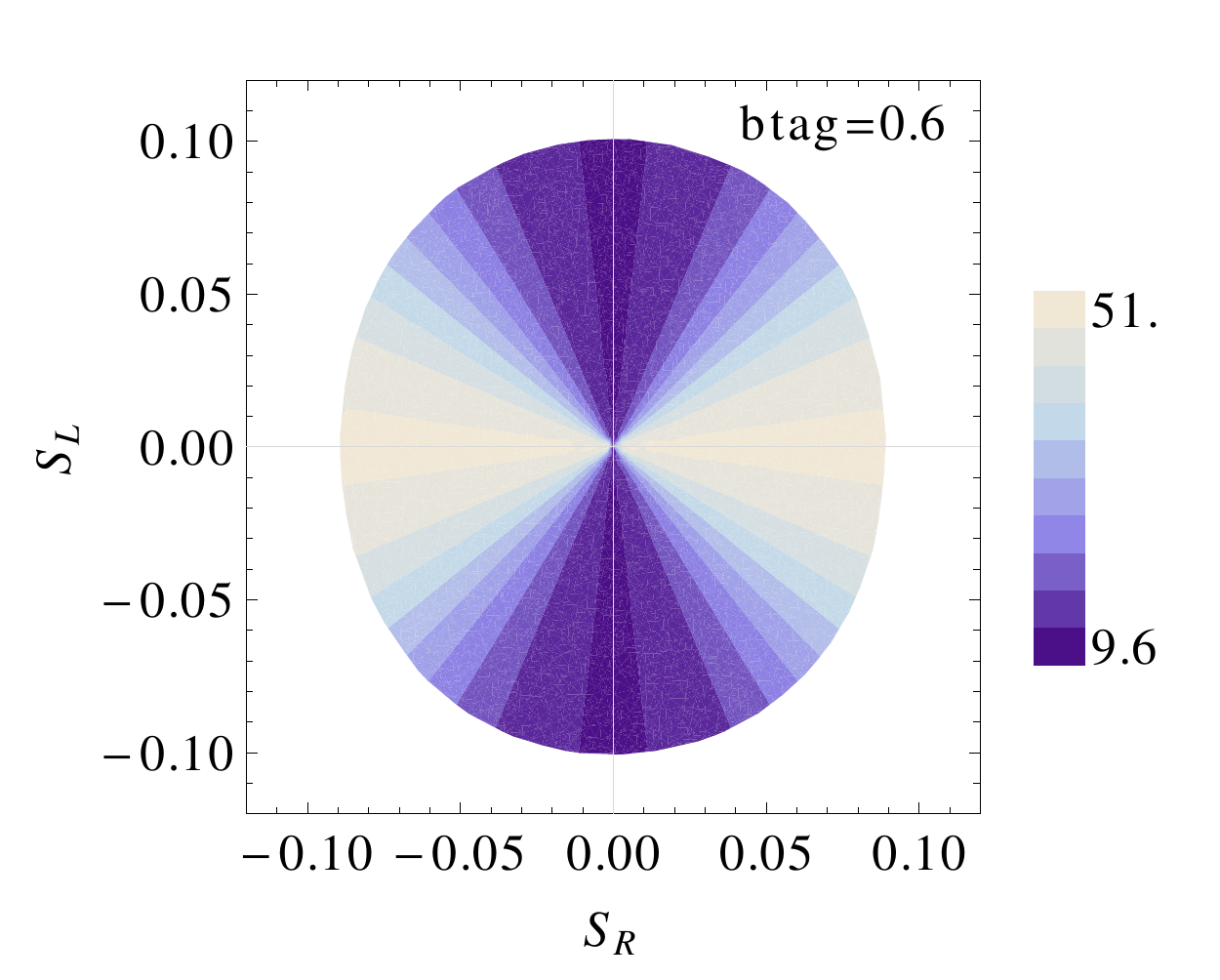}
% \vspace{-5mm}
 \hfill{}
% \vspace{0.5cm}
 \includegraphics[scale=\imgsize]{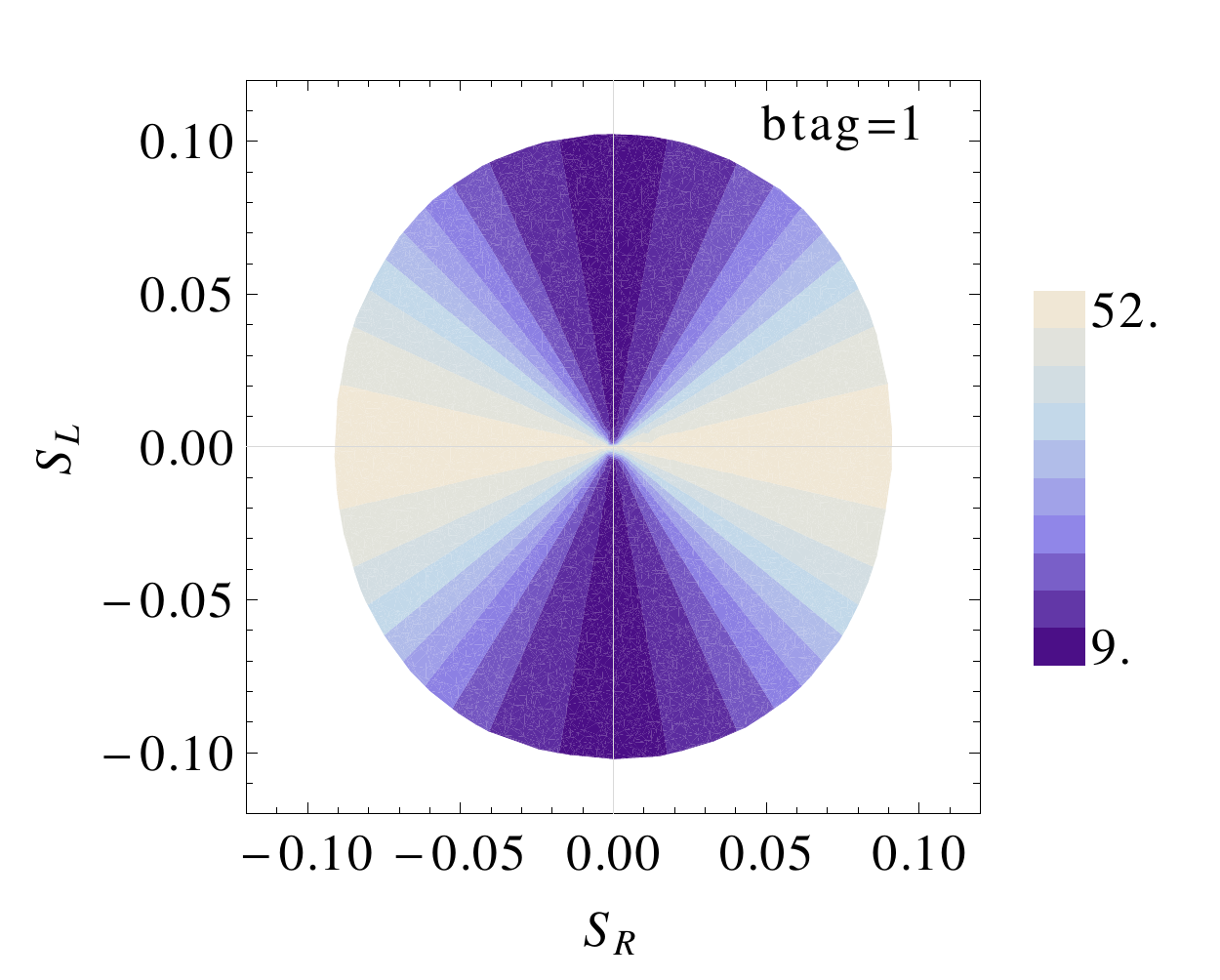}
% }
 \caption{Contribution from the helicity sensitive observables $p_T(\ell)$
 and $\cos\theta_\ell$ to the total $\chi^2$~value (in percent) inside the 
 $1\sigma$ region from $tj$ selection in the scalar plane.
 \label{scalar_cont}}
\end{figure*}

In the vector plane (top left in Fig.~\ref{cc_contact_scatter_w}),
the relative matrix element sizes of the two channels do not deviate that
much, so that the sensitivities generally drop at detector level.
Varying all four couplings independently,
the best limits are obtained by a combination of both channels,
numerically resulting in
$0.90\lesssim V_L\lesssim 1.09$ and $-0.6\lesssim\VO\times10^3\lesssim 0.55$,
the latter being almost exactly 2~orders of magnitude
better than the result of~\cite{Bach2012} and corresponding to a sensitivity
on the respective Wilson coefficient of $\OO(0.01\text{--}0.1)$
at $\Lambda\sim\unit[3]{TeV}$.
\renewcommand{\imgsize}{0.63}
\begin{figure*}[t]
% \vspace{-3mm}
 \includegraphics[scale=\imgsize]{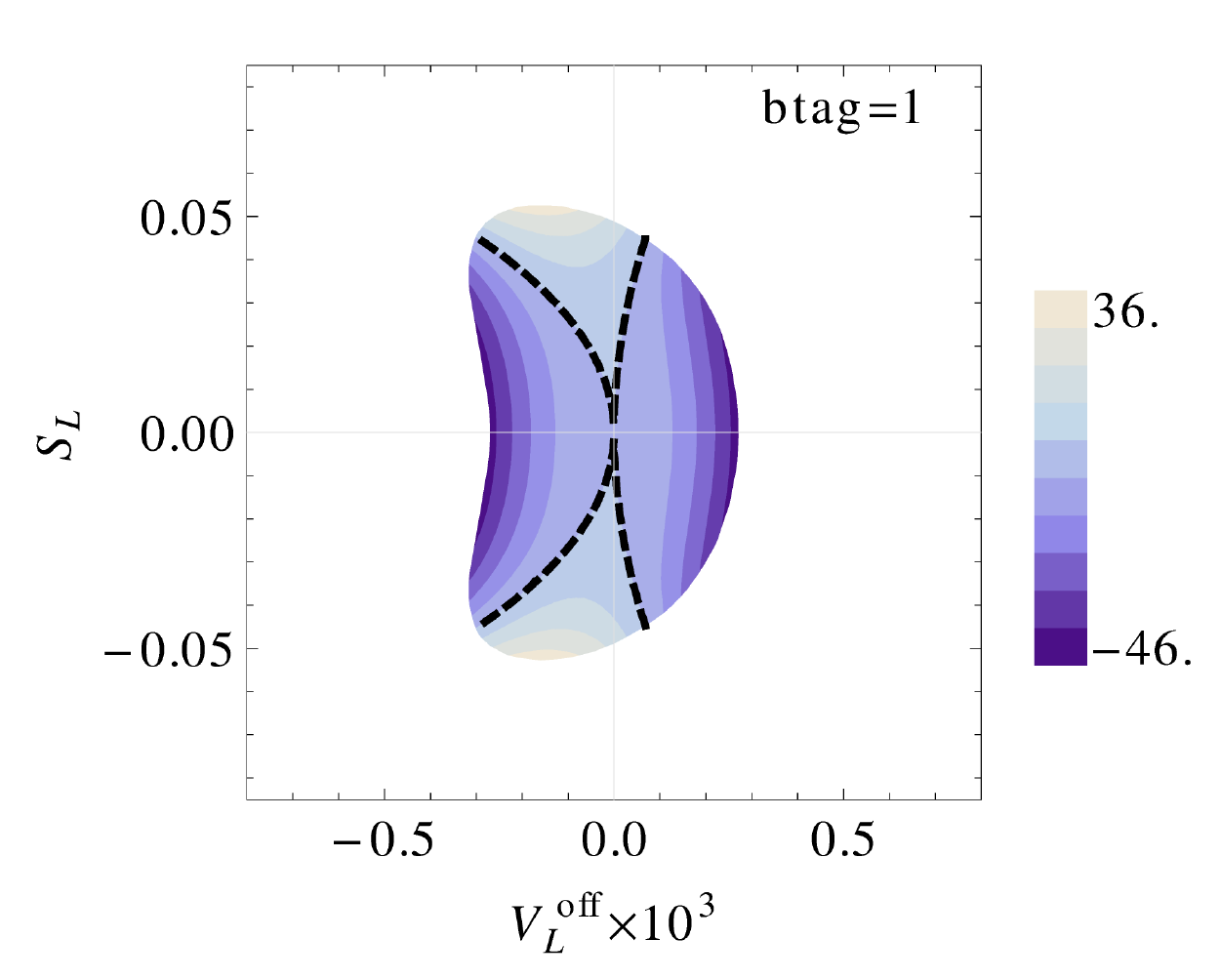}
% \hspace{0.3cm}
% \vspace{-1.3mm}
 \hfill{}
 \includegraphics[scale=\imgsize]{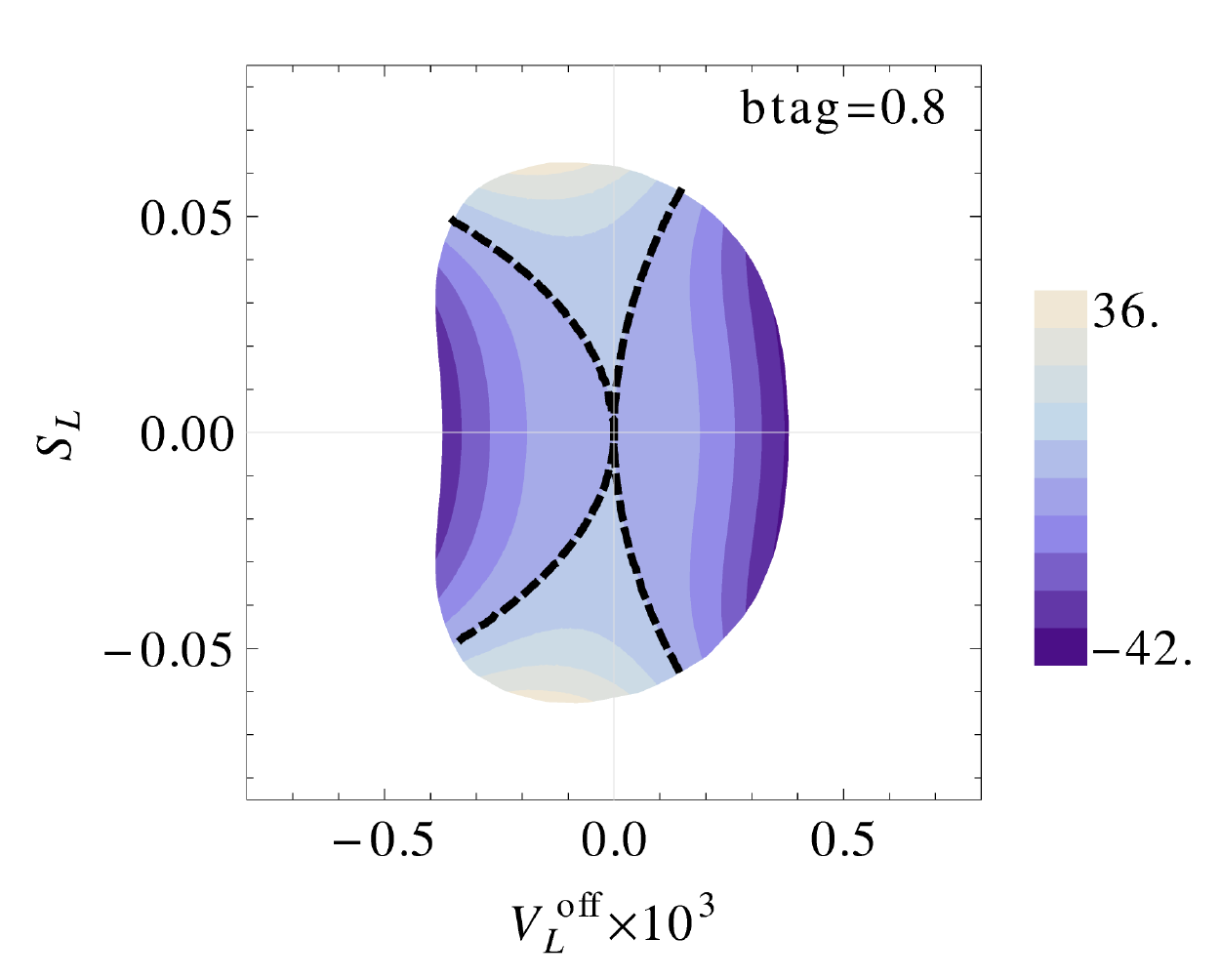}
% \vspace{-8mm}
 \includegraphics[scale=\imgsize]{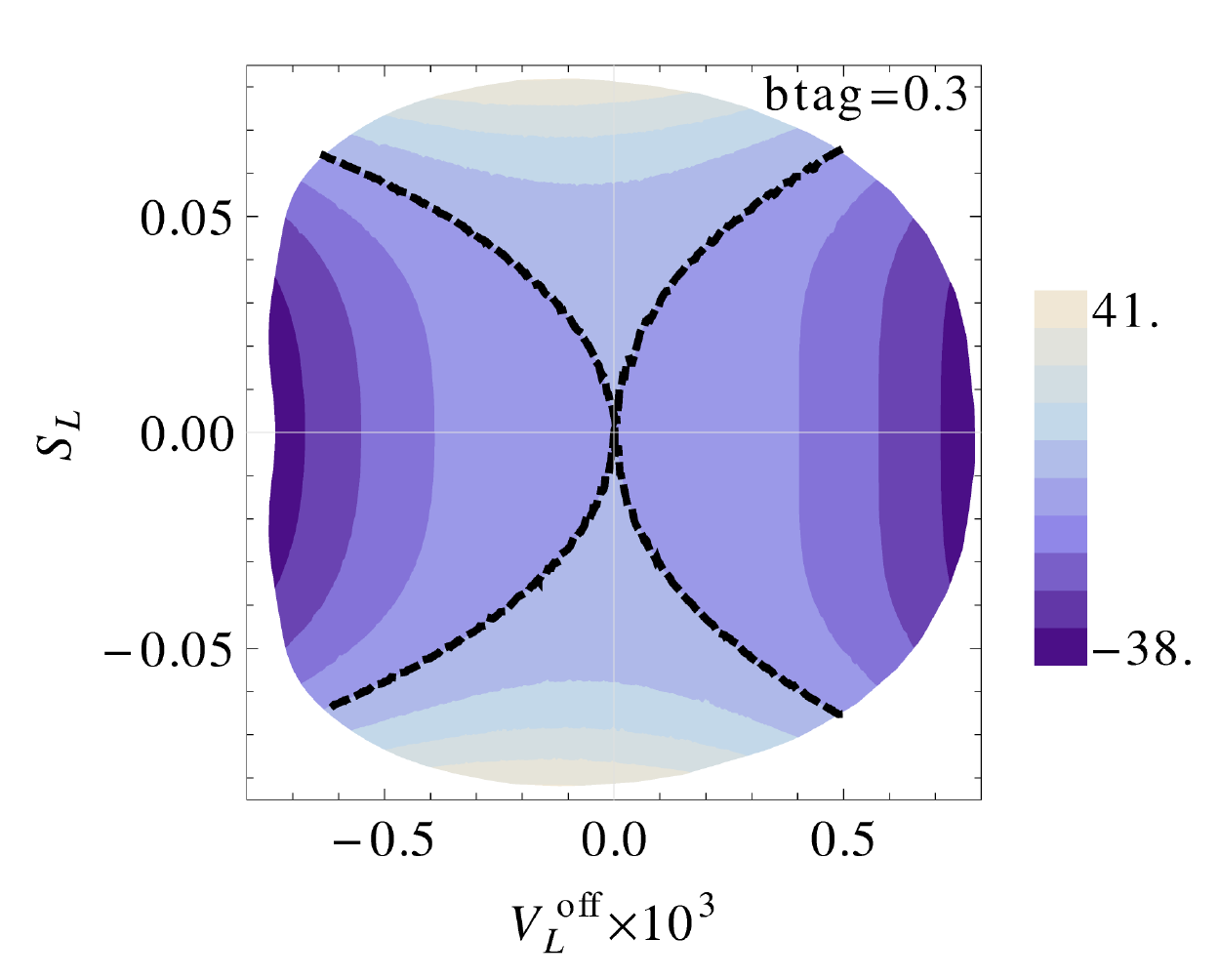}
% \hspace{0.3cm}
% \vspace{-1.5mm}
 \hfill{}
 \includegraphics[scale=\imgsize]{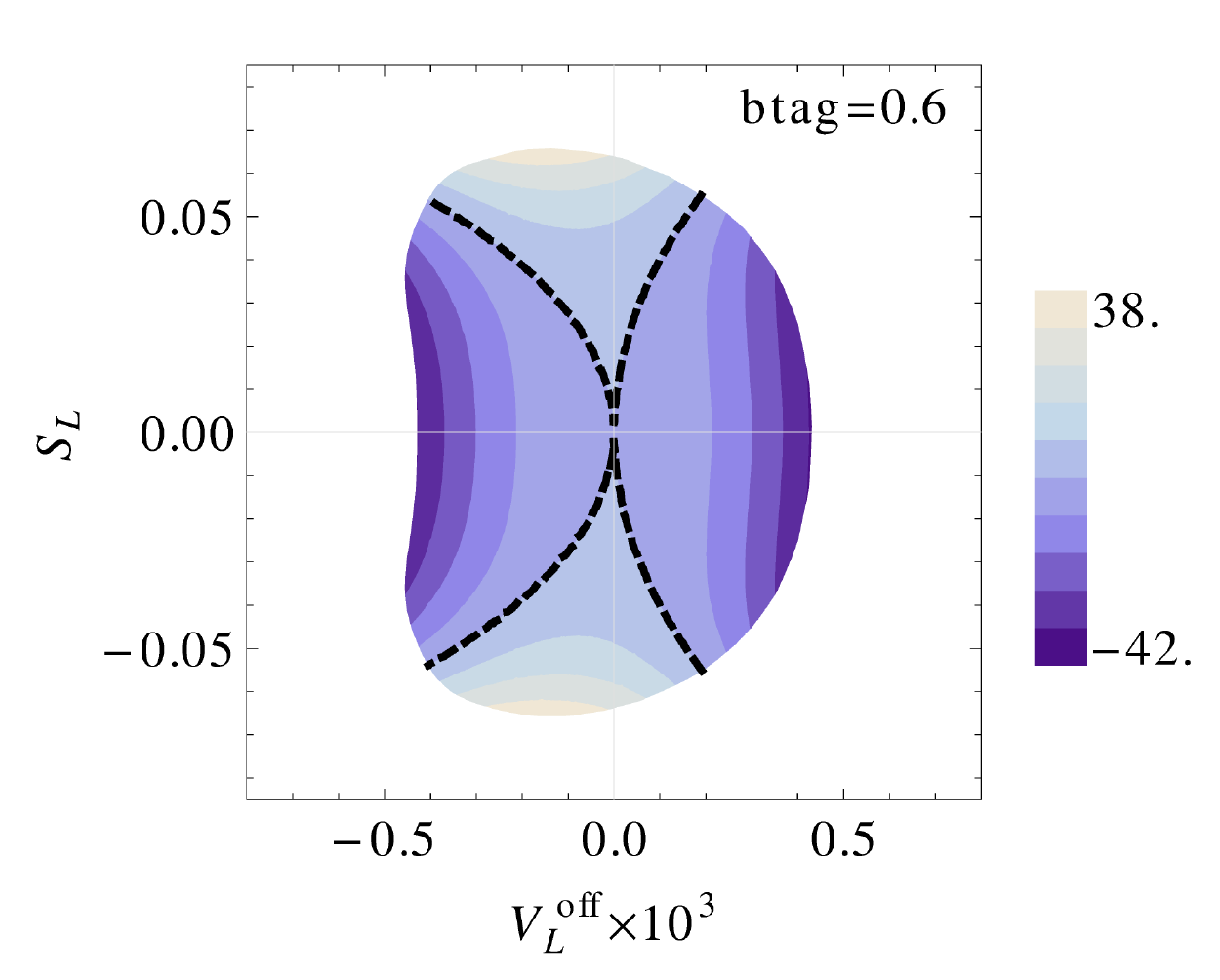}
 %\hspace{0.5cm}
 \caption{
 Value of the vector--scalar discriminant $P$ inside the $1\sigma$
 region from combination of $tb$ and $tj$ selections in the
 $\VO$--$S_L$ plane, with varying $b$-tagging setups.
 The 0-contour is highlighted by black dashed lines.
 \label{scalar_vector_pm}}
\end{figure*}
It remains to address the discrimination of $\VO$ vs~scalar NP contributions.
As already hinted in Sec.~\ref{pheno_p}, this may be achieved by
correlating $s$~channel with $t$~channel observations, thus exploiting the
$V_L$--$\VO$ interference which switches sign between the channels
(cf.~bottom left in Fig.~\ref{cc_contact_scatter_part}, respecitvely, center
left in
Fig.~\ref{cc_contact_scatter_w}).  However, as also becomes clear once more
from these plots, the final state selections suffer from mutual admixture,
essentially governed by the $b$-tagging performance because the main
difference of the selections is the tag of the hardest spectator jet.
Nonetheless, defining a sign sensitive channel-specific pull
\begin{align}
 P_k(\vec g) &= \sum_{i\in k} \left(
  \frac{w^\text{exp}_i -w^\text{th}_i(\vec g)}{\delta_i} \right)
  \;,\quad k=s,t\text{ channel}
\end{align}
one can construct a discriminant $P\equiv P_s\cdot P_t$, which is negative
(positive) in a $\VO$ ($S_{L,R}$) enriched sample.
The result is shown in Fig.~\ref{scalar_vector_pm}, resolved by various
$b$-tagging setups apart from the default one described above,
namely perfect tagging (unit efficiency and vanishing impurity) as well as
several reference points taken from the characteristic efficiency/impurity
curve as estimated
in~\cite{Aad2009}.
One observes that while the general correlation between the sign of $P$ and
the parameter point~$\vec g$ is retained, the $b$~tagging affects not only
the overall bounds but also the curve of the 0~contour
(black dashed lines in Fig.~\ref{scalar_vector_pm}) particularly in the
positive $\VO$ hemisphere:
this systematics is essential for the accuracy of delimiting
the relative admixture of $\VO$ and $S_{L,R}$.
On the other hand, note that the helicity disambiguation shown in
Fig.~\ref{scalar_cont} can be carried out entirely in the $t$~channel
selection, so that the outcome is stable
against the $b$-tagging setup.

In summary, it can be stated that generally both the absolute limits as well
as the disambiguation of vector
and scalar structures require digging out the tiny amount of
$s$~channel events among the $t$~channel ones to a very high precision
at the detector level.
In the absence of the forward jet tag in the $t$~channel (which would kill the
NP~signal as illustrated in Fig.~\ref{cc_contact_tj_part}, bottom left),
this reduces to an accurate understanding of the $b$-tagging performance.

\section{Conclusions}\label{sum}

In this study, we have addressed the impact of
including anomalous four-fermion terms $tbf\! f^\prime$ into the anomalous
top charged-current coupling basis,
and point out the possibilities to assess their size and specifically resolve
the remaining ambiguities experimentally, by analyzing single top production
in the $s$ and $t$~channels at the LHC.
The strategy is to employ a binned likelihood test over a
set of sensitive kinematic observables, thus exploiting the different
kinematic behavior of the four-fermion interactions.
In this context, we have first discussed the minimal set of independent
minimally flavor violating contact interactions,
originally consisting of one vector and eight scalar couplings
emerging from dimension six operators.
However, after exploiting some kinematic degeneracies, the parameter space
%for this study
is reduced to two scalar couplings $S_{L,R}$ in addition to the vector
coupling $\VO$, while the trilinear $tbW$ coupling $V_L\sim V_{tb}$
normalizing the SM part is kept also
because of the $V_L$--$\VO$ interference, and the insensitivity of
$W$~helicity fractions.
Detector effects have been taken into account for each bin, including
off-diagonal elements among the $s$ and $t$~channel from selection impurity.
It turns out that this mutual signal pollution is crucial for the
quality of the resulting bounds, because particularly in the scalar
directions the $s$~channel sensitivity exceeds the $t$~channel one by orders
of magnitude, with the immediate consequence that any gain in the $s$~channel
signal purity is directly reflected in improved limits on the scalars.

Finally, numerical bounds were obtained from the
analysis normalized to $\int\! L=\unit[100]{fb^{-1}}$ at
$\sqrt s=\unit[14]{TeV}$, illustrating the excellent sensitivity reach,
of $\OO(0.01\text{--}1)$ on the Wilson coefficients for $\Lambda=\unit[3]{TeV}$,
to such charged-current contact interactions
in single top production.
Particularly, the interference in the $V_L$-$\VO$ plane which caused the
bounds to leak out to small $V_L$ values in correlation with $\VO$
still allowed by the total cross section analysis
of~\cite{Bach2012} is cleanly cut away here.
In terms of disambiguating the various contact couplings in the case that
a deviation from the SM should become significant, two differential
observables are identified which are sensitive to anomalous
right-handed top production, namely the charged lepton momentum $p_T(\ell)$
as well as the spin analyzer angle $\cos\theta_\ell$, which becomes a clean
window to right-handed top production once the values of the spin analyzers
are fixed with respect to the trilinear couplings $V_R$, respecitvely,
$g_{L,R}$ from $W$~helicity fractions.
On the other hand, if the experimental signal purities permit it,
scalar and vector contributions can potentially be told apart by
exploiting the sign change in the $V_L$--$\VO$ interference
in the two production channels,
which poses an interesting challenge to the LHC experiments for the upcoming
high energy, high luminosity run.

\begin{acknowledgments}
F.~B.~was partly supported by
Deutsche Forschungsgemeinschaft through the Research Training Group GRK\,1147
\textit{Theoretical Astrophysics and Particle Physics}.
\wz\ development is supported by the Helmholtz Alliance
\textit{Physics at the Terascale}.
Parts of this work
are supported by the German Ministry of Education and Research (BMBF) under
Contract No.~05H09WWE.
\end{acknowledgments}

\appendix*
\section{Variation of the analysis setup}

\renewcommand{\imgsize}{0.62}
\begin{figure*}
 \vspace{-3mm}
 \includegraphics[scale=\imgsize]{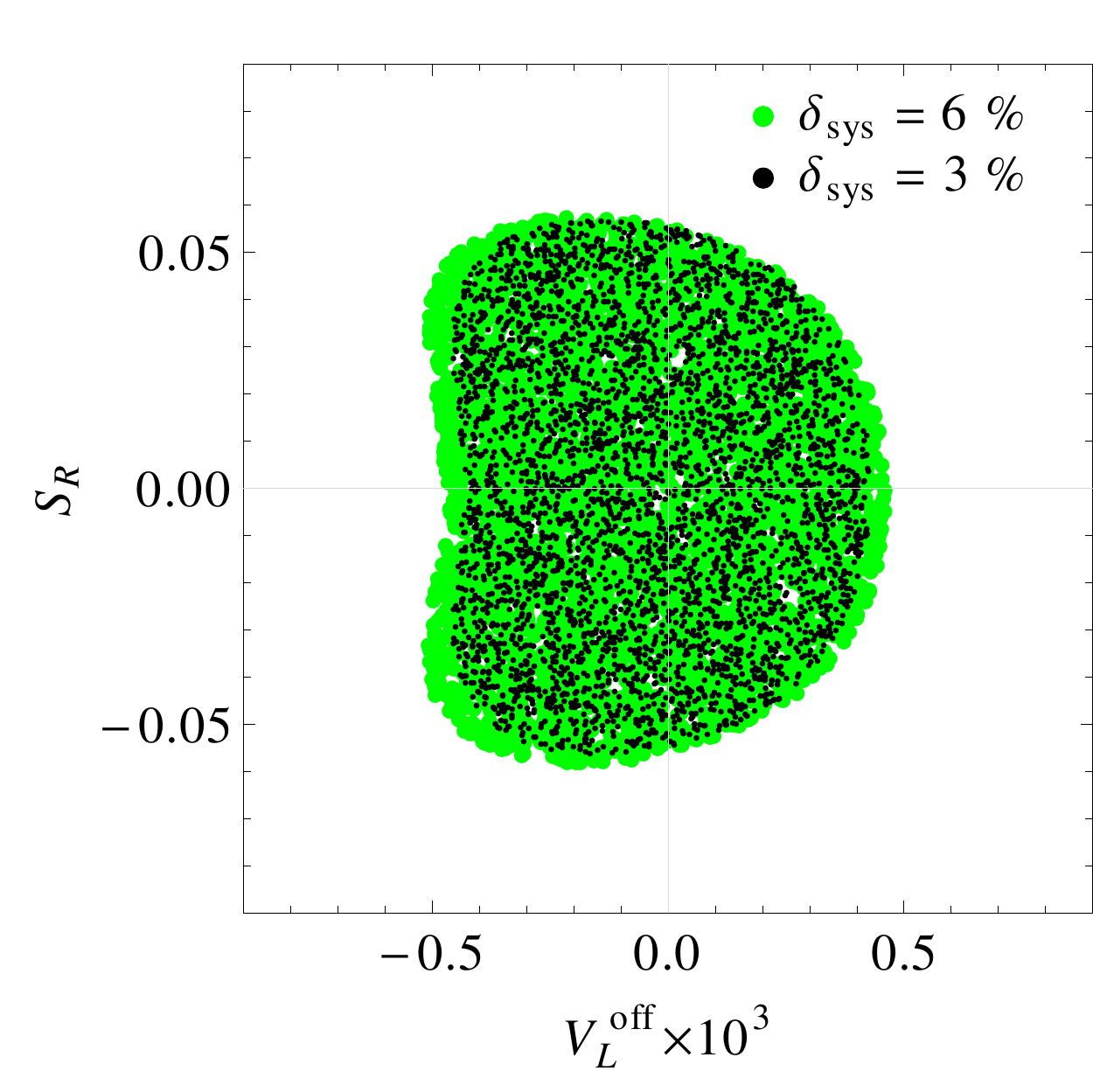}
% \hspace{0.3cm}
 \vspace{-1.5mm}
 \hfill{}
 \includegraphics[scale=\imgsize]{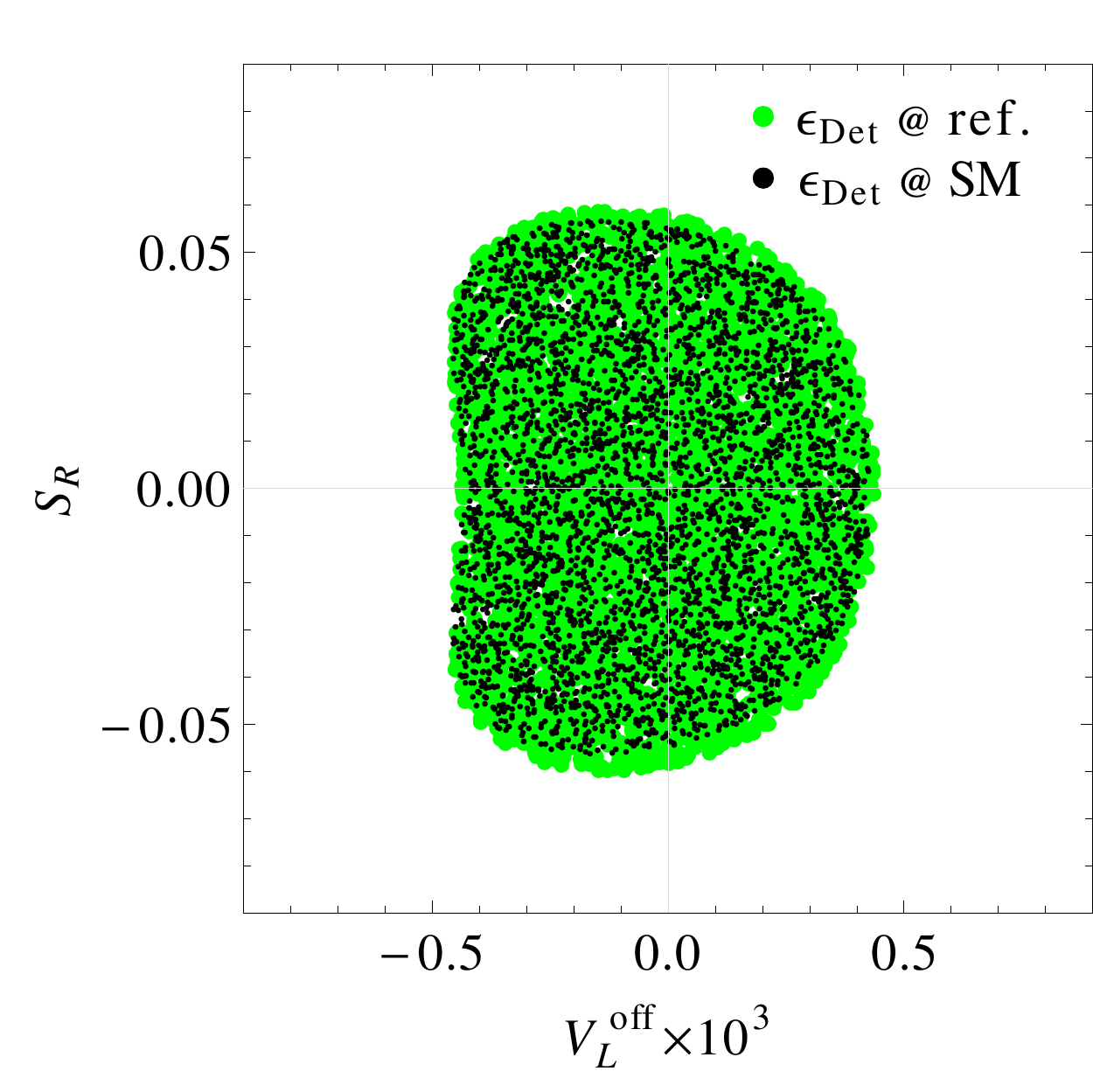}
 %\hspace{0.5cm}
 \includegraphics[scale=\imgsize]{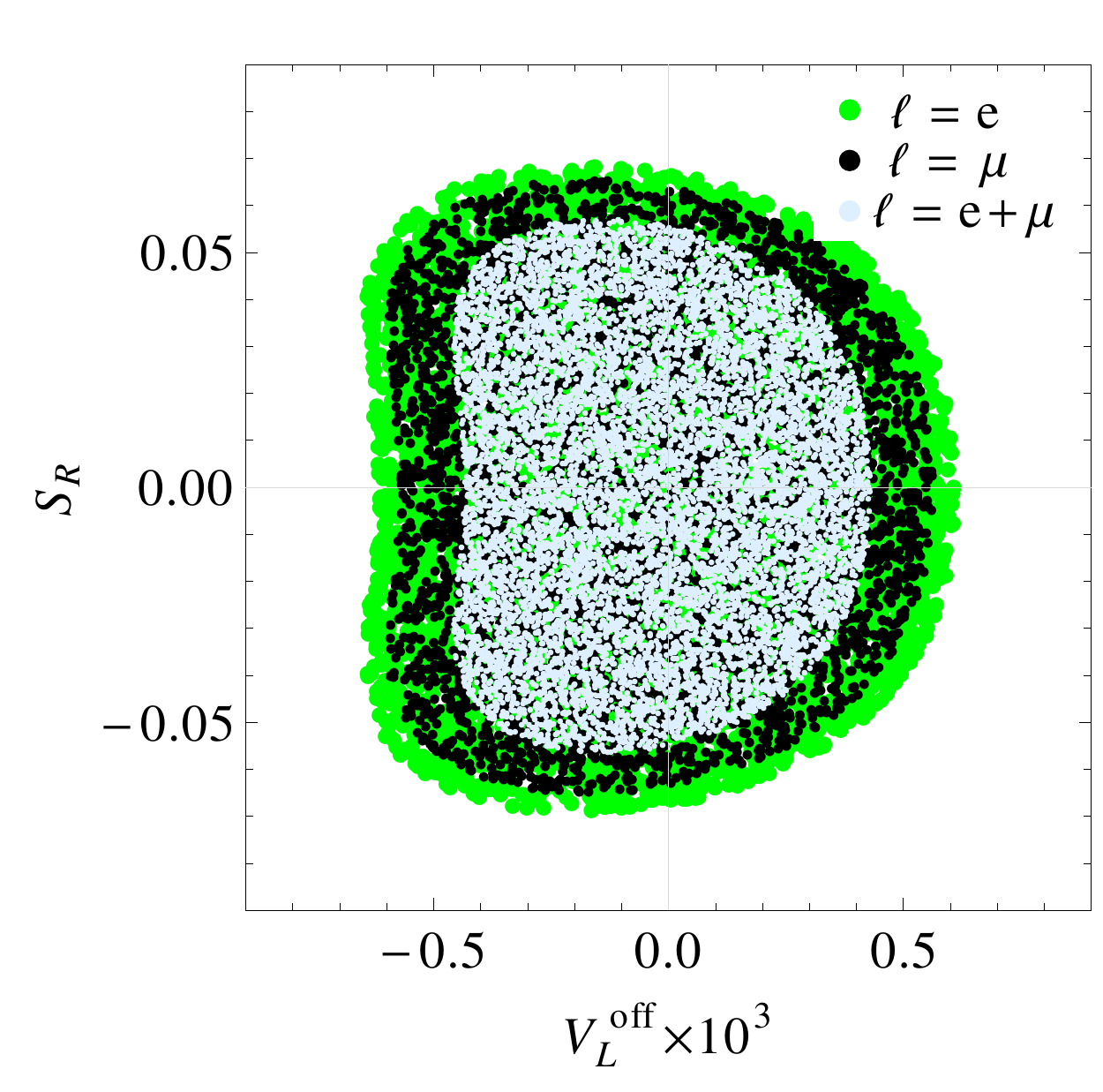}
% \hspace{0.3cm}
 \vspace{-5mm}
 \hfill{}
 \includegraphics[scale=\imgsize]{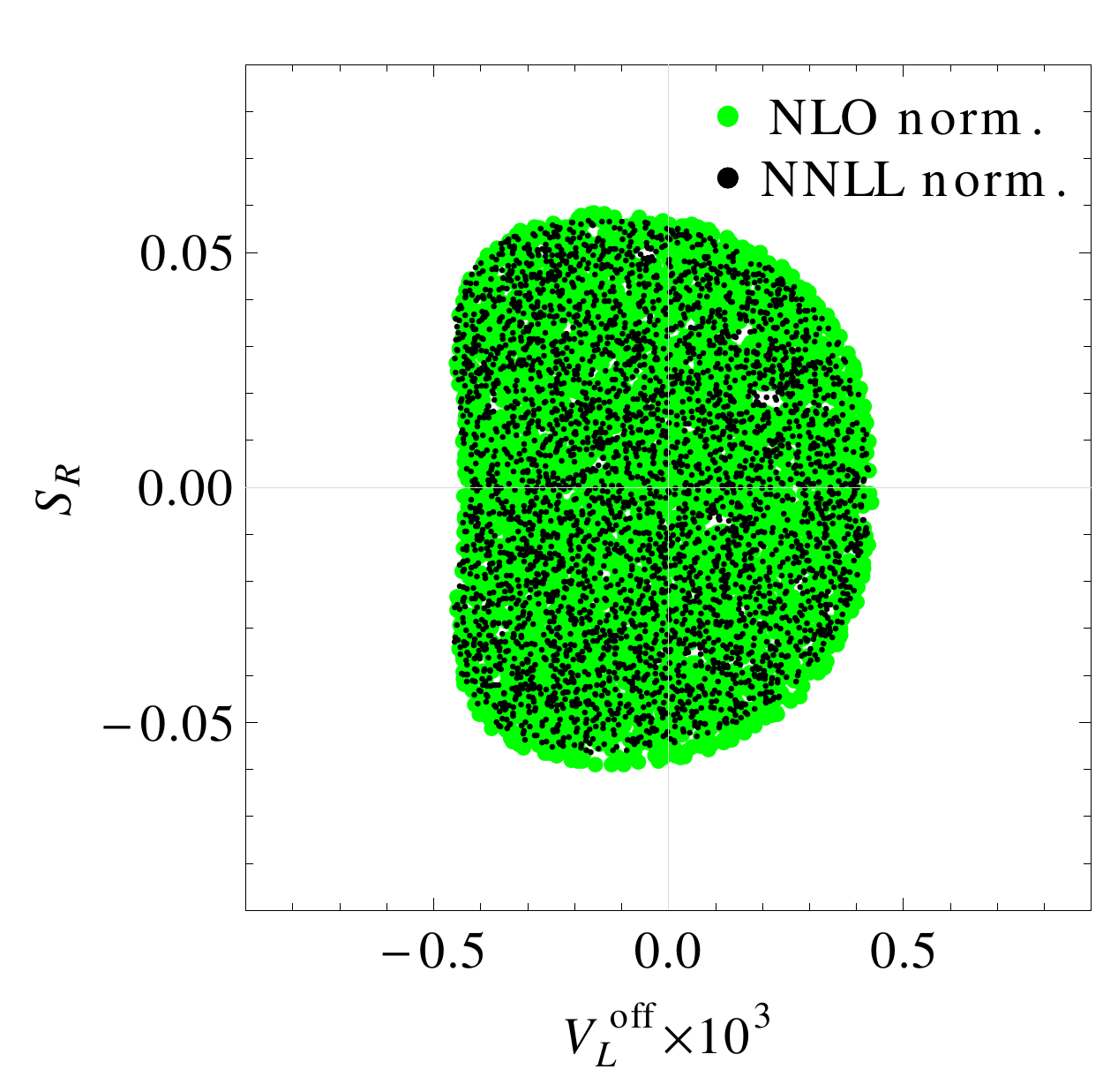}
% \vspace{-5mm}
 \caption{Impact of various parameter variations on the combined $tb$ and $tj$ selection
 $1\sigma$ bounds in the $\VO$-$S_R$ at detector level (cf.~text).
 \label{consistency_tests}}
\end{figure*}

In this short appendix we illustrate the stability of the coupling
sensitivities obtained in Sec.~\ref{pheno} against various variations of
the analysis setup.
Respective plots are shown in Fig.~\ref{consistency_tests} in the $\VO$--$S_R$
coupling plane containing the vector and one scalar direction for reference,
while stating that the observed effects are of the same size also along
the other coupling directions.  In detail, we have varied
(clockwise in Fig.~\ref{consistency_tests}, beginning top left)
the following:
\begin{itemize}

 \item the tentative systematic error $\delta_\text{sys}$ assumed
 in Eq.~\eqref{bin_error}, illustrating that the $s$~channel sensitivity
 driving the scalar bounds is still statistics dominated at the reference
 luminosity $\int\! L=\unit[100]{fb^{-1}}$ (i.e.~no visible effect from
 $\delta_\text{sys}$), while the $t$~channel sensitivity contributing to the
 $\VO$ bound does change slightly with the value of~$\delta_\text{sys}$.

 \item the parameter point from which the detector response matrix
 $\epsilon$ was inferred, namely once at the SM point and once at the
 reference point $S_R=0.1$ introduced in Sec.~\ref{pheno_d}.
 This generically tests the impact of populating the high energy bins
 on the selection efficiencies and
 impurities, while also including effects on the helicity-sensitive
 observables $p_T(\ell)$ and $\cos\theta_\ell$
 (cf.~Fig.~\ref{cc_contact_rp_d}).

 \item the higher order SM cross section normalization,
 which is nontrivial despite the normalized bin widths
 because the $K$~factors, though assumed global in the
 kinematic sense, are channel specific and hence potentially influence the
 mutual signal pollution at detector level.  For instance, the effect shown
 in Fig.~\ref{consistency_tests} is mainly along the scalar direction,
 whose sensitivity is driven by the $s$~channel, which in turn generally has
 larger $K$~factors than the $t$~channel. Therefore, going from NLO to NNLL
 slightly improves the sensitivity estimate on the scalars.

 \item the charged lepton flavor.  Note that the discrepancy
 thus obtained is mainly because of the different detection efficiencies of
 electrons and muons as modeled by~{\sc Delphes}
 version~3~\cite{Favereau2013}.
 For the discussion in Sec.~\ref{pheno}, both flavors were summed up.
\end{itemize}
In general, one can state that the systematic effects obtained by these
variations are marginal and particularly leave the conclusions of
Sec.~\ref{pheno_d} unaffected, with the largest uncertainty
actually coming from
the $b$~tagging which was already discussed in Sec.~\ref{pheno_d}.

\providecommand{\href}[2]{#2}
\renewcommand{\eprint}[1]{\href{http://arxiv.org/abs/#1}{{[arXiv:#1]}}}
\bibliography{references}

\end{document}